\documentclass[a4paper,11pt]{article}
\pdfoutput=1
\usepackage{jheppub}
\usepackage[T1]{fontenc} 
\usepackage{amsmath,amssymb,graphicx,bbm,mathrsfs,nicefrac}
\usepackage{tikz}
\usepackage{amsmath}
\usepackage{amssymb}
\usepackage{graphicx,textcomp,float,gensymb,wrapfig, enumitem,comment,dsfont,framed,slashed,appendix,wrapfig,xcolor}
\usepackage{ bbold }
\usepackage{bm}
\usepackage{filecontents}
\usepackage{braket} 
\usepackage{hhline}
\usepackage{ mathrsfs }
\usepackage{rotating}
\usepackage[small]{caption}
\usepackage{subcaption}
\usepackage{etoolbox,hyperref,makecell}
\usepackage{feynmp}
\usepackage{accents}
\usetikzlibrary{arrows.meta}
\usepackage{empheq}
\usepackage{filecontents}

\usepackage{multirow}

\usepackage{braket} 

\patchcmd{\abstract}{\null\vfil}{}{}{}
\newcommand{\g}{(g-2)_\mu}
\newcommand{\deltaaexp}{\Delta a_\mu^\mathrm{obs}}
\newcommand{\deltaa}{\Delta a_\mu}

\newcommand{\NBSM}{N_{\rm BSM}}
\newcommand{\MBSM}{M_{\rm BSM}}

\usepackage{pifont}

   \DeclareMathOperator{\gev}{GeV} \DeclareMathOperator{\tev}{TeV}

\newcommand{\beq}{\begin{equation}} \newcommand{\eeq}{\end{equation}}
\newcommand{\bea}{\begin{eqnarray}} \newcommand{\eea}{\end{eqnarray}}

\newcommand{\fref}[1]{Figure~\ref{#1}}
\newcommand{\eref}[1]{Eqn.~(\ref{#1})}
\newcommand{\sref}[1]{Section~\ref{#1}}
\newcommand{\tref}[1]{Table~\ref{#1}}

\newcommand{\be}{\begin{eqnarray}} \newcommand{\ee}{\end{eqnarray}}
\newcommand{\Eq}[1]{Eq.~(\ref{#1})}

\usepackage{hhline}
\usepackage{float}
\usepackage{caption}

\usepackage[utf8]{inputenc}

\preprint{FERMILAB-PUB-21-012-T}

\title{A No-Lose Theorem for Discovering the New Physics of $\g$ at Muon Colliders}

\author[a,b]{Rodolfo Capdevilla,}
\author[a]{David Curtin,}
\author[c,d]{Yonatan Kahn,}
\author[e,f]{Gordan Krnjaic}

\affiliation[a]{Department of Physics, University of Toronto, Canada}
\affiliation[b]{Perimeter Institute for Theoretical Physics, Waterloo, Ontario, Canada} 
\affiliation[c]{University of Illinois at Urbana-Champaign, Urbana, IL,  USA}
\affiliation[d]{Illinois Center for Advanced Studies of the Universe, University of Illinois at Urbana-Champaign, Urbana, IL, USA}
\affiliation[e]{Fermi National Accelerator Laboratory, Batavia, IL, USA}
\affiliation[f]{Kavli Institute for Cosmological Physics, University of Chicago, Chicago, IL, USA}

\emailAdd{rcapdevilla@perimeterinstitute.ca}
\emailAdd{dcurtin@physics.utoronto.ca}
\emailAdd{yfkahn@illinois.edu}
\emailAdd{krnjaicg@fnal.gov}

\date{\today}

\abstract{
We perform a model-exhaustive analysis of all possible beyond Standard Model (BSM) solutions to the $\g$ anomaly to study production of the associated new states at future muon colliders, and formulate a no-lose theorem for the discovery of new physics if the anomaly is confirmed and weakly coupled solutions below the GeV scale are excluded.
Our goal is to find the highest possible mass scale of new physics subject only to perturbative unitarity, and optionally the requirements of minimum flavour violation (MFV) and/or naturalness. 
We prove that a 3 TeV muon collider is guaranteed to discover all  BSM scenarios in which $\deltaa$ is generated by SM singlets with masses above $\sim \gev$; lighter singlets will be discovered by upcoming low-energy experiments. 
If new states with electroweak quantum numbers contribute to $\g$, the minimal requirements of perturbative unitarity guarantee  new charged states below  $\mathcal{O}(100 \tev)$, but this is strongly disfavoured by stringent constraints on charged lepton flavour violating (CLFV) decays.
Reasonable BSM theories that satisfy CLFV bounds by obeying Minimal Flavour Violation (MFV) and avoid generating two new hierarchy problems require the existence of at least one new charged state below $\sim 10 \tev$.
This strongly motivates the construction of high-energy muon colliders, which are guaranteed to discover new physics: either by producing these new charged states directly, or by setting a strong lower bound on their mass, which would empirically prove that the universe is fine-tuned and violates the assumptions of MFV while somehow not generating large CLFVs.
The former case is obviously the desired outcome, but the latter scenario would perhaps teach us even more about the universe by profoundly revising our understanding of naturalness, cosmological vacuum selection, and the SM flavour puzzle. 
}

\begin{document}

\maketitle

\newpage
\section{Introduction and Executive Summary}
\label{s.intro}

The magnetic moments of leptons have spurred the development of quantum field theory (QFT) and provided the most precise comparison between theory and experiment in the history of science. While the measured anomalous magnetic moment of the electron, ${(g-2)_e}$, agrees with the Standard Model (SM) prediction to better than one part per billion \cite{Aoyama:2017uqe}\footnote{While there is a  $\sim -2.5 \sigma$($+1.6 \sigma$) discrepancy between the theoretical prediction of  $(g-2)_e$ \cite{Aoyama:2017uqe} and the experimental measurement \cite{Parker191} (\cite{Morel:2020dww}) (with the difference between the two measurements arising from a discrepancy in the measurement of the fine-structure constant), in this paper we proceed under the assumption
that this is not evidence of new physics. See \textit{e.g.}
Refs. \cite{Marciano:2016yhf, Cesarotti:2018huy,Jana:2020pxx} for a discussion of possible BSM implications. 
}
 the analogous quantity for the muon, $\g$, has been discrepant between theory and experiment at a statistically significant level for nearly two decades \cite{Bennett_2006}. Since the muon mass is much closer to the QCD scale than the electron mass, hadronic contributions to $\g$ are an important part of the calculation, and a recent tour-de-force effort \cite{Aoyama:2020ynm} combining lattice calculations with quantities extracted from experimental data \cite{Aoyama:2012wk,Aoyama:2019ryr,Czarnecki:2002nt,Gnendiger:2013pva,Davier:2017zfy,Keshavarzi:2018mgv,Colangelo:2018mtw,Hoferichter:2019gzf,Davier:2019can,Keshavarzi:2019abf,Kurz:2014wya,Melnikov:2003xd,Masjuan:2017tvw,Colangelo:2017fiz,Hoferichter:2018kwz,Gerardin:2019vio,Bijnens:2019ghy,Colangelo:2019uex,Blum:2019ugy,Colangelo:2014qya} has recently confirmed the discrepancy to be
\begin{equation}
\label{delta-amu}
\deltaaexp = a^{\rm exp}_\mu - a^{\rm theory}_\mu = (2.79 \pm 0.76) \times 10^{-9}~,
\end{equation}
with a statistical significance of $3.7\sigma$.\footnote{Some lattice calculations \cite{Borsanyi:2020mff} find no discrepancy with the measured $\g$, but are discrepant with $R$-ratio measurements. The source of this tension may lie in electroweak precision observables \cite{Crivellin:2020zul,Passera:2008jk,Keshavarzi:2020bfy}, preserving the $\g$ anomaly.} The Muon $g-2$ experiment at Fermilab \cite{fienberg2019status} is expected to surpass the statistics of the previous Brookhaven experiment in the coming months, which would further reduce the uncertainty on the experimental result. If the discrepancy persists after this measurement (and if it is also confirmed by JPARC~\cite{Sato:2017sdn}) 
it would be the first terrestrial discovery of physics beyond the Standard Model (BSM).

Whenever a discrepancy is found in a low-energy precision measurement, it is imperative to understand the implications for other experiments, both to confirm the anomaly and because such a discrepancy could point to the existence of new particles at higher but accessible energy scales. Direct production and observation of new states is, after all, the gold standard for discovering  new physics.
In the long history of the $\g$ anomaly, many such studies were performed. Examples include investigations of complete theories like supersymmetry \cite{Cho:1999km,Cho:2011rk,Kowalska:2015zja}; minimal low-energy scenarios involving only very light states \cite{Pospelov_2009,Chen_2017}; or various simplified model approaches to study the generation of $\g$ at higher energy scales 
\cite{Dermisek:2013gta,Freitas:2014pua,Queiroz:2014zfa,Keus:2017ioh}, which can include additional considerations like the existence of a viable dark matter (DM) candidate \cite{Cox:2018qyi,Calibbi:2018rzv,Agrawal:2014ufa,Kowalska:2017iqv,Barducci:2018esg,Kowalska:2020zve,Jana:2020joi}.

However, in all these past investigations, a simple question was left unanswered: \emph{What is the highest mass that new particles could have while still generating the measured BSM contribution to $\g$?}
In this paper, we answer that crucial question in a precise yet model-exhaustive way, relying only on gauge invariance and perturbative unitarity, and optionally on well-defined tuning or flavour considerations, without making any detailed assumptions about the complete underlying theory.

We provide a detailed description of our model-exhaustive 
approach in Section~\ref{s.modelindependent}, but it can be briefly summarized as follows. 
We assume that one-loop effects involving BSM states are responsible for the anomaly,\footnote{We work under the assumption that the $\g$ anomaly is due to new physics which genuinely affects the value of $g_\mu$ in vacuum, rather than its measurement being sensitive to other BSM effects on the muon spin, for example ultralight scalar dark matter \cite{Janish:2020knz}. The latter case is also eminently testable in upcoming experiments.}
 since scenarios where new contributions only appear at higher loop order require a lower BSM mass scale to generate the required new contribution.
We can, thus, organize all possible one-loop BSM contributions to $\deltaa$ into two classes: 
\begin{itemize}
\item {\bf Singlet Scenarios:} in which each BSM $\g$ contribution only involves a muon and a new SM singlet boson that couples to the muon (analyzed in Section~\ref{s.singletmodels}); 

\item{\bf Electroweak (EW) Scenarios:} 
in which new states with EW quantum numbers contribute to $\g$  (analyzed in Section~\ref{s.ewmodels}). 
\end{itemize}
Singlet Scenarios generate $\deltaa$ contributions proportional to  $m_\mu y_\mu v/M_{\rm BSM}^2$, where $y_\mu \sim 10^{-3}$ is the small SM muon Yukawa coupling. 
Electroweak Scenarios can generate the largest possible $\g$ contributions without the additional $y_\mu$ suppression. In particular, we carefully study two simplified models denoted SSF and FFS with new scalars and fermions that yield the \emph{largest possible BSM mass scale} able to account for the anomaly.
Careful analysis of these two EW Scenarios allows us to derive our model-exhaustive upper bound on BSM particle masses for scenarios that resolve the $\g$ anomaly. 
We also account for the possibility of many new states contributing to $\deltaa$ by considering $N_{\rm BSM} \geq 1$ copies of each BSM model being present simultaneously, allowing us to understand how the maximum possible BSM mass scales with BSM state multiplicity in each case. 

We find that if $\deltaaexp$  is generated in a Singlet Scenario, the maximum mass of the BSM singlet particle(s) is 3 TeV regardless of BSM multiplicity $N_{\rm BSM}$.
For EW Scenarios, we find that there must always be at least one new charged state lighter than the following upper bound:
\begin{equation}
\label{e.MchargedmaxXresultpreview}
M^\mathrm{max, X}_\mathrm{BSM, charged} 
\approx
\left( \frac{2.8 \times 10^{-9}}{\deltaaexp} \right)^{\frac{1}{2}} \times
\left\{
\begin{array}{rcl}
(100 \tev)  \ N_{\rm BSM}^{1/2} &   \mathrm{for}  & X = \mbox{(unitarity*)}
\\ \\ 
(20 \tev)  \ N_{\rm BSM}^{1/2} &   \mathrm{for} & X = \mbox{(unitarity+MFV)} 
\\ \\ 
(20 \tev)  \ N_{\rm BSM}^{1/6} & \mathrm{for}    & X =  \mbox{(unitarity+naturalness*)}
\\ \\ 
(9 \tev)  \ N_{\rm BSM}^{1/6} &  \mathrm{for} & X =  \mbox{(unitarity+naturalness+MFV)}
\end{array}
\right.
\end{equation}
where this upper bound is evaluated under four assumptions that the BSM solution to the $\g$ anomaly must satisfy: perturbative unitarity only; unitarity + Minimal Flavour Violation (see e.g.~\cite{Buras:2003jf, Csaki:2011ge}); unitarity + naturalness (specifically, avoiding two new hierarchy problems); and unitarity + naturalness + MFV. 
The unitarity-only bound represents the very upper limit of what is possible within QFT, but realizing such high masses requires severe alignment tuning or another unknown mechanism to avoid stringent constraints from charged lepton flavour-violating (CLFV) decays~\cite{Calibbi:2017uvl, Aubert:2009ag}.
We have therefore marked  every scenario without MFV with a star (*) above, to indicate additional tuning or unknown flavour mechanisms that have to also be present.

Our results have profound implications for the physics motivation of \emph{future muon colliders} (MuC), which have recently garnered renewed attention as an appealing possibility for the future of the high energy physics program \cite{Buttazzo:2020uzc,Costantini:2020stv,Delahaye:2019omf,Buttazzo:2018qqp,han2020wimps,Bartosik_2020,Yin:2020afe,Huang:2021nkl}. 
Muon colliders still face significant technical challenges \cite{Delahaye:2019omf}, but are in many ways ideal BSM discovery machines: compared to electron colliders, the suppressed synchrotron radiation loss might make it easier to reach high energies in excess of 10 TeV; unlike in proton collisions, the entire center-of-mass energy is available for the pair-production of new charged particles with masses up to $m \sim \sqrt{s}/2$ \cite{Delahaye:2019omf}; and finally they collide the actual particles that exhibit the $\g$ anomaly.

These features 
enable us to formulate a \emph{no-lose theorem for a future muon collider program.}
We presented our first investigation of this issue in~\cite{Capdevilla:2020qel}. Here, we supply important additional details, perform detailed muon collider studies, and generalize our original derivation to include crucial flavour considerations and present all possible EW Scenarios that maximize BSM masses, all of which reinforce the robustness of our  conclusions. Since our original study appeared, there have also been additional investigations of indirect probes of $\g$ at future muon colliders~\cite{Buttazzo:2020eyl, Yin:2020afe}. The results of these studies, despite their different technical approach, agree with our overall conclusions and strengthen them in important ways, as we explain below.

We give a detailed description of this no-lose theorem in Section~\ref{s.implications}, but its most important final points are as follows, broken down in chronological progression:
\begin{enumerate}

\item \textbf{Present day confirmation:} 

Assume the $\g$ anomaly is real.

\item \textbf{Discover or falsify low-scale Singlet Scenarios $\mathbf{\lesssim}$ GeV:}

 If Singlet Scenarios with BSM masses below $\sim \gev$ generate the required $\deltaaexp$ contribution~\cite{Pospelov_2009}, multiple fixed-target and $B$-factory experiments are projected to discover new physics in the coming decade \cite{Gninenko_2015,Chen_2017,Kahn_2018,kesson2018light,Berlin_2019,Tsai:2019mtm,Ballett_2019,Mohlabeng_2019,Krnjaic_2020}. 

\item \textbf{Discover or falsify all Singlet Scenarios $\mathbf{\lesssim}$ TeV:} 

If fixed-target experiments do not discover new BSM singlets that account for $\deltaaexp$, a 3 TeV muon collider with $1~\mathrm{ab}^{-1}$ would be guaranteed to directly discover these singlets if they are heavier than $\sim 10 \gev$.

Even a lower-energy machine can be useful: a  215 GeV  muon collider with $0.4~\mathrm{ab}^{-1}$ could directly observe singlets as light as 2 GeV under the conservative assumptions of our inclusive analysis, while indirectly observing the effects of the singlets for all allowed masses via Bhabha scattering. 

Importantly, for singlet solutions to the $\g$ anomaly, only the muon collider is guaranteed to discover these signals since the only required new coupling is to the muon.

\item   \textbf{Discover non-pathological Electroweak Scenarios ($\mathbf\lesssim$  10 TeV):}

If TeV-scale muon colliders do not discover new physics, the $\g$ anomaly \emph{must} be generated by EW Scenarios. 
In that case, all of our results indicate that in most reasonably motivated scenarios, the mass of new charged states cannot be higher than few $\times$ 10 TeV. However, such high masses are only realized by the most extreme boundary cases we consider. 
Therefore, a muon collider with $\sqrt{s} \sim 10 \tev$ is highly motivated, since it will have excellent coverage for EW Scenarios in most of their reasonable parameter space.

A very strong statement can be made for future muon colliders with $\sqrt{s} \sim 30 \tev$: such a machine can discover via pair production of heavy new charged states \emph{all} EW Scenarios that avoid CLFV bounds by satisfying MFV and avoid generating two new hierarchy problems, with $N_{\rm BSM}  \lesssim 10$.

\item \textbf{Unitarity Ceiling ($\mathbf\lesssim$  100 TeV):}

Even if such a high energy muon collider does not produce new BSM states directly, the recent investigations by~\cite{Buttazzo:2020eyl, Yin:2020afe} show that a 30 TeV machine would detect deviations in $\mu^+ \mu^- \to h \gamma$, which probes the same effective operator generating $\g$ at lower energies. This would provide high-energy confirmation of the presence of new physics.

In that case,  our results guarantee the presence of new states below $\sim 100 \tev$ by perturbative unitarity, and the lack of direct BSM particle production at $\sqrt{s} \sim 30 \tev$ will prove that the universe violates MFV and/or is highly fine-tuned to stabilize the Higgs mass and muon mass, all while suppressing CLFV processes. 
\end{enumerate}
Even the most pessimistic final case would profoundly reshape our understanding of the universe by providing new information about the nature of fine-tuning, flavour and cosmological vacuum selection.
If no new states are discovered at 30 TeV, the renewed confirmation of the $\g$ anomaly at these higher energies and the associated guaranteed presence of new states below the unitarity bound with deep implications for naturalness and flavour means finding the solution to all these puzzles will surely provide impetus for pushing our knowledge of the energy frontier to even greater heights.

If the $\g$ anomaly is confirmed, our analysis and the results of~\cite{Buttazzo:2020eyl, Yin:2020afe} show that finding the origin of this anomaly should be regarded as one of the most important physics motivations for an entire muon collider \emph{program}. Indeed, a  series of colliders with energies from the test-bed-scale $\mathcal{O}(100 \gev)$ to the far more ambitious but still imaginable $\mathcal{O}(10 \tev)$ scale and beyond has excellent prospects to discover the new particles necessary to explain this mystery.  Regardless of what these direct searches find, each will make invaluable contributions to allow us to understand the precise nature of the new physics that must  be present. Therefore, this truly is a no-lose theorem for the discovery of new physics, the greatest imaginable motivation for a heroic undertaking like the construction of a revolutionary new type of particle collider.\footnote{While we argue in this work that muon colliders are sufficient for discovery, they are not the only such probe: proton-proton colliders, electron linear colliders, and even photon colliders have strong potential for observing new TeV-scale EW states. That said, muon-specific singlets will likely be challenging to observe at any collider not utilizing muon beams, and discovering EW-charged states at the 10 TeV scale may not be as straightforward with a 100 TeV $pp$ collider due to PDF factors and a noisier detector environment~\cite{Costantini:2020stv}, while reaching such energies could be challenging in an electron machine. Of course, all these cases deserve a dedicated analysis.}

We now present the details necessary to fill out this argument. Our model-exhaustive approach is explained in Section~\ref{s.modelindependent}; Singlet Scenarios and EW Scenarios are analyzed in detail in Sections~\ref{s.singletmodels} and~\ref{s.ewmodels}; the implications for a future muon collider program and the no-lose theorem for discovery of new physics is fully outlined in Section~\ref{s.implications}.

\section{Model-Exhaustive Approach}
\label{s.modelindependent}

In this paper, we aim to address a simple question: how could we discover \emph{all possible BSM solutions} to the $\g$ anomaly? 
Specifically, how could we \emph{directly} discover at least some of the BSM particles that play a role in generating $\deltaaexp$? 
The bewildering plethora of possible BSM solutions to the anomaly make answering this question very challenging; by construction, our answer cannot depend on the particular choice of BSM model.

Very light, weakly coupled solutions to $\g$ near or below the scale of the muon mass will be exhaustively tested by low energy experiments, and we focus on all other BSM possibilities. 
In that case,
at the low energies at which the $\g$ measurement is performed, we can parameterize the deviation from the SM expectation as a BSM contribution to the anomalous magnetic moment operator. Taking into account electroweak gauge invariance, in two-component fermion notation this is 
 \begin{equation}
 \label{leff}
 {\cal L}_{\rm eff} =  C_{\rm eff}   \frac{ v}{M^2}  ({\mu}_L  \sigma^{\nu \rho}  \mu^c )  F_{\nu \rho}   + {\rm h.c.}~,
 \end{equation}
where $\mu_L$ and $\mu^c$ are the two-component muon fields, $v = 246$ GeV is the SM Higgs 
 vacuum expectation value (VEV), 
 and $C_{\rm eff}$ is a constant. The factor of $v$ arises from the fact that coupling left- and right-handed muon fields requires a Higgs insertion, so the electroweak-symmetric operator is dimension-6, $H^\dagger L  \sigma^{\nu \rho} \mu^c F_{\nu \rho}$,
 and thus must be suppressed by two powers of a mass scale $1/M^2$.
Unfortunately, such \emph{model-independent} EFT analyses are limited to \emph{indirect signatures} of the new physics, making this approach unsuitable to answer the question of how to directly discover the new states.

\begin{figure}
\begin{center}
\includegraphics[width=0.99\textwidth]{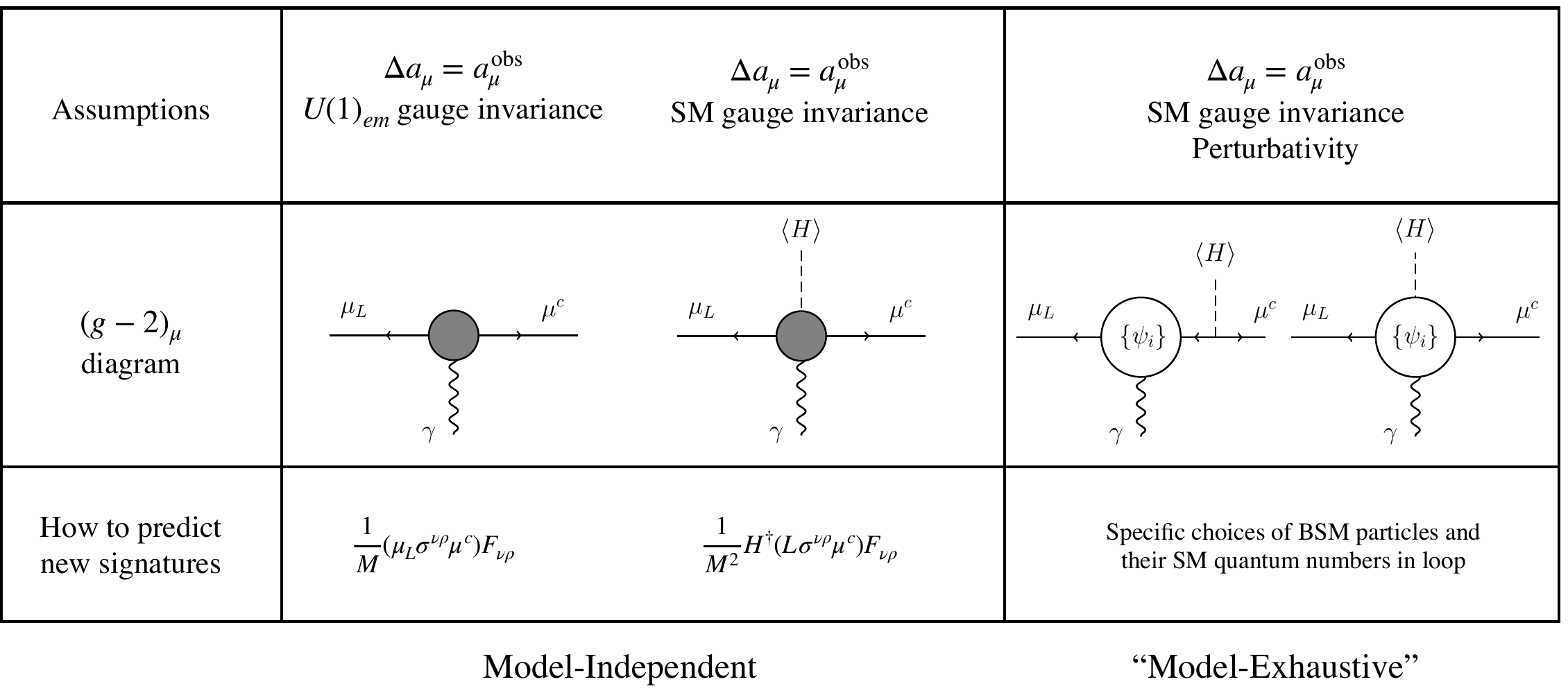}
\end{center}
\caption{
The philosophy of our ``model-exhaustive'' analysis. Traditional model-independent analyses express the new physics contribution to $\g$ as a non-renormalizable operator, either in the low-energy theory after EW symmetry breaking (left) or in the full SM gauge invariant formulation (middle). 
This makes no assumptions about the new physics but is limited to indirect signatures of the new physics produced by the same operator. 
Since we want to probe direct signatures of the BSM physics which solves the $\g$ anomaly, we add the single assumption of perturbativity to the traditional model-independent analysis, which resolves the new $\deltaa$ contributions into explicit loop diagrams of new states $\{\psi_i \}$ carrying specific SM quantum numbers (right). 
If the Higgs insertion lies on the external muon, $\Delta a_\mu$ is suppressed by $y_\mu$, while $\deltaa$ can be significantly enhanced if the Higgs couples to new particles in the loop.
By exhaustively analyzing all possible choices of new states, we can derive predictions for direct signatures that are as universal as the traditional model-independent predictions for indirect signatures.
}
\label{f.cartoon}
\end{figure}

To study high-energy direct signatures of new physics, we instead adopt a ``model-exhaustive'' approach. As illustrated in \fref{f.cartoon}, this simply involves adding the assumption that the new physics is perturbative, which resolves the new $\g$ contributions into individual loop diagrams involving various possible BSM particles in different SM gauge representations. In principle, if all possibilities were considered, one could study direct signatures of new physics in the same full generality that model-independent EFT analyses afford for indirect signatures.\footnote{While our analysis is formally limited to perturbative BSM solutions of the $\g$ anomaly, our results nonetheless end up parametrically covering the case of strongly coupled BSM scenarios as well, as we argue in \sref{s.mBSMupperbound}.}

 The idea of a model-exhaustive analysis is not, of course, a new one. However, the challenge lies in systematically covering all possibilities of BSM particles, or at least those possibilities relevant to answering a specific phenomenological question. We now explain how to perform this analysis for the $\g$ anomaly, with an eye towards direct signatures at future muon colliders.\footnote{For a philosophically similar approach to the Hierarchy Problem, see~\cite{Curtin:2015bka}.}

We limit ourselves to those perturbative BSM scenarios where the required $\deltaa$ is generated at \emph{one-loop order}. 
There are certainly many possibilities for BSM physics that solves the $\g$ puzzle by generating only new higher-loop contributions \cite{Marciano:2016yhf,Chang:2000ii,Cheung:2001hz}
 (e.g. from $\mathbb{Z}_2$ preserving interactions with the muon),
but such models necessarily require lower mass scales, which must be accessible via pair production at the collider energies we consider here. We therefore omit a detailed discussion of these scenarios without loss of  generality. However, we note that even if such signals were to be ultimately elusive to direct searches due to complicated, high-background decay channels, a future muon collider would still 
detect their presence through enhanced $\mu \mu\to\gamma h$ production~\cite{Buttazzo:2020uzc} and $\mu \mu \to \mu \mu$ Bhabha scattering~\cite{Capdevilla:2020qel}.

\begin{figure}
\begin{center}
\includegraphics[width=0.9\textwidth]{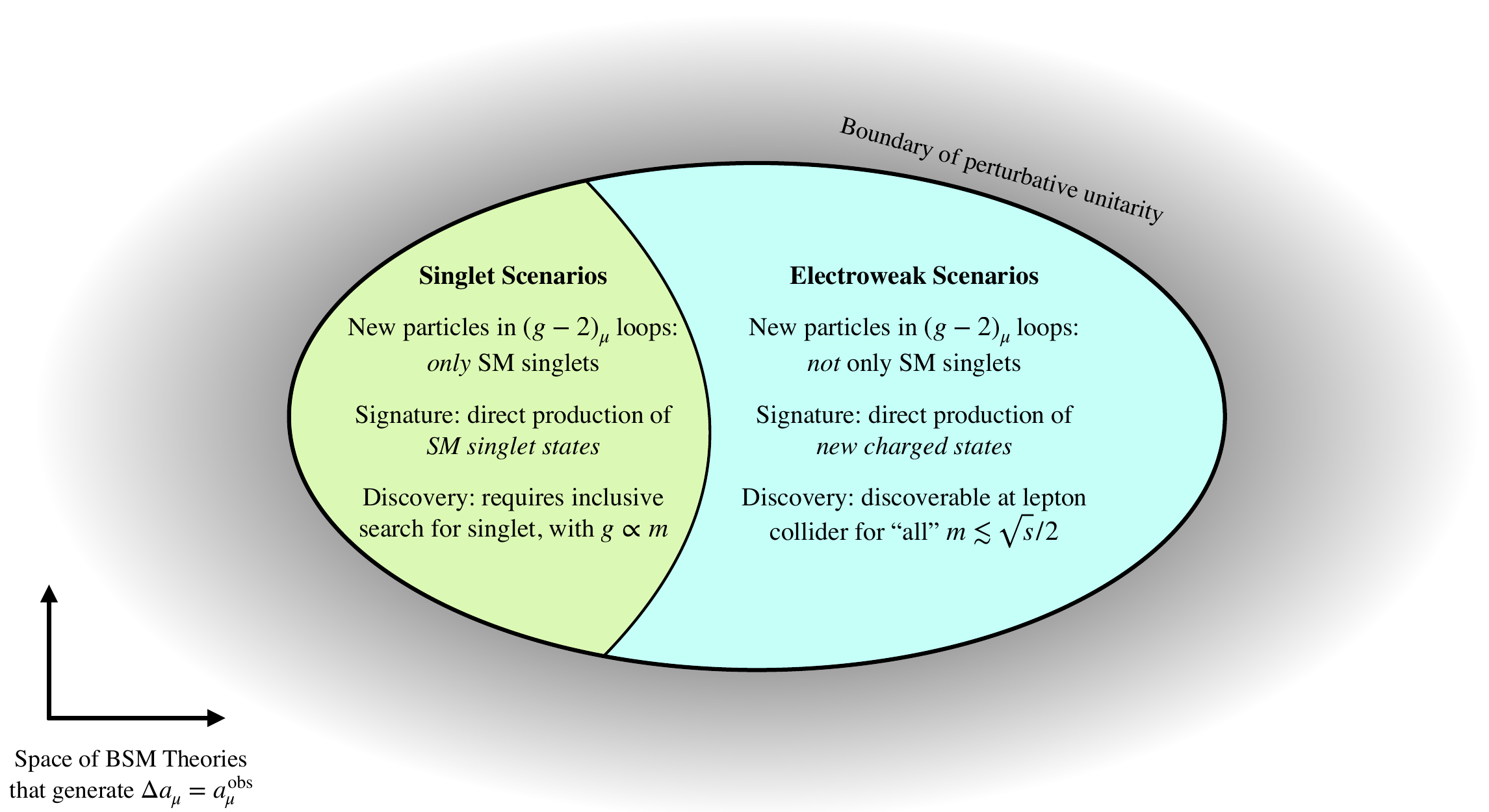}
\end{center}
\caption{Schematic representation of the model-exhaustive space of BSM theories that can solve the $\g$ anomaly, and our mutually exclusive and collectively exhaustive categorization into Singlet Scenarios and Electroweak Scenarios. For these two classes of theories, the phenomenological questions are distinct. To understand how to discover Singlet Scenarios, we have to not only find the heaviest possible mass of the singlet(s), but also how to discover this singlet for all possible masses, since its phenomenology depends on its stability and decay mode, and lighter singlets have weaker coupling. Electroweak Scenarios predict new charged states, and since those have to produce visible final states in a collider and are efficiently produced at lepton colliders for $m \lesssim \sqrt{s}/2$, we only have to find the maximum mass the lightest new charged state in the BSM theory can have. (We limit ourselves to scenarios that generate $\deltaaexp$ at one-loop, since higher-loop solutions have lower BSM mass scales.)}
\label{f.modelspace}
\end{figure}

Our exhaustive coverage of candidate BSM theories for $\g$ is informed by the 
characteristic 
experimental signatures available in each class of scenarios. For this reason, we divide up the space of possibilities into two classes, illustrated schematically in \fref{f.modelspace}:
\begin{enumerate}
\item \emph{\bf Singlet Scenarios}: defined as BSM solutions to the $\g$ anomaly in which the only new particles in the $\g$ loop are SM gauge singlets. 
This selects the first type of diagram in \fref{f.cartoon} (right box) with the Higgs VEV insertion on the external muon leg, such that the chirality flip and the Higgs coupling both come from the muon, and hence $\deltaa \propto m_\mu y_\mu v /M_{\rm BSM}^2$. 
Their singlet nature means these particles could be very light ($\lesssim \gev$) while evading present constraints~\cite{Pospelov_2009}, but they could also be much heavier. 

For Singlet Scenarios, our task is to find the largest possible mass these singlets could have, and determine how a muon collider could produce and observe them for all possible masses, regardless of how or if they decay in the detector.

\item \emph{\bf  Electroweak (EW) Scenarios}:
defined as all BSM solutions that are not Singlet Scenarios. 
This necessarily implies that $\g$ receives contributions from loops involving BSM states with EW quantum numbers, which in turn implies the existence of \emph{new heavy charged states} with masses $\gtrsim 100 \gev$ to evade LEP bounds. These charged particles could contribute to $\g$ directly, or be  new states that must exist due to gauge invariance.
The new charged states will be our focus, since any lepton collider with $\sqrt{s} \gtrsim 2 m$ can directly pair-produce such states of mass $m$, and as they have to either be detector-stable or decay into charged final states, they should be discoverable in a clean detector environment regardless of their detailed phenomenology. 
For EW Scenarios, our task is therefore to find the largest possible mass that the \emph{new charged states} could have.

EW Scenarios can generate diagrams of both types shown in \fref{f.cartoon} (right). Of particular interest is the second type where the Higgs insertion and chirality flip belong to BSM particles in the loop, which would give  $\deltaa \propto m_\mu g_{\rm BSM} v /M_{\rm BSM}^2$ without the suppression of the small muon Yukawa. This can result in much heavier BSM mass scales than Singlet Scenarios. 

\end{enumerate}
If we examine both of these possibilities exhaustively, we will have completed our model-exhaustive analysis.

Singlet Scenarios are relatively straightforward to analyze. In the next \sref{s.singletintro} we define simplified models that cover all possibilities for this singlet. These models have few parameters, and the parameter space can be explored in full generality. 
Electroweak Scenarios present more of a challenge. To find the minimum muon collider energy that would guarantee direct production and discovery of at least one BSM charged state, we have to find the heaviest 
possible charged state consistent with resolving the anomaly. This amounts to
finding the  following quantity:
\begin{equation}
\label{e.Mchargedmaxgeneral}
M^\mathrm{max}_\mathrm{BSM, charged}  \ \ \ 
\equiv \ \ \ 
\max_{
\begin{array}{c} 
\scriptstyle \mathrm{BSM\ theory\ space} 
\\ 
\scriptstyle \deltaa = \deltaaexp
\end{array}
} \ \  \left\{ \ \  \min_{\tiny i  \ \in \  \mathrm{BSM\ spectrum}} \left( m_{\rm charged}^{(i)}   \right)  \ \ \right\} \ .
\end{equation}
This can be understood in the following algorithmic way. The outer maximization scans over \emph{all possible BSM theories and possible values of their parameters} that give $\Delta a_\mu = \Delta a_\mu^{\rm obs}$ while satisfying the constraints of perturbative unitarity.
For each specific theory and given values of its parameters, we find the lightest new charged state (inner bracket) and add it to a list. The outer maximization then picks the maximum value from this list, giving the heaviest possible mass of the lightest new charged state that must exist to resolve the $\g$ anomaly, and therefore the minimum energy of a muon collider that is guaranteed to produce these particles.
The difficulty obviously arises in performing the first theory space maximization.
 In \sref{s.EWmodeldefinition} we explain how this maximization can be performed, allowing our model-exhaustive analysis to determine the heaviest possible masses of new charged states with the generality of a traditional model-independent analysis.

\subsection{Singlet Scenarios}
\label{s.singletintro}

\begin{figure}
\center
\includegraphics[width=4.5in]{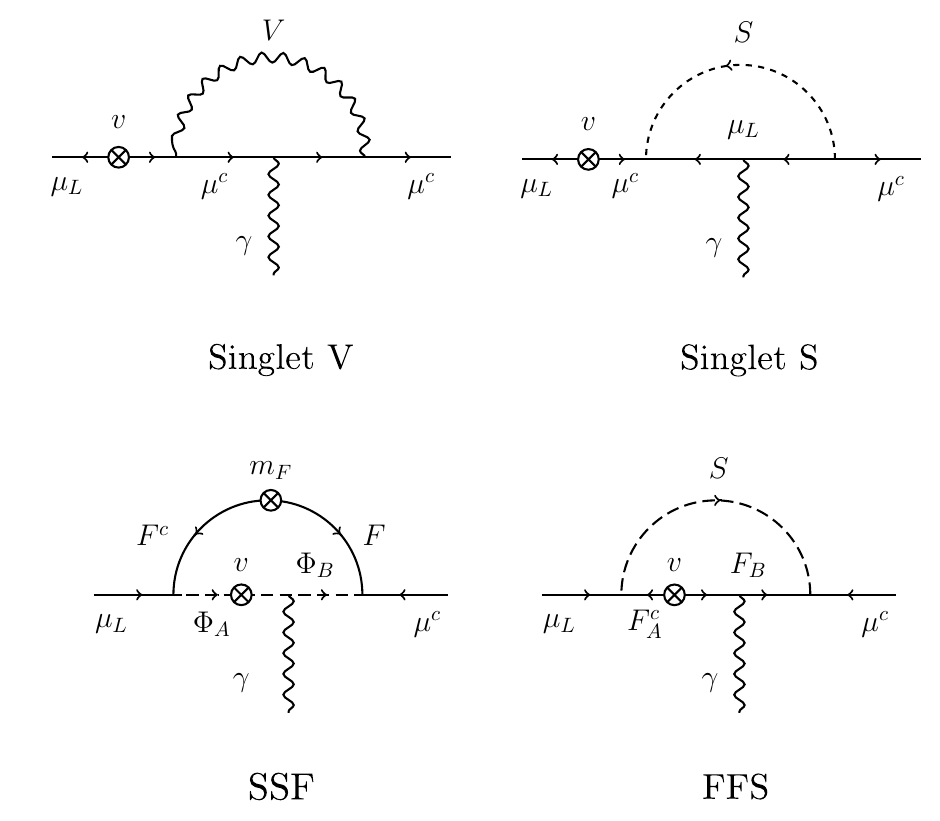}
\caption{ Representative 1-loop contributions to $\g$ in the simplified models we consider. Top row: Singlet Scenarios with a SM neutral  vector $V$ or scalar $S$ that couple to the muon. Note that the Higgs VEV on the muon line gives both the chirality flip and the EW breaking insertions in these models.  Bottom left: EW Scenario of SSF type, with one BSM fermion and two BSM scalars that mix via a Higgs insertion. Bottom right: EW Scenario of FFS type, with one BSM scalar and two BSM fermions that mix via a Higgs insertion.
\label{f.feynmang-2}
}
\end{figure}

In this case, SM singlets that could be below the GeV scale (or much heavier) generate the new one-loop contributions to $\g$. 
The singlet could either be a scalar, vector, or fermion.
Our focus will be the case of a new real scalar $S$ or vector $V$.
The relevant Lagrangian terms for the real scalar case are
\begin{eqnarray}
\label{e.LS}
\mathcal{L}_\mathrm{S} &\supset&
- \left( g_S S  \mu_L \mu^c + {\rm h.c.}\right)- \frac{1}{2} m_S^2 S^2 
\ .
\end{eqnarray}
Note that the Yukawa coupling of the real scalar to muons $g_S$ is not gauge invariant. This implies that either the interaction arises from the non-renormalizable operator $\frac{1}{\Lambda} c_S \mu_L \mu^c H S,$ in which case $g_S \propto v/(\sqrt{2} \Lambda)$, or the interaction comes from a singlet-Higgs mixing, in which case $g_S \sim  y_\mu \sin\theta,$ where $\theta$ is the mixing angle.
We briefly discuss the consequences of consistent embedding in the full electroweak theory in \sref{s.singletmodels}.
For the vector case, the relevant Lagrangian terms are
\begin{eqnarray}
\label{e.LV}
\mathcal{L}_\mathrm{V} &\supset&
 g_V V_\alpha ( \mu_L^\dagger \bar \sigma^\alpha \mu_L +
 \mu^{c \, \dagger} \bar \sigma^\alpha \mu^c)
 + 
 \frac{m_V^2}{2} V_\alpha V^\alpha \ .
\end{eqnarray}
  These two scenarios are representative of muophilic new gauge forces or scalars that have been extensively studied in the literature \cite{Bauer:2018onh,Ilten:2018crw,Batell:2017kty,Chen_2017} and
  their contributions to $\g$ are shown in  \fref{f.feynmang-2}.
  
As discussed in \sref{s.singletmodels}, the only viable anomaly-free vector model is gauged $L_\mu - L_\tau$, which   can still resolve $\g$ for $m_V \in (10 \, {\rm MeV}, 2 m_\mu)$ \cite{Escudero:2019gzq,Krnjaic:2019rsv}.
Bounds on muon-philic singlet scalars are more model dependent and can, in principle, resolve $\g$ with any mass   between the MeV scale and the perturbative unitarity limit $\sim$ few TeV.
For both scalars and vectors, the lower limit is set by cosmological constraints, most importantly bounds on $\Delta N_{\rm eff}$, the effective number of relativistic species
  at big bang nucleosynthesis \cite{Krnjaic:2019rsv, Nollett:2014lwa}.
Thus, the scalar Singlet Scenario will be of most interest to us, but we keep the vector case in our discussions for completeness since the analyses are very similar.

These Singlet Scenarios are the most minimal BSM solutions to the $\g$ anomaly, featuring new particles required only to couple to the muon and no other SM particles. 
Consequently, muon colliders and muon-beam fixed-target experiments might be the only guaranteed way to probe all Singlet Scenarios.
Given that fixed-target experiments and $B$-factories will exhaustively probe Singlet Scenarios with  masses below $\sim \gev$~\cite{Gninenko_2015,Chen_2017,Kahn_2018,kesson2018light,Berlin_2019,Tsai:2019mtm,Ballett_2019,Mohlabeng_2019,Krnjaic_2020}, we will particularly focus on Singlet Scenarios above the GeV scale in our muon collider physics analyses.

Of course, it is possible that more than one new degree of freedom contributes to $\g$. We account for this possibility by considering $N_{\rm BSM} \geq 1$ copies of each SM Singlet Scenario in Eqns.~(\ref{e.LS}) or (\ref{e.LV}), and analyzing how the various higher-energy signatures scale with BSM multiplicity. Note that the assumption that all $N_{\rm BSM}$ copies of the simplified model have equal masses and couplings is the most pessimistic one with regards to high-energy signatures, since non-degenerate masses and couplings always lead to larger signatures due to the non-linearity of the associated cross sections and amplitudes. If couplings or masses are highly unequal, the phenomenology will be dominated by just a few new states. Considering degenerate $N_{\rm BSM} \geq 1$ copies therefore covers the signature space of possibilities.

Finally we note that, in principle, one could also consider the case of a neutral fermion $N$ contributing to $\g$. This would essentially be a right-handed-neutrino-type scenario (see e.g.~\cite{Drewes:2013gca} for a review), where the new $\g$ contribution consists of a loop of a $W$ boson and the neutral $N$ that mixes with the muon neutrino. 
However, in the presence of a unitary neutrino mixing matrix, such contributions would cancel up to 
corrections of order $\sim (m_\nu/m_W)^2$, which are inadequate to explain $\Delta a_\mu^{\rm obs}$.
We therefore restrict our focus to scalar and vector singlets.

\subsection{Electroweak Scenarios}
\label{s.EWmodeldefinition}

We now move on to discuss  the most general class of BSM solutions to the $\g$ anomaly, \emph{Electroweak Scenarios}.
This includes an overwhelmingly large number of possibilities, but fortunately, we do not need to study all of them. To perform the maximization over all of BSM theory space in \eref{e.Mchargedmaxgeneral}, we merely need to study those models which are guaranteed to give the \emph{largest possible} BSM mass scales. This will be sufficient to model-exhaustively determine the heaviest possible mass for new charged states.

Which EW Scenarios maximize the BSM mass scale? Consider the most general new one-loop diagrams that could contribute to $\g$. To make sure the relevant masses and couplings are maximally unconstrained, we consider the cases where \emph{all} fields in the loop are BSM fields. Furthermore,  the chirality flip and the Higgs VEV insertion necessary to generate \Eq{leff} should both come from these BSM fields to avoid additional suppression by the small muon Yukawa. 
The minimal ingredients are therefore: 
\begin{enumerate}
\item at least 3 BSM fields, either two bosons and one fermion or one boson and two fermions;
\item a pair of these fields undergo mass-mixing with each other via a Higgs coupling after electroweak symmetry breaking (EWSB);
\item all new fermions are vector-like under the SM to maximize allowed masses and avoid constraints on new 4th generation fermions \cite{Kumar:2015tna};
\item 
no VEVs for any new scalars with EW charge. Since we are primarily interested in BSM states
above the TeV scale, any new VEVs that break electroweak symmetry will exceed the measured
value $v \approx 246$ GeV for perturbative scalar self couplings. 
\end{enumerate}

As in our analysis for Singlet Scenarios, 
our default focus is on the most experimentally pessimistic case in which these new BSM states only couple to the SM through their muonic (and gauge) interactions. 
We find that scenarios with new vectors generate smaller $\deltaa$ contributions than the analogous scenario with a new scalar, and likewise for Majorana fermions or real scalars. 
Since this results in a lower BSM mass scale that would be easier to probe, we focus on EW Scenarios with new complex scalars and vector-like fermions only. 
This leaves just two classes of models, which we label  SSF and FFS by their field content. 

The \textbf{SSF simplified model} is defined by  two complex scalars $\Phi_A, \Phi_B$ in $SU(2)_L$ representations $R^A, R^B$ with hypercharges $Y^A, Y^B$ and a single vector-like fermion pair $F (F^c)$ in $SU(2)_L$ representation $R$ ($
\bar R$) with hypercharge $Y$ ($-Y$):
\begin{eqnarray}
\label{e.SSF}
\mathcal{L}_\mathrm{SSF}  &\supset& -y_1 F^c L_{(\mu)}  \Phi_A^*  - y_2 F \mu^c  \Phi_B 
- \kappa H \Phi_A^* \Phi_B 
\nonumber \\  &&
- m^2_A |\Phi_A|^2 - m^2_B |\Phi_B|^2 - m_F F F^c
 + {\rm h.c.} \ .
\end{eqnarray}
Here $y_1, y_2$ are new Yukawa couplings and $\kappa$ is a trilinear coupling with dimensions of mass.
$L_{(\mu)} = (\nu_L, \mu_L)$ and $\mu^c$ are the two 2-component 
second-generation SM lepton fields, and $H$ is the Higgs doublet. 
A typical SSF contribution to $\g$ is shown in \fref{f.feynmang-2}~(b). Note that the chirality flip comes from the heavy vector-like fermion $F$ while the Higgs VEV insertion arises due to mixing of the new scalars.

The \textbf{FFS simplified model} is analogously defined but reverses the role of fermions and scalars, featuring two vector-like fermion pairs $F_A, F_B$ ($F_A^c, F_B^c$)
in $SU(2)_L$ representations $R^A, R^B$ ($\bar R^A, \bar R_A$) with hypercharges $Y^A, Y^B$ $(-Y_A, -Y_B)$ and a single complex scalar $S$ in $SU(2)_L$ representation $R$ with hypercharge $Y$: 
\begin{eqnarray}
\label{e.FFS}
\mathcal{L}_\mathrm{FFS} &\supset& -y_1 F_A^c L_{(\mu)}   \Phi^*  - y_2 F_B \mu^c  \Phi 
- y_{12} H F_A^c F_B - y_{12}^\prime H^\dagger F_A F_B^c
\nonumber \\ 
&& - m_A F_A F_A^c - m_B F_B F_B^c - m_S^2 |\Phi|^2
 + {\rm h.c.}
 \end{eqnarray}
There are now two renormalizable Yukawa couplings $y_{12}, y_{12}^\prime$ which control the mixing of the $A$ and $B$ fermions via the Higgs. 
A typical FFS contribution to $\g$ is shown in \fref{f.feynmang-2}~(c). The chirality flip and Higgs VEV insertion both arise in the loop due to the Higgs couplings of the new fermions.

These two simplified models generate the largest possible BSM particle masses that could account for $\deltaaexp$. Therefore, the maximization over theory space in \eref{e.Mchargedmaxgeneral} can be replaced by a maximization over the SSF and FFS parameter spaces:
\begin{equation}
\label{e.MchargedmaxSSFFFS}
M^\mathrm{max}_\mathrm{BSM, charged}  \ \ \ 
\equiv \ \ \ 
\max_{
\begin{array}{c} 
\scriptstyle \mathrm{SSF,\ FFS\ models} 
\\ 
\end{array}
} \ \  \left\{ \ \  \min_{\tiny i  \ \in \  \mathrm{BSM\ spectrum}} \left( m_{\rm charged}^{(i)}   \right)  \ \ \right\} \ .
\end{equation}
Note that one could in principle consider extensions of the SM Higgs sector with additional scalar contributing to EWSB. In that case, the $\kappa$ and $y_{1,2}$ terms in the above Lagrangians could arise from coupling to these new scalars rather than a SM-like Higgs doublet, which might change the allowed EW representations of the BSM states. However, current constraints already dictate that most of the observed EWSB arises from the VEV of a single doublet \cite{Bernon_2016,craig2013searching}, which means that relying only on BSM scalars to generate the required EWSB insertions in the above Lagrangians would lead to smaller effective mixings and hence smaller  $\deltaa$ and BSM masses. We therefore do not have to consider such extended scenarios to perform the maximization of the lightest new charged particle mass over BSM theory space.

 In both SSF and FFS models, the choices of representations must satisfy
 \begin{eqnarray}
\label{e.reprequirement}
 \mathbf{1} &\subset& R^A \otimes R \otimes \mathbf{2}\\
 \nonumber R^B &=&\bar R\\
 \nonumber Y^A &=& -\frac{1}{2} - Y\\
 \nonumber Y^B &=& -1 - Y,
 \end{eqnarray}
with $Y$ chosen to make the electric charges integer-valued. We will explore all choices of representations involving $SU(2)_L$ singlets, doublets and triplets, and all choices of $Y$ that ensure that all electric charges satisfy $|Q| \leq 2$. As we discuss, this is sufficient to perform the above maximization.
The possibility of a high multiplicity of new BSM states is again taken into account by considering the trivial generalizations where there are $N_{\rm BSM}$ identical copies of the above fields contributing to $\deltaa$.
The Lagrangians in \Eq{e.SSF} and \Eq{e.FFS} only show the
 interactions necessary to form new one-loop contributions to $\g$.
Depending on the choice of $SU(2)_L \otimes U(1)_Y$ representations, additional couplings between the new fermions/scalars and the muon or Higgs may be allowed by gauge invariance. However, these couplings will not contribute to $\g$ at leading order, at most supplying a small correction to the 
leading terms generated by the couplings in Eqns.~(\ref{e.SSF}) and (\ref{e.FFS}), or slightly modifying the mass spectrum of the fermions/scalars that couple to the Higgs after EWSB by $\lesssim \tev$, which does not meaningfully affect our results or discussion. We can therefore neglect these additional couplings in our analysis. We also assume that the new BSM states do not couple to any other SM fermions (except when discussing leptonic flavour violation bounds). Both of these assumptions are conservative in that they minimize additional experimental signatures arising from the new physics responsible for the $\g$ anomaly.

Depending on the choice of representations, some of the EW Scenarios we consider were previously studied in Refs.~\cite{Calibbi:2018rzv, Agrawal:2014ufa, Kowalska:2017iqv, Barducci:2018esg, Calibbi:2019bay,Calibbi:2020emz,Crivellin:2018qmi}.
There have also been previous attempts to define simplified model dictionaries for generating $\deltaaexp$~\cite{Freitas:2014pua, Calibbi:2018rzv, Lindner:2016bgg, Kowalska:2017iqv, Barducci:2018esg, Kelso:2014qka, Biggio:2014ela, Queiroz:2014zfa, Biggio:2016wyy}, but none took our completely model-exhaustive approach and none aimed to find the highest possible mass of new BSM charged states that could account for $\deltaaexp$. We also make no assumptions about e.g. the existence of a viable DM candidate, or any couplings of the new degrees of freedom (dof's) that are not required for resolving the $\g$ anomaly (except optionally considering flavour). 
Other possible simplified models for $\g$, such as adding fewer than 3 new BSM particles with non-trivial EW representations (see \textit{e.g.}~\cite{Freitas:2014pua}),  require smaller masses for the new charged particles than the SSF and FFS models,  and their inclusion does not affect the outcome of the maximization over theory space of \eref{e.Mchargedmaxgeneral}. We demonstrate this  explicitly in Section~\ref{s.edgecases}.

\subsection{Upper Bounds on BSM Couplings}

The size of the $\g$ contribution is controlled by BSM couplings and masses, and the largest possible BSM masses that can account for the anomaly depend on the largest possible BSM couplings. 
In \sref{s.unitarity} we describe first how perturbative unitarity supplies an absolute upper bound on the new couplings. This will inform our baseline analysis, but more careful consideration of how these simplified models must arise as part of  a more complete BSM theory suggests that an upper bound based on unitarity alone is likely far too conservative, especially in light of stringent CLFV bounds. In Sections~\ref{s.unitarityMFV} and~\ref{s.unitarityNAT}  we therefore consider the additional constraints on the new muon couplings arising by assuming either Minimal Flavour Violation (MFV) or requiring the absence of large, explicitly calculable new tunings.

\subsubsection{Unitarity}
\label{s.unitarity}

To define the boundaries of parameter space in our simplified models we  appeal to tree-level partial-wave unitarity, expressed in terms of helicity amplitudes so that we can apply the constraints to fermions as well as bosons \cite{Jacob:1959at}. (See e.g.~\cite{Lee:1977yc,Haber:1994pe,Goodsell:2018tti,Biondini:2019tcc,Endo:2014mja,Castillo:2013uda} for more recent studies.)
We begin from the partial-wave expansion of the (azimuthally symmetric) scattering amplitude for the $2\to2$ process ${i \to f \equiv} \{a,b\}\to\{c,d\}$:
\begin{equation}
\label{e.pwd}
\mathcal{M}_{i\to f}(\theta)=8\pi\sum_{j=0}^{\infty}(2j+1)T_{i\to f}^{j}d_{\lambda_{f}\lambda_{i}}^{j}(\theta),
\end{equation}
where $d_{\lambda_{f}\lambda_{i}}^{j}(\theta)$ are the Wigner d-functions, $T^j_{i \to f}$ is the $j$-th partial wave of the tree-level scattering amplitude, $\lambda_{i}=\lambda_{a}-\lambda_{b}$ and $\lambda_{f}=\lambda_{c}-\lambda_{d}$ are the helicities of the initial and final states, and $j$ is the eigenvalue of the total angular momentum. The coefficients $T_{i\to f}^{j}$ can be found by using the orthogonality condition of the d-functions
\begin{equation}
\frac{1}{16\pi}\int\mathcal{M}_{i\to f}(\theta)d_{\lambda_{f}\lambda_{i}}^{j}(\theta)d(\cos\theta)=T_{i\to f}^{j}.
\end{equation}
From the optical theorem one can get the partial-wave unitarity condition of an inelastic process for each $j$ 
\begin{equation}
\beta_{i}\beta_{f}|T_{i\to f}^{j}|^{2}\leq1,
\label{UniCond}
\end{equation}
where the phase space factors for states of mass $m_1$ and $m_2$ are
\begin{equation}
\beta(m_1, m_2) = \frac{1}{s}\sqrt{[s - (m_1+m_2)^2][s-(m_1-m_2)^2]},
\end{equation}
and $s$ is the squared center of mass energy.
For a given set of mass eigenstates which appear in our theory, we will require that the lowest partial-wave tree-level 2-to-2 scattering amplitudes between initial states $i$ and final states $f$ satisfy the unitarity condition (\ref{UniCond}). We will consider boson-boson ($j=0$), boson-fermion ($j=\pm1/2$), and fermion-fermion ($j=0,1$) scattering; fermion-vector scattering ($j=1/2$) will always lead to weaker constraints for large $\NBSM$. The relevant Wigner d-functions are given in Table \ref{d-functions}.

\begin{table}
\begin{center}
\begin{tabular}{|l|l|l|}
\hline 
{\rm Scalar-Scalar} & $T_{0\to0}^{j=0}$ & $d_{00}^{0}(\theta)=1$\tabularnewline
\hline 
\multirow{2}{*}{{\rm Scalar-Fermion}} & $T_{+\to+}^{j=1/2}$ & $d_{\frac{1}{2}\frac{1}{2}}^{1/2}(\theta)=\cos(\theta/2)$\tabularnewline
\cline{2-3} \cline{3-3} 
 & $T_{+\to-}^{j=1/2}$ & $d_{\frac{1}{2}-\frac{1}{2}}^{1/2}(\theta)=\sin(\theta/2)$\tabularnewline
\hline 
\multirow{3}{*}{{\rm Fermion-Fermion}} & $T_{++\to\pm\pm}^{j=0}$ & $d_{00}^{0}(\theta) = 1$  \tabularnewline
\cline{2-3} \cline{3-3} 
 & $T_{+-\to\pm\pm}^{j=1}$ & $d_{01}^{1}(\theta)=-\sqrt{2}\sin(\theta/2)\cos(\theta/2)$\tabularnewline
\cline{2-3} \cline{3-3} 
 & $T_{+-\to+-}^{j=1}$ & $d_{11}^{1}(\theta)=\cos^{2}(\theta/2)$\tabularnewline
\hline 
\end{tabular}
\end{center}
\vspace{-3ex}
\caption{The Wigner d-functions used in our partial-wave unitarity calculations.}
\label{d-functions}
\end{table}

Note that the partial wave decomposition in \Eq{e.pwd}  requires specifying the angular momenta of the initial and final states, so in principle the different helicity amplitudes for $j = 1/2$ can give independent constraints. 
Note also that these partial-wave constraints are valid at any kinematically allowed value of $s$, as the phase space factors vanish at kinematic thresholds and enforce physical kinematics.

The constraints
obtained from (\ref{UniCond})
amount to the requirement that loop contributions to scattering amplitudes are smaller than tree-level contributions at scales up to a factor of a few above $m_\mathrm{max}$, where $m_{\rm max}$ is the largest mass eigenvalue in the model under consideration.\footnote{Specifically, in some processes we take the $s \to \infty$ limit to obtain our constraint, but numerically the constraint asymptotes rapidly for energies a factor of a few times above threshold.}
 The violation of these constraints would require nonperturbative physics to appear at an energy scale close to $m_{\rm max}$ to unitarize the theory, so restricting to parameter space which satisfies tree-level unitarity amounts to the following statement: either a theory with masses up to $m_{\rm max}$ is perturbatively calculable, or new physics appears at the scale  $s_{\rm max}$. 

In some processes, we may encounter singularities either in the scattering amplitude itself in the form of $s$-channel poles, or after integrating the amplitude as demanded by Eqn. (\ref{UniCond}). The latter appear in $t$- and $u$-channel diagrams. In Ref.~\cite{Goodsell:2018tti}, these singularities are treated by removing values of the CM energy $\sqrt{s}$ around the singularities. We avoid such a complication by studying processes where $t$- and $u$-channel amplitudes do not appear, and where $s$-channel singularities correspond to poles at energies below the threshold where the cross section is nonvanishing. This will become clear when we discuss the perturbative unitarity constraints for specific processes in the sections below.

Note that somewhat stronger constraints could be achieved by considering a coupled-channel analysis where the full scattering matrix between all initial and final states is diagonalized, by considering higher partial waves, and/or by relaxing the constraints on poles; our constraints are thus conservative, but will suffice for the statement of our no-lose theorem.

\subsubsection{Unitarity and Minimal Flavour Violation}
\label{s.unitarityMFV}

Proposing new scalars with Yukawa couplings to the muon prompts us to ask how these new degrees of freedom couple to the other lepton generations. 
The physics which solves the $\g$ anomaly would have to be embedded in whichever UV-complete framework explains the flavour structure of the SM fermions. 
From a bottom-up perspective, this is most relevant since flavour-changing neutral currents (FCNCs) in the lepton sector, most importantly charged-lepton flavour violating (CLFV) decays $\ell_i \to \ell_j \gamma$, are tightly constrained \cite{Calibbi:2017uvl, Aubert:2009ag}: 
\begin{eqnarray}
\mathrm{Br}(\mu \to e \gamma) &<& 4.2 \times 10^{-13}  \ \ 
\\
\mathrm{Br}(\tau \to \mu \gamma) &<& 4.4 \times 10^{-8} \ \ 
\\
\mathrm{Br}(\tau \to e \gamma) &<& 3.3 \times 10^{-8} \ \
\end{eqnarray}
It is well known that CLFV constraints impose stringent requirements on BSM solutions to the $\g$ anomaly (see e.g.~\cite{Lindner:2016bgg, Freitas:2014pua}).
We can demonstrate this by considering a flavour-anarchic version of the scalar Singlet Scenario:
\be
\label{scalar-shit}
-{\cal L} \supset   S( g^{ee}_S  e_L e^c + g^{\mu\mu}_S  \mu_L \mu^c + g^{\tau \tau}_S  \tau_L \tau^c +   g^{e\mu}_S  \mu_L e^c  + g^{\mu e}_S  e_L \mu^c \ldots) ~.
\ee
where ``\ldots'' indicates the additional off-diagonal terms.
This would generate flavour-violating versions of the low-energy operator  \eref{leff}
 \begin{equation}
 \label{CLFVop}
 {\cal L}_{\rm eff} =  C^{(ij)}_{\rm eff}   \frac{ v}{M^2}  (\ell^{(j)}_L  \sigma^{\nu \rho}  {\ell^{(i)}}^c )  F_{\nu \rho}   + {\rm h.c.}~,
 \end{equation}
 where $i,j$ are lepton generation indices. The assumption that the above scalar Singlet Scenario resolves the $\g$ anomaly fixes the $C_{\rm eff}^{\mu\mu}$  Wilson coefficient.
 Assuming for simplicity that $C^{\mu \mu}_{\rm eff}$ is fully determined by $g^{\mu \mu}_S$, this determines all the other operators up to ratios of $g_S^{ij}$ couplings: 
 \begin{equation}
\label{CLFVcouplingratiosSinglet}
 C_{\rm eff}^{ij} \approx \frac{\max(m_{\ell_i}, m_{\ell_j})}{m_\mu}  \sum_k \frac{g_S^{ik}}{g_S^{\mu\mu}} \frac{g_S^{kj}}{g_S^{\mu\mu}} \ ,
 \end{equation}
 where we have set $g^{ij}_S = g^{ji}_S$, again for simplicity.
It is straightforward to obtain CLFV branching ratios from this low-energy description, which can be used to constrain ratios of the singlet scalar couplings to different fermion generations:
\begin{eqnarray}
\sum_\ell \frac{g_S^{\mu \ell}}{g_S^{\mu\mu}} \frac{g_S^{\ell e}}{g_S^{\mu\mu}}
\lesssim 1 \times 10^{-5} 
\ \ , \ \
\sum_\ell \frac{g_S^{\tau \ell}}{g_S^{\mu\mu}} \frac{g_S^{\ell \mu}}{g_S^{\mu\mu}}
 \lesssim 7 \times 10^{-3}
\ \ , \ \
\sum_\ell \frac{g_S^{\tau \ell}}{g_S^{\mu\mu}} \frac{g_S^{\ell e}}{g_S^{\mu\mu}}
\lesssim 6 \times 10^{-3},~~
\end{eqnarray}
from  $\mu \to e \gamma$ , $\tau \to \mu \gamma$ and $\tau \to e \gamma$ decays respectively. 
We emphasize that these bounds assume that $g_S^{\mu \mu}$ is fixed to generate $\deltaaexp$.
Clearly, flavour-universal couplings of the singlet scalar are excluded, and flavour-anarchic couplings  are severely disfavoured by CLFV bounds. 

The situation is similar for EW Scenarios. Consider  flavour anarchic versions of the SSF and FFS models:
\be
\label{e.FFS-flavor}
- \mathcal{L}_\mathrm{SSF}  &\supset& y^i_1 F^c L_{i}  \Phi_A^*  + y^i_2 F \ell^c_i  \Phi_B   + \kappa H \Phi^*_A \Phi_B \\
- \mathcal{L}_\mathrm{FFS} &\supset& y^i_1 F_A^c L_i  S^*  + y^i_2 F_B \ell_i^c  S + y_{12} H F_A F^c_B ~.
\ee
Again, in this anarchic ansatz, the same new fermions and scalars that account for the $\g$ anomaly generate the flavour violating operators in \eref{CLFVop}, and $C_{\rm eff}^{ij}$ is determined by $\deltaaexp$ up to coupling ratios:
 \begin{equation}
\label{CLFVcouplingratiosEW}
 C_{\rm eff}^{ij} \approx \frac{g_S^i}{g_S^\mu} \frac{g_S^j}{g_S^\mu} \ ,
 \end{equation}
where we again assumed for simplicity that $y_1^i = y_2^i$ and that $C_{eff}^{\mu\mu}$ is fully determined by $y_{1,2}^\mu$. 
 The only difference to scalar Singlet Scenarios is the absence of the lepton mass ratio in \eref{CLFVcouplingratiosSinglet}, since for FFS and SSF models, the chirality flip and Higgs coupling insertion now lie on the propagators of the BSM particles in the loop.
 %
Repeating the estimates for CLFV decay branching ratios, we obtain the following bounds on the lepton coupling ratios:
 \begin{eqnarray}
 \frac{y_{1,2}^e}{y_{1,2}^\mu} \lesssim 10^{-5} 
 \ \ \ \ , \ \ \ \ 
\frac{y_{1,2}^\tau}{y_{1,2}^\mu} \lesssim 10^{-1}
\ \ \ \ , \ \ \ \ 
\frac{y_{1,2}^\tau}{y_{1,2}^\mu} \frac{y_{1,2}^e}{y_{1,2}^\mu} \lesssim 10^{-1}
\ ,
\end{eqnarray}
from $\mu \to e \gamma$ , $\tau \to \mu \gamma$ and $\tau \to e \gamma$ decays respectively if $y_{1,2}^\mu$ is fixed by resolving the $\g$ anomaly.

Clearly CLFV constraints, in particular $\mu \to e \gamma$, exclude flavour-universal BSM solutions to the $\g$ anomaly (that involve new scalars), and severely constrain flavour-anarchic ones.  
It is of course possible that a flavour anarchic model evade the above constraints by some coincidence (perhaps all the more unlikely given that the above coupling ratio constraints have to be satisfied in the lepton mass basis after PMNS diagonalization, not the lepton gauge basis).
However, it seems much more reasonable to take the absence of observed CLFVs as evidence of some protection against FCNCs in whatever UV-complete theory solves the SM flavour puzzle, and that the physics of $\g$ has to respect that protection.

A robust model-independent framework that encompasses many possible flavour embeddings and provides strong protection against FCNCs is the Minimal Flavour Violation (MFV) ansatz (see e.g.~\cite{Buras:2003jf, Csaki:2011ge}).
In MFV, the SM Higgs Yukawa matrices couplings are assumed to be the only spurions of global $U(3)_L \times U(3)_{\ell^c} \to  U(1)_{\rm lepton}$
 flavour breaking, so that all BSM flavour violation is aligned with the SM Yuwakas.
Such a structure naturally emerges if the SM Yukawa matrices arise as the VEVs of heavy UV fields responsible for breaking a larger flavour group.

The MFV ansatz does not specify the representations of BSM fields under the flavour group, but it does require all Lagrangian terms to be flavour-singlets (with the Yukawa matrices as spurions). 
This would, for example, forbid off-diagonal terms in \eref{scalar-shit}, avoiding large CLVFs while still providing a viable explanation for $\g$ over a wide range of scalar masses~\cite{Chen:2017awl}.
For EW Scenarios, the muon-scalar-fermion index has to involve a Yukawa coupling factor and the scalar and fermion together have to contract into triplets of $U(3)_L$ or $U(3)_{\ell^c}$. This automatically forbids interactions of the form \eref{e.FFS-flavor} since there would have to be at least one separate BSM fermion (or scalar) for each lepton flavour and the CLFV diagrams are not generated.\footnote{
This statement is strictly true only for massless
neutrinos, in which case the lepton Yukawa matrices are spurions of $U(3)_L \times U(3)_{\ell^c} \to 
U(1)_{e} \times U(1)_{\mu} \times U(1)_{\tau}$ flavor breaking and 
lepton flavors are are separately conserved. However, for nonzero
neutrino masses, there will still be some CLFV contributions from these models, but they involve
diagrams with virtual $W$ exchange and are further suppressed by powers of $m_\nu/m_W$ relative
to the leading diagrams that resolve $\g$,
so we do not consider them here.}

Imposing MFV has several important consequences. 
First, non-trivial flavour representations of BSM fields in EW Scenarios can give rise to more than one set of BSM states coupling to the muon and contributing to $\g$. In effect, this  corresponds to $N_{\rm BSM}  > 1$, which is covered by our analysis. 
Second, MFV requires that some of the muonic BSM couplings in the scalar singlet, SSF and FFS models have a tau-like equivalent that is at least a factor $m_\tau/m_\mu \approx 17$ larger. 
This larger tau-like coupling will therefore have to satisfy the bounds of perturbative unitarity, effectively lowering the upper bound from unitarity on the relevant muonic coupling that generates $\deltaaexp$ by a factor of $\approx 17$. 
This leads to a dramatic reduction in the maximum allowed BSM mass scale compared to imposing unitarity alone (and implicitly assuming that CLFV decays are suppressed by accidentally small flavour-anarchic BSM couplings in the lepton mass basis). 

Precisely which muonic BSM couplings have a tau-equivalent can depend on the 
$ U(3)_{L} \times U(3)_{\ell^c}  \to U(1)_{\rm lepton}$ 
representation of the BSM fields. 
The situation is simple for the scalar Singlet Scenario, since the $g_S$ coupling must be in the same representation as the SM Yukawas, and therefore $g_S^\mu/g_S^\tau = m_\mu/m_\tau$. 
For EW Scenarios there is more ambiguity. An example of a minimal choice for the flavour representation of the BSM fields in the SSF model (the discussion is similar for FFS) is
\begin{equation}
F \sim (3,1) \ \ , \ \ 
F^c \sim (\bar 3, 1) \ \ , \ \ \ 
S_{A,B} \sim (1,1)
\end{equation}
Since $L \sim (3,1)$ and $e^c \sim (1, \bar 3)$ this implies that $y_2$ must transform like the SM electron Yukawa while $y_1$ can be a flavour singlet:
\begin{equation}
y_2 \sim y_e \sim (\bar 3, 3) \ \ , \ \ y_1 \sim (1,1)
\end{equation}
Therefore, the MFV assumption implies $y_2^\mu/y_2^\tau = m_\mu/m_\tau$ and the $y_2^\mu$ coupling effectively has a smaller perturbativity bound, while the upper bound for $y_1$ is unaffected since that coupling is flavour-universal. 
Other minimal choices can make $y_2$ a flavour singlet and $y_1$ a bifundamental, but at least one of the two muonic $y_{1,2}$ couplings has its perturbativity bound reduced by  $m_\mu/m_\tau$.
Non-minimal flavour representations for the BSM fields may introduce additional coupling ratios and hence even tighter perturbativity bounds, but for the purposes of our conservative estimates, we only make the minimal assumption.

\subsubsection{Unitarity and Naturalness in Electroweak Scenarios}
\label{s.unitarityNAT}

The hierarchy problem in the SM is often formulated using an estimate of loop corrections to the Higgs mass regulated with a finite momentum cutoff $\Lambda_{\rm UV}$:
\begin{equation}
\label{e.deltamhcutoff}
\Delta m_H^2 \sim \frac{y_t^2}{4 \pi^2} \Lambda_{\rm UV}^2 \ , 
\end{equation}
where $y_t$ is the SM top Yukawa, which dominates this estimate. Avoiding fine tuning of the Higgs mass parameter in the Lagrangian requires either cancellation of the above quadratically divergent correction (SUSY) or new physics far below the Planck or GUT scale (i.e. a low UV cutoff). 
This is simple and intuitive, appealing to the physical interpretation of unknown physics at some high scale in a Wilsonian picture. 
The cutoff argument is also ``morally correct'' in that it accurately indicates the quadratic sensitivity of the Higgs mass to UV corrections, whatever they may be.
However, without knowledge of what the new physics is, one could argue that the specific cutoff-dependent quantity in \eref{e.deltamhcutoff} has no physical meaning. While it might seem unlikely or even absurd that quantum gravity corrections at the Planck scale contribute nothing to $\Delta m_H^2$, without explicit knowledge of (1) new physics between the weak scale and the Planck scale, and (2) the precise nature of quantum gravity,  one cannot be absolutely sure that the hierarchy problem does, in fact, refer to a real tuning of our universe's parameters.

The situation is entirely different when explicit new states with high mass and sizeable couplings to the Higgs are introduced, as is the case for the EW Scenarios we examine. These models have been engineered to account for the $\g$ anomaly with the highest possible BSM particle masses in order to perform the theory-space maximization of \eref{e.Mchargedmaxgeneral} and identify the experimental worst-case scenario and the minimum energy of future colliders required for discovery. Realizing these high-mass scenarios requires unavoidably large couplings to the Higgs, which in turn leads to \emph{large} but \emph{finite} and \emph{calculable} corrections to the Higgs mass; this makes the hierarchy problem explicit. 

Specifically, we can calculate the one-loop contributions of  the new $S, F$ fields  to the Higgs mass using dimensional regularization (DR) as a regulator in the $\overline{MS}$ renormalization scheme. 
This gives contributions of the schematic form
\begin{equation}
\Delta m_H^2 \sim \frac{1}{4 \pi^2} \left( c_1 M_{\rm BSM}^2 + c_2 M_{\rm BSM}^2 \log \frac{\mu_R^2}{M_{\rm BSM}^2} \right),
\end{equation}
where in this instance $M_{\rm BSM}$ stands for various combinations of BSM masses in each term, and $\mu_R$ is the renormalization scale. The quadratic UV sensitivity of the Higgs mass is illustrated by the first term, with the size of the correction given by the scale of new physics as expected.

Naively, one might worry that the dependence of the second term on the renormalization scale invalidates such a straightforward physical interpretation. One might in principle choose $\mu_R$ to set the above correction to zero. However, this would not be physically meaningful, since for such a choice of $\mu_R$, the perturbative expansion would be invalid. Restoration of perturbativity by inclusion of higher-loop diagrams would restore the large size of $\Delta m_H$. 
Therefore, the most reasonable physical interpretation of this correction is obtained setting $\mu_R$ to \emph{optimize the validity of the perturbative expansion}, in which case the above one-loop result is the best possible approximation for the total size of the Higgs mass correction to all orders. This is why one typically choose $\mu_R \sim m$ in $\overline{MS}$ calculations that are dominated by physics at scale $m$. 
In that case, the $\mu_R$ dependence becomes minor and simply corresponds to the fact that in a truncated perturbative expansion, there are unknown higher-order terms that could slightly modify the one-loop result. 

With this in mind, we fix $\mu_R \sim \mathcal{O}(M_{BSM})$ to the value that sets the log terms to zero. This gives the following expressions for the Higgs mass corrections in SSF and FFS models: \begin{eqnarray}
\label{e.deltamh}
\Delta m_H^2  &=&  C_1 N_{\rm BSM} \frac{\kappa^2}{16 \pi^2} \ \ \ \ \ \ \ \ \ \ \ \ \ \ \ \ \ \ \ \ \ \ \ \ \ \ \ \ \ \ \ \  \ \ \ \ \ \ \ \ \ \ \ \ \ \ \ \ \ \ \ \  \ \ \ \, \ \  \ \ \ \  \mbox{(SSF)}
\\
\Delta m_H^2 &=&  C_2 N_{\rm BSM} \frac{1}{8 \pi^2} \left(
(y_{12}^2 + {y_{12}'}^2)(m_A^2 + m_B^2) + 2 y_{12} y_{12}' m_A m_B
\right)  \ \ \ \ \mbox{(FFS)}
\end{eqnarray}
where $C_{1,2} \sim \mathcal{O}(1)$ depend on the  gauge representations of the new scalars and fermions in the SSF/FFS model.
The required presence of such corrections in BSM theories that solve the $\g$ anomaly with the highest possible BSM mass scale makes the hierarchy problem explicit.

What is more surprising, if not entirely unfamiliar \cite{Cesarotti:2018huy, Calibbi:2020emz, Capdevilla:2020qel}, is that these same theories actually lead to a \emph{second hierarchy problem} for the \emph{muon mass}. Fermion masses are usually technically natural, but the required muon coupling to new heavy fermions $F$ means their chiral symmetry is shared in the limit where both are massless. Corrections to the muon mass therefore no longer scale with the muon Yukawa $y_\mu$.\footnote{Indeed, if the new physics is not so heavy it can modify the muon Yukawa \cite{Crivellin:2020tsz}.}
Following the same procedure as the calculation of Higgs mass corrections we obtain corrections to the muon Yukawa due to loops of heavy fermions and scalars in  EW Scenarios: 
\begin{eqnarray}
\Delta y_\mu   &\sim&   N_{\rm BSM} \frac{y_1 y_2 }{16 \pi^2} \  \frac{\kappa\, m_F}{M_{\rm BSM}^2}  \    \ \ \ \ \mbox{(SSF)}
\\
\Delta y_\mu   &\sim&   N_{\rm BSM} \frac{y_1 y_2  y_{12}^{(\prime)} }{4 \pi^2}  \ \ \ \  \ \ \ \ \ \ \ \ \mbox{(FFS)}
\label{e.deltaymu}
\end{eqnarray}
For large BSM couplings and masses, $|\Delta y_\mu| \gg y_\mu$, necessitating tuning of the Lagrangian parameters. 
This hierarchy problem of the muon Yukawa arises due to large, calculable corrections from new states present in the theory, making it just as explicit as the Higgs hierarchy problem above.

It is therefore reasonable to consider BSM scenarios that avoid adding two explicit hierarchy problems to the SM by keeping such a dual-fine-tuning to a reasonable minimum, \textit{e.g.} $1\%$ each for the muon and Higgs mass. Similar to the MFV ansatz, this shrinks the viable parameter space by reducing the maximum allowed size of BSM couplings, thereby reducing the maximum BSM mass scale.\footnote{Any lower-scale new physics that somehow cancels this fine-tuning would lead to new experimental signatures and hence also lead to a discovery.}

\subsection{Upper Bound on the BSM Mass Scale}
\label{s.mBSMupperbound}

The analysis of Singlet and EW Scenarios is discussed in detail in Sections~\ref{s.singletmodels} and~\ref{s.ewmodels}. 
In each scenario, the viable parameter space of BSM masses and  couplings is compact, since we require the new states to explain the $\g$ anomaly, and the couplings cannot exceed the limit set by perturbative unitarity, or unitarity + MFV, or unitarity + naturalness.  
Therefore, each scenario has well-defined maximum BSM particle masses for a given BSM multiplicity $N_{\rm BSM}$. 
We then analyze the signatures of these models at future muon colliders. The details are slightly different for Singlet and EW Scenarios due to their different collider signatures.

Singlet Scenarios feature new SM singlets which can be invisible. Lighter singlets are more weakly coupled to account for the $\g$ anomaly, so the scenarios with the heaviest BSM particles are not necessarily the hardest to discover.
Furthermore, the sensitivity of collider searches can depend  on whether the new singlets are stable or how they decay.
We therefore have to map out the complete parameter space of the simplified Singlet Scenarios. 
Fortunately, with the muon coupling $g_{S,V}$ determined by the requirement of accounting for the observed $\deltaaexp$, the model has just two parameters, singlet mass $m_{S,V}$ and multiplicity $N_{\rm BSM}$ (as well as the choice of singlet being a scalar or vector). 
As a function of mass and multiplicity we then analyze the sensitivity of a completely inclusive search for the production of the BSM singlets at muon colliders regardless of their decays. We also analyze the reach of an indirect search based on deviations in Bhabha scattering to explore the physics potential of a muon collider Higgs factory.
We find that the singlet BSM states cannot be heavier than about 3 TeV, and can be directly discovered at a 3 TeV muon collider with $1~\mathrm{ab}^{-1}$ for masses $\gtrsim 10 \gev$ in singlet + photon production processes. A 215 GeV muon collider that might be used as a Higgs factory can directly discover singlets as light as 2 GeV in our conservative inclusive analysis with $0.4~\mathrm{ab}^{-1}$ of luminosity. Heavier singlets up to the 3 TeV maximum can be probed with Bhabha scattering.

The parameter space of the SSF and FFS simplified models that allow us to perform the EW Scenario theory space maximization of BSM charged particle mass in \eref{e.Mchargedmaxgeneral} is much more complex, featuring three masses, several BSM couplings, the number of BSM flavours $N_{\rm BSM}$, and the choice of EW gauge representations for the BSM states. 
However, since we only need to find the heaviest possible BSM masses, for each SSF/FFS model with a given choice of $N_{\rm BSM}$ and EW gauge representation we can simply find the boundaries of the parameter space defined by the maximum possible BSM masses that still allow BSM couplings below the unitarity (or unitarity + MFV/naturalness) limit to account for the $\g$ anomaly. 

For EW Scenarios we find that requiring only perturbative unitarity allows the lightest charged states to sit at the 100~TeV scale, but this assumption is disfavoured by CLFV bounds.
Requiring either consistency with MFV to avoid CLFVs, or avoiding two explicit new tunings worse than 1\%, predicts new charged states at the 10~TeV scale or below. Encouragingly, these states are in reach of some muon collider proposals.

It is worth noting that at the very boundaries of the BSM parameter spaces we explore, with couplings set at the upper limit set by perturbative unitarity, the theory itself strictly speaking has already lost predictivity, by definition. If the couplings actually had this value, we would have to regard the theory as a strongly coupled one, requiring different analysis tools. 
This is suitable for deriving upper bounds on the BSM mass scale, but it is interesting to note these  bounds could actually be saturated by strongly coupled BSM solutions to the $\g$ anomaly (which would still have to feature new states with EW gauge charges). One feature of composite theories is a large multiplicity of states, which we include by considering $N_{\rm BSM} > 1$, with $N_{\rm BSM} = 10$ serving as a ``high-multiplicity benchmark'' for our analyses. Therefore, while our quantitative predictions are unlikely to apply precisely to strongly coupled BSM solutions of the $\g$ anomaly, by including couplings up to the unitarity limit and considering large numbers of BSM flavours we parametrically include the signature space swept out by these strongly coupled theories. The statements we make about the discoverability of new physics should, broadly speaking, apply to those scenarios as well. 
 That being said, it would be interesting to undertake a dedicated investigation of high-scale composite BSM solutions to the $\g$ anomaly within our framework. We leave this for future work.

 While CLFV constraints strongly favour the existence of some kind of flavour protection mechanism, the degree to which the precise assumptions of MFV would have to be satisfied is obviously up for debate. Similarly, the precise degree of tuning depends on the tuning measure, and it is difficult to define exactly at what point a theory becomes ``un-natural'' in a meaningful sense. 
However, our model-exhaustive approach has the advantage of throwing 
these issues into stark relief: \emph{Solving the $\g$ anomaly with BSM masses up to $\sim$ 10 TeV is apparently relatively ``easy'', while pushing the masses of new states to the maximum 100 TeV scale limited only by unitarity appears to require some extreme form of tuning \emph{and} violation of MFV while \emph{somehow} suppressing CLFV decays.}

In particular, if the 10 TeV scale were exhaustively probed without direct detection of new states while the $\g$ anomaly is confirmed, this would confirm empirically  that nature is fine-tuned\footnote{A similar observation was made in connection with electron EDM measurements~\cite{Cesarotti:2018huy} and in~\cite{Calibbi:2020emz}. On a similar ground, see \cite{Baker:2020vkh} for the implications of tuning in the context of models with radiative leptonic mass generation.} and does not obey the assumptions of the MFV ansatz but still suppresses CLFV decays in some way.
An analogy would be the discovery of split supersymmetry~\cite{Giudice:2004tc,ArkaniHamed:2004yi}, where the lightest new physics states are heavy and couple to the Higgs; in our case, the situation is even more severe since heavy states in EW Scenarios make the muon mass radiatively unstable as well, and very heavy BSM states also preclude  MFV solutions to the SM flavour puzzle.

Our analysis generalizes and reinforces  our earlier results in~\cite{Capdevilla:2020qel} by including a more complete basis for the relevant EW Scenarios, considering consistent electroweak embeddings of Singlet Scenarios, addressing flavour physics considerations, and supplying important technical details. 
Subsequent studies have employed an effective field theory (EFT) approach to explore \emph{indirect} signatures of the new physics causing the $\g$ anomaly at muon colliders~\cite{Buttazzo:2020eyl, Yin:2020afe}. 
While this EFT approach would not allow us to ask detailed questions about the BSM physics -- like studying direct particle production, tuning, and flavour considerations -- it is nonetheless extremely useful due to its maximal model-independence and simplicity. As we discuss in \sref{s.implications}, the results of these analyses are highly complementary to our own and help flesh out the muon collider no-lose theorem.

\section{Analysis of Singlet Scenarios}
\label{s.singletmodels}

\subsection{$\g$ in Singlet Scenarios}

As defined in Eqns.~(\ref{e.LS}) and (\ref{e.LV}), if BSM singlet scalars or vectors are responsible for the $\g$ anomaly, the relevant muonic interactions are 
\be
\label{singlet-lag}
  \left( g_S \mu_L \mu^c S + h.c.\right)
  ~~,~~   g_V V_\alpha  (\mu^\dagger_L \bar \sigma^\alpha \mu_L +
\mu^{c \, \dagger} \bar \sigma^\alpha \mu^c) ~.
\ee
The contribution of $N_{\rm BSM}$ scalar singlets to $\g$ is
\be
\label{aS}
\Delta a^{S}_{\mu}  =  N_{\rm BSM}  \frac{g_S^2}{16\pi^2} \int_0^1 dz \frac{m^2_\mu (1-z) (1-z^2)}{ m_\mu^2 (1-z)^2 +  m_S^2 \, z} 
\approx  2 \times 10^{-9}  N_{\rm BSM}  \, g_S^2   \left( \frac{700 \, \rm GeV}{  m_S} \right)^{2},
\ee
where in the last step we have taken the 
$m_{S} \gg m_\mu$ limit. 
For vectors, the corresponding  $\g$ contribution is 
\be
\label{aV}
\Delta a^{V}_{\mu}  = N_{\rm BSM}   \frac{g_V^2 }{4\pi^2} \int_0^1\! dz  \frac{  m^2_\mu z (1-z)^2}{ m_\mu^2 (1-z)^2 +  m_V^2 \, z} \approx 2\times 10^{-9}   N_{\rm BSM} \, g_V^2   \left( \frac{200 \,\rm GeV}{  m_V} \right)^{2},
\ee
where again we have taken the $m_{V} \gg m_\mu$ limit. 
It is known in the literature
that pseudo-scalar or pseudo-vector  contributions 
to $\g$ have the wrong sign to explain the anomaly \cite{Freitas:2014pua}, so we do not consider these scenarios here. 
Note also that in both cases $\Delta a_\mu\propto m^2_\mu$, which
implies a low  $(\lesssim $ TeV) mass scale for any choice of perturbative couplings that 
yield $\Delta a_\mu \sim 10^{-9}$ required to explain the anomaly (see discussion in Sec. \ref{singlet-unitarity}).
Therefore, any TeV-scale collider with sufficient luminosity will produce the $S$ or $V$ states
on shell via $\mu^+ \mu^- \to \gamma S/V$. Our challenge in the remainder of this section is not just to identify the highest singlet masses of interest, but rather to demonstrate that a plausible muon collider would unambiguously discover the signatures associated with these states regardless of their mass or how they decay.

\subsection{Constraining the BSM mass scale with Perturbative Unitarity}
\label{singlet-unitarity}

In our analysis, we first calculate the perturbative unitarity constraints on singlet couplings $g_S$ and $g_V$ that arise from the amplitude $\mu^- \mu^+ \to \mu^- \mu^+$ with an intermediate $S$ or $V$. We then calculate how the singlet mass is determined by the coupling to explain $\g$, up to the maximum allowed values of these couplings. This will give a maximum possible mass for the singlet(s).

The amplitude for the process $\mu^-(p_1) \mu^+(p_2) \to S/V \to \mu^-(p_3) \mu^+(p_4)$ is given by (note that we have temporarily switched to 4-component fermion notation for convenience)
\begin{equation}
\mathcal{M}_{S}=\bar{u}_{3}(-ig_{S})v_{4}\frac{i}{s-m_{s}^{2}}\bar{v}_{2}(-ig_{S})u_{1}-\bar{u}_{3}(-ig_{S})u_{1}\frac{i}{t-m_{s}^{2}}\bar{v}_{2}(-ig_{S})v_{4},
\end{equation}
\begin{multline}
\mathcal{M}_{V}=\bar{u}_{3}(-ig_{V}\gamma^{\alpha})v_{4}\frac{i}{s-m_{V}^{2}}\left[-g_{\alpha\beta}+\frac{(p_{1}+p_{2})_{\alpha}(p_{1}+p_{2})_{\beta}}{m_{V}^{2}}\right]\bar{v}_{2}(-ig_{V}\gamma^{\beta})u_{1}\\
-\bar{u}_{3}(-ig_{V}\gamma^{\alpha})u_{1}\frac{i}{t-m_{V}^{2}}\left[-g_{\alpha\beta}+\frac{(p_{1}-p_{3})_{\alpha}(p_{1}-p_{3})_{\beta}}{m_{V}^{2}}\right]\bar{v}_{2}(-ig_{V}\gamma^{\beta})v_{4}.
\end{multline}
We calculated the constraints on the scalar and vector singlets by calculating Eqn. \ref{UniCond} for different $j$. For scalars, the strongest constraint was obtained from the process $\mu^-(\lambda_+) \mu^+(\lambda_+)$ $\to$ $\mu^-(\lambda_-) \mu^+(\lambda_-)$, where $\lambda_\pm$ represents positive/negative helicities. For vectors, the strongest constrain was obtained for the process $\mu^-(\lambda_+) \mu^+(\lambda_-) \to \mu^-(\lambda_-) \mu^+(\lambda_+)$. 
Using the procedures outlined in \sref{s.unitarity}  we get the following constraints:
\begin{equation}
g_S^2 \leq \frac{4 \pi}{N_{\rm BSM}} \ \ \ \ \  , \ \ \ \ \ \ 
g_V^2 \leq \frac{12 \pi}{N_{\rm BSM}} \ \ \ ,
\label{unit_singlet}
\end{equation}
where  $N_{\rm BSM}$ is the number of singlets with common masses and couplings in the theory. 
For $N_{\rm BSM}=1(10)$  the upper bound on the scalar singlet coupling is $g_{S}\leq3.54 \ (1.12)$ and on the vector singlet coupling is $g_{V}\leq6.14 \ (1.94)$.\footnote{The process $\mu^-(\lambda_+) V(\lambda_+) \to \mu^-(\lambda_+) V(\lambda_+)$ can provide stronger constraints for singlet vectors with $N_{\rm BSM}=1.$ However, because this process is $N_{\rm BSM}$ independent, for larger values of $N_{\rm BSM}$ the strongest constraint is provided by Eqn. \ref{unit_singlet}. We omit this constraint from our analysis for simplicity since it does not change our final result.}

\begin{figure}
\begin{center}
\includegraphics[width=7cm]{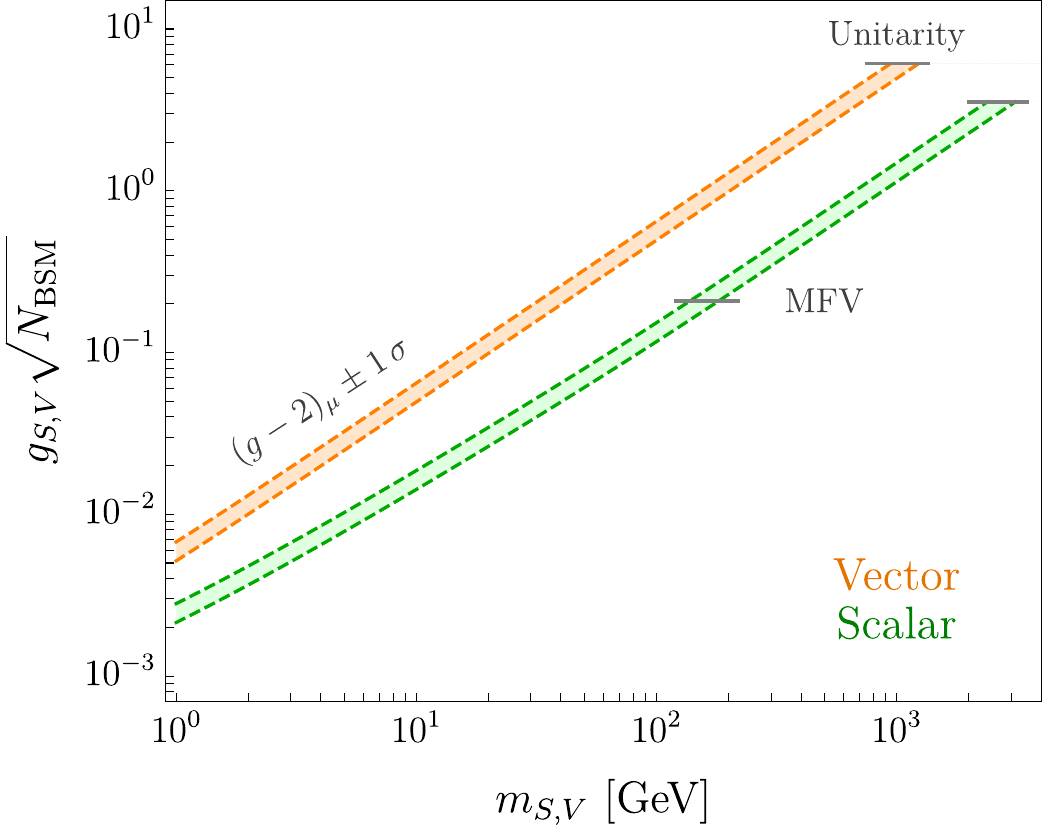}
\end{center}
\vspace{-3ex}
\caption{The coupling of the singlet scalar ($g_S$) and vector ($g_V$) required to account for the $\g$ anomaly as a function of its mass $m_{S, V}$ and multiplicity. For $N_{\rm BSM}=1$, perturbative unitarity imposes $g_S\leq3.5$ and $g_V\leq6.1$, which implies an upper bound on the masses needed for $\g$ of $m_s\leq2.7\tev$ and $m_V\leq1.1\tev$, respectively. If one imposes MFV in the scalar couplings, the upper bounds for scalars become $(g_S,m_s) \leq (0.2,155\gev)$. Note that the $N_{\rm BSM}$-dependence of the singlet mass drops out by requiring $\deltaa = \deltaaexp$.}
\label{f.singletmodelmassvscoupling}
\end{figure}

In \fref{f.singletmodelmassvscoupling} we show the singlet scalar or vector coupling required for a given mass to account for the $\g$ anomaly.  The upper bounds are $m_s\leq2.7\tev$ and $m_V\leq1.1\tev$, for scalar and vector singlets respectively. 
Even though the upper bound on the singlet couplings decreases as the number of BSM flavours increases, the upper bound on the singlet masses does not change, since the $N_{\rm BSM}$ dependence drops out by imposing $\deltaa = \deltaaexp$.

\subsection{Flavour Considerations}
\label{singlet-flavor}

As discussed in \sref{s.unitarityMFV}, CLFV constraints exclude flavour-universal couplings of the scalar to leptons, and severely disfavour anarchic ones. 
This serves as strong motivation for the MFV ansatz in scalar Singlet Scenarios, resulting in a lower maximum mass scale than unitarity alone. \fref{f.singletmodelmassvscoupling} shows that the scalar should be no heavier than 200 GeV if MFV is satisfied.

The vector interaction $V_\alpha ( \mu_L^\dagger \bar \sigma^\alpha \mu_L +
 {\mu^c}^\dagger \bar \sigma^\alpha \mu^c)$  
must arise from a new $U(1)$ gauge extension to the SM, which is spontaneously broken at low energies.
If $V$ is a ``dark photon" whose SM interactions arise from  $V$-$\gamma$ kinetic mixing, the parameter space
for explaining $\g$ has been fully excluded for both visibly and invisibly decaying $V$ \cite{Ilten:2018crw,Bauer_2018}; 
some viable parameter space still exists for semi-visible cascade decays, but 
this will be tested in with upcoming low energy experiments \cite{Mohlabeng_2019}.
 If, instead,  $V$ couples directly to muons, the 
only\footnote{Other $U(1)$ options may also be viable if additional electroweak charged BSM states are included to cancel anomalies, but these models  are phenomenologically similar for the purpose of our $\g$ analysis and are further subject to strong bounds at scales below  the masses of these new particles \cite{Kahn:2016vjr,Dror_2017}.} 
 anomaly free options for this gauge group are 
\be
U(1)_{B-L}~~,~~U(1)_{L_i-L_j}~~,~~U(1)_{B-3L_i},
\ee
where $B$ and $L$ are baryon and lepton number respectively, and $L_i$ is a lepton flavour with $i = e, \mu,\tau$.
Importantly, all of these options require couplings to first generation SM particles and are, therefore,
excluded as explanations for $\g$ by the same bounds that rule out dark photons 
\cite{Ilten:2018crw,Bauer_2018}, see also~\cite{Dror:2017nsg}. The sole exception  is gauged $L_\mu -L_\tau$ which can
still explain the anomaly for $m_V$, but in that case the vector mass is constrained to lie in the narrow range  $\sim$ (1-200) MeV. This scenario  will soon be tested with a variety of low-energy and cosmological probes \cite{Altmannshofer:2014pba,Escudero:2019gzq,Krnjaic:2019rsv,Ballett:2019xoj}.
Therefore, singlet vector scenarios are  less relevant to our discussion of high energy muon collider signatures, but we include them since their phenomenology is nearly identical to that of singlet scalars.

\subsection{Muon Collider Signatures}

We now discuss the collider signatures of Singlet Scenario explanations for the $\g$ anomaly. In particular, here we focus on the region of masses above $\sim \gev$, with the understanding that low energy experiments will cover the lower mass region. The first signal we discuss is direct production of the singlets in association with a photon. The presence of a photon is important because we will consider the possibility that the singlets decay invisibly, in which case the MuC can look for monophoton signatures. This $\gamma+X$ signal is particularly important for low masses. The second signal that we will discuss is Bhabha scattering. The process $\mu^- \mu^+ \to \mu^- \mu^+$ receives contributions via singlet exchange. This process is particularly important for high singlet masses in a low-energy collider. An important question that we want to address is at which luminosity a given signal can be detected at $5\sigma$ significance for a given collider energy.

We consider two possible muon colliders: a high energy 3 TeV collider with $1\,{\rm ab}^{-1}$ of integrated luminosity and a low energy 215 GeV collider (a potential Higgs factory) with $0.4\,{\rm ab}^{-1}$ of luminosity. These benchmark luminosities are discussed by the international muon collider collaboration at CERN \cite{Long:2020wfp}. As opposed to conventional colliders, MuC has the extra complication of Beam-Induced Background (BIB) due to muon decay-in-flight. For this reason the detector design includes two tungsten shielding cones along the direction of the beam. The opening angle of these cones should be optimized as a function of the energy of the MuC. In order to be conservative, our simulations assume that the detector cannot reconstruct particles with angles to the beamline below $10^{\circ}$ ($20^{\circ}$) for the higher (lower) energy muon collider~\cite{Kahn:2011zz}.

\subsubsection{Inclusive Analysis of Singlet Direct Production}

\begin{figure}[t]
\begin{center}
\includegraphics[width=6cm]{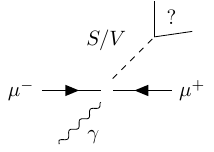}
\end{center}
\vspace{-3ex}
\caption{Single production of the singlet in association with a photon at a muon collider. The singlets can be stable and constitute missing energy, or decay to any SM final states. The search is defined by the search for the recoiling photon, as well as any possible SM final states (including missing energy) inside the singlet decay cone.}
\label{monoA}
\end{figure}

Here we focus on single production of the singlets in association with a photon. 
In principle, to study direct production of the singlets one would need to make an assumption about how they decay to optimally search for them at the collider. 
We want to avoid such a model dependence by implementing an inclusive analysis for singlet + photon production with the following signal topology for a given singlet mass $m_S$, illustrated in \fref{monoA}:
\begin{enumerate}
\item A nearly monochromatic photon with $E_\gamma \sim \sqrt{s}/2$ (with some mild dependence on the singlet mass) in one half of the detector.
\item No other activity anywhere else in the detector, except inside of a ``singlet decay cone'' of angular size $\phi_\mathrm{max}$ around the assumed singlet momentum vector $\vec p_S = - \vec p_\gamma$.
\item For each singlet mass, $\phi_\mathrm{max}$ is defined as the opening angle within which $\sim 95\%$ of singlet decay products must lie, regardless of decay mode. This is determined from simulation under the assumption that the singlet decays to two massless particles, which gives the largest possible opening angle of any decay mode. 
\item There are no requirements of any kind on what final states are found inside the singlet decay cone. This gives near-unity signal acceptance for stable singlets (resulting in missing energy) as well as all possible visible or semi-visible decay modes. 
\end{enumerate}
The veto on detector activity anywhere except the monochromatic photon and inside the singlet decay cone would have to be adjusted for a realistic analysis due to the presence of BIB and initial- and final-state radiation. However, the former is likely to be subtractable and the latter are small corrections at a lepton collider, not greatly reducing signal acceptance. We therefore ignore this complication with the understanding that a more complete treatment would not significantly change our results.

This inclusive analysis allows us to remain as model-independent as possible, something that is necessary when scanning over a large range of singlet masses with only the coupling to the muon known, without paying any branching fraction penalty that would arise by perhaps trying to exploit some minimum decay rate to muons.
For instance, for $m_S \gtrsim 200 \gev$, the muon coupling is $> 1$, making it natural for the dominant decay mode to yield two muons, although other visible or invisible decay modes could be co-dominant. For smaller masses, e.g. close to 1 GeV, the muon coupling is 2-3 orders of magnitude smaller, and the singlet could decay to invisible particles, electrons, quarks, or photons.

Note that instead of searching for bumps in the invariant mass distribution of candidate singlet decay products inside the decay cone, we analyze the photon energy distribution. This takes advantage on the fact that producing an on-shell particle in association with a photon forces the latter to be nearly monochromatic in a lepton collider. 
For a given singlet mass, the photon energy is determined within a bin ($\Delta E_\gamma^{{\rm bin}}$) whose width is correlated with the decay width of the singlet. We calculated $\Delta E_\gamma^{{\rm bin}}$ assuming a decay width of $30\%$ around the mass of the singlet, which is near the upper bound from perturbative unitarity and very conservative. 
For small singlet masses that result in a very narrow photon energy distribution, we instead define the bin size $\Delta E_\gamma^{{\rm bin}}$ to be equal to the energy resolution of the electromagnetic calorimeter (ECAL). We assume an ECAL resolution similar to that of the Large Hadron Collider (LHC) main detectors \cite{Chatrchyan:2013dga}, again a very conservative assumption that takes into account the most important detector effects.
Tables \ref{Ebins1} and \ref{Ebins2} show the assumed photon energy bins $\Delta E_\gamma^{{\rm bin}}$ for a few values of the singlet mass at a 3 TeV and 215 GeV MuC.

\begin{table}[t]
\begin{center}
\begin{tabular}{|c|c|c|c|c|c|c|}
\hline 
\multirow{2}{*}{
\begin{tabular}{c}
Mass\tabularnewline
(GeV)\tabularnewline
\end{tabular}
} & \multirow{2}{*}{%
\begin{tabular}{c}
$E_\gamma$ bin\tabularnewline
(GeV)\tabularnewline
\end{tabular}} & \multirow{2}{*}{%
\begin{tabular}{c}
$\Delta E_\gamma$\tabularnewline
$(\Gamma_{{\rm singlet}})$\tabularnewline
\end{tabular}} & \multirow{2}{*}{%
\begin{tabular}{c}
$\Delta E_\gamma$\tabularnewline
(ECAL)\tabularnewline
\end{tabular}} & 
\multirow{2}{*}{%
\begin{tabular}{c}
Background\tabularnewline
(fb)\tabularnewline
\end{tabular}
} & \multicolumn{2}{c|}{Signal (fb)}\tabularnewline
\cline{6-7} \cline{7-7} 
 &  &  &  &  & Scalar & Vector\tabularnewline
\hline 
\hline 
10 & (1492, 1508) & 0.02 & 16.17 & 3.23 & 0.22 & 4.31\tabularnewline
\hline 
100 & (1490, 1506) & 2.0 & 16.15 & 3.65 & 14.1 & 391\tabularnewline
\hline 
500 & (1433, 1483) & 50 & 15.75 & 2.51 & 372 & 11,177\tabularnewline
\hline 
1000 & (1233, 1433) & 200 & 14.50 & 3.18 & 1,636 & 52,074\tabularnewline
\hline 
\multicolumn{7}{|c|}{Muon Collider Energy: 3 TeV}\tabularnewline
\hline 
\end{tabular}
\end{center}\vspace{-3ex}
\caption{Photon energy bins as well as background and signal cross sections
for different singlet masses. The width
of the energy bin corresponds to the maximum of
the third and fourth columns for a given row. Values in this table correspond to a MuC with $\sqrt{s} = 3$ TeV.}
\label{Ebins1}
\end{table}

\begin{table}[t]
\begin{center}
\begin{tabular}{|c|c|c|c|c|c|c|}
\hline 
\multirow{2}{*}{
\begin{tabular}{c}
Mass\tabularnewline
(GeV)\tabularnewline
\end{tabular}
} & \multirow{2}{*}{%
\begin{tabular}{c}
$E_\gamma$ bin\tabularnewline
(GeV)\tabularnewline
\end{tabular}} & \multirow{2}{*}{%
\begin{tabular}{c}
$\Delta E_\gamma$\tabularnewline
$(\Gamma_{{\rm singlet}})$\tabularnewline
\end{tabular}} & \multirow{2}{*}{%
\begin{tabular}{c}
$\Delta E_\gamma$\tabularnewline
(ECAL)\tabularnewline
\end{tabular}} & \multirow{2}{*}{%
\begin{tabular}{c}
Background\tabularnewline
(fb)\tabularnewline
\end{tabular}} & \multicolumn{2}{c|}{Signal (fb)}\tabularnewline
\cline{6-7} \cline{7-7} 
 &  &  &  &  & Scalar & Vector\tabularnewline
\hline 
\hline 
1 & (106, 108) & 0.01 & 2.07 & 2.56 & 0.247 & 1.58\tabularnewline
\hline 
10 & (106, 108) & 0.28 & 2.07 & 9.14 & 10.86 & 147.4\tabularnewline
\hline 
50 & ( 98, 105) & 6.98 & 2.02 & 77.9 & 172.7 & 3356\tabularnewline
\hline 
100 & ( 90, 96) & 28 & 1.96 & 5.78 & 6.821 & 100.8\tabularnewline
\hline 
\multicolumn{7}{|c|}{Muon Collider Energy: 215 GeV}\tabularnewline
\hline 
\end{tabular}
\end{center}\vspace{-3ex}
\caption{Similar to Table \ref{Ebins1} but for a 215 GeV MuC.}
\label{Ebins2}
\end{table}

We assume singlet production for each possible scalar or vector mass is determined only by the coupling $g_S, g_V$ to the muon, which is in turn fixed by $\deltaa = \deltaaexp$. We then calculated the production cross section by coding up the Singlet Scenarios as simplified models in FeynRules \cite{Degrande:2011ua} and generating tree-level signal events with \texttt{MadGraph5\_aMC@NLO} \cite{Alwall:2014hca}. 
We confirmed that with the above cuts, signal acceptance for singlet decays is close to 1 regardless of decay mode.
The background was calculated by simulating $\gamma+\bar{f}f$ (including neutrinos) and $\gamma+\gamma\gamma$ final states at tree-level and imposing the above cuts in an offline analysis. 
Background contributions involving additional SM states would either fail one of the vetos or cut on additional states outside of the decay cone, or supply small corrections to the lowest-order background rates we calculate in our signal region. Our analysis should therefore reliably estimate the sensitivity of a realistic inclusive singlet search.
Table \ref{Ebins1} shows the total background cross section after imposing analysis cuts for a few values of the singlet mass and compares them to signal. 

\begin{figure}
\begin{center}
\hspace{-1cm} \includegraphics[width=7cm]{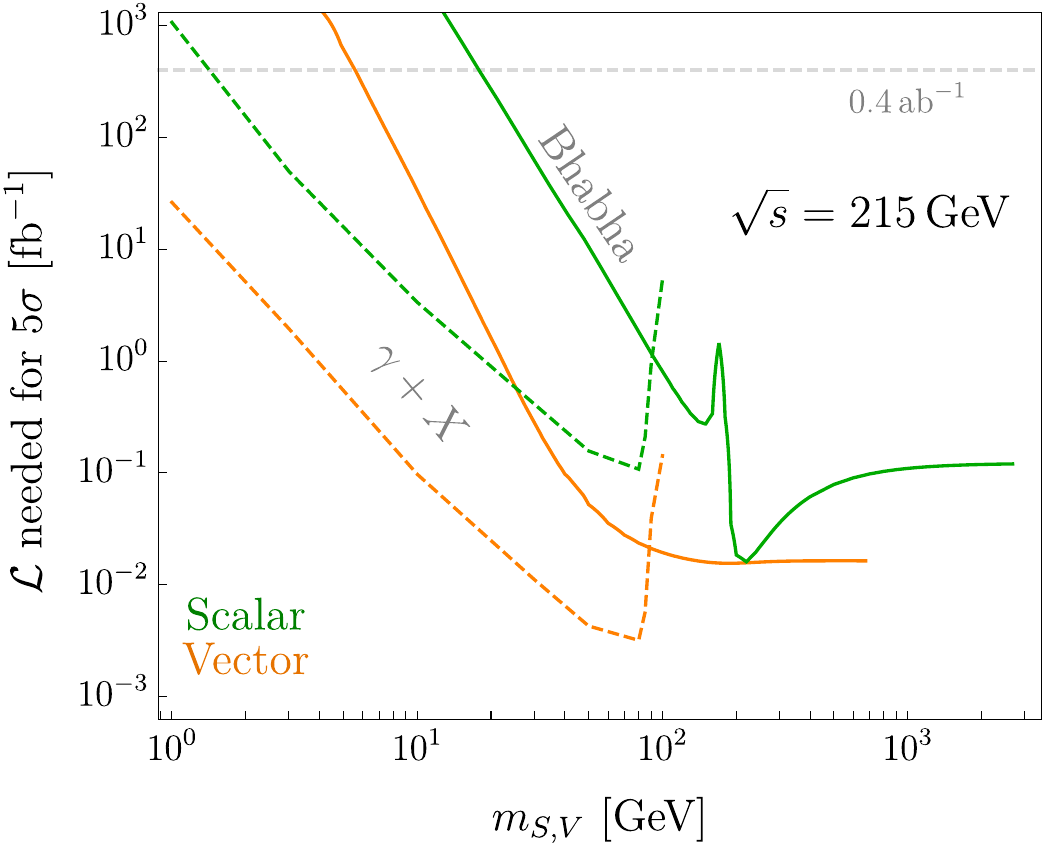}~~
\includegraphics[width=7cm]{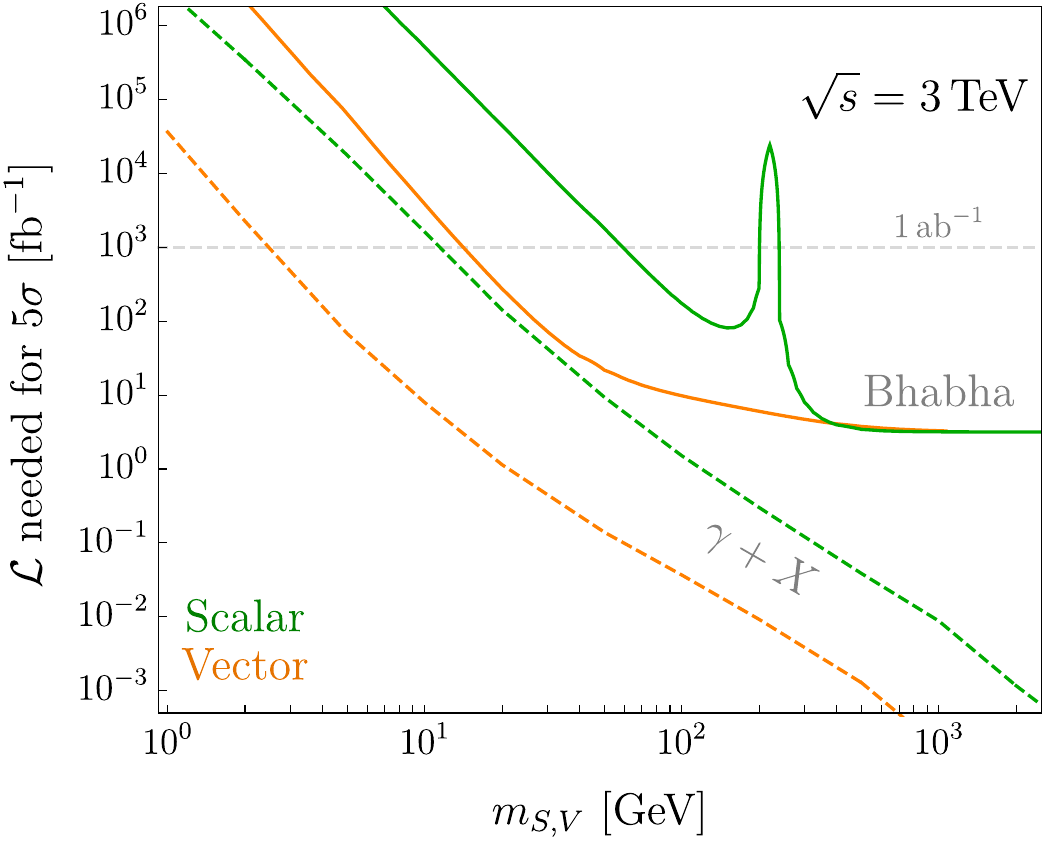}
\end{center}
\vspace{-3ex}
\caption{Luminosity needed for $5\sigma$ discovery significance of  inclusive Singlet Scenario searches at a 215 GeV and 3 TeV muon collider for singlet scalars (green) and singlet vectors (orange). This is shown for singlet masses up to the perturbativity limit calculated in Section \ref{singlet-unitarity}. Dashed lines (solid lines) show the results from the inclusive direct $\gamma+X$ analysis (Bhabha scattering analysis). Note that these sensitivities do not depend on $N_{\rm BSM}$.}
\label{lumi}
\end{figure}

In the right panel of Fig. \ref{lumi}, dashed lines show that a 3 TeV MuC with $1~ \mathrm{ab}^{-1}$ of luminosity will be able to probe singlet masses above 11 GeV for scalars and 2.4 GeV for vectors through $\gamma+X$ events. 
Note that these sensitivities do not depend on $N_{\rm BSM}$, since signal rates at the MuC and $\deltaa$ both scale as $N_{\rm BSM} \, g_{S,V}^2$.

In order to probe smaller masses, one could use a lower energy MuC. In the left panel of Fig. \ref{lumi} we see that a 215 GeV MuC with $0.4~\mathrm{ab}^{-1}$ will probe masses above 1.4 GeV for scalars and sub-GeV masses for vectors, owing to the larger production rate for light states at lower collider energies.
Such a lower-energy collider might be built as a MuC test-bed or Higgs factory, and while it would not  be able to directly produce singlets at the heaviest possible masses allowed by unitarity, it would cover most of the scalar parameter space allowed under the most motivated MFV assumption. Furthermore, as we show in the next section, it will be able to indirectly discover the effects of the Singlet Scenarios by detecting deviations in Bhabha scattering.

\subsubsection{Bhabha Scattering}

\begin{figure}[t]
\begin{center}

\begin{tabular}{lcc}
  \begin{tabular}{l}
    \includegraphics[width=6cm]{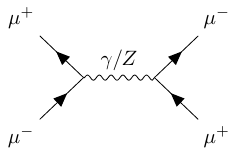}
  \end{tabular} &
  \begin{tabular}{l}
    \includegraphics[width=5.2cm]{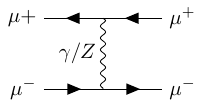}
  \end{tabular}
\end{tabular} \\

\begin{tabular}{lcc}
  \begin{tabular}{l}
    \includegraphics[width=6cm]{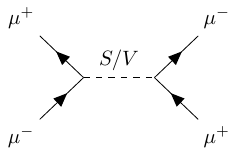}
  \end{tabular} &
  \begin{tabular}{l}
    \includegraphics[width=5.2cm]{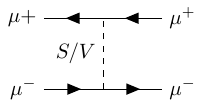}
  \end{tabular}
\end{tabular}

\end{center}
\vspace{-5ex}
\caption{Feynman diagrams for Bhabha scattering in the SM (top) and contributions from singlet scalars or vectors (bottom). (Note that the arrows in this diagram represent charge flow, not helicity.)}
\label{diag_bhabha}
\end{figure}

\begin{figure}
\begin{center}
\hspace{-1cm}
\includegraphics[width=7cm]{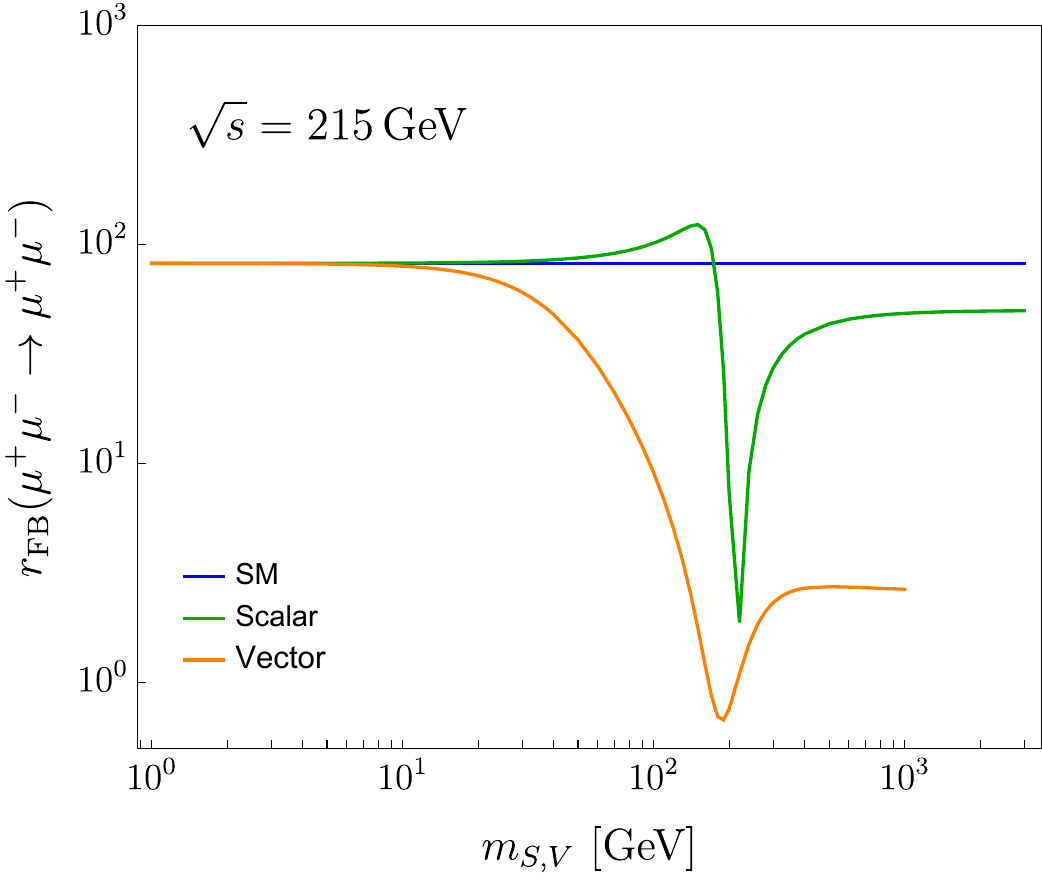}~~~~~
\includegraphics[width=7cm]{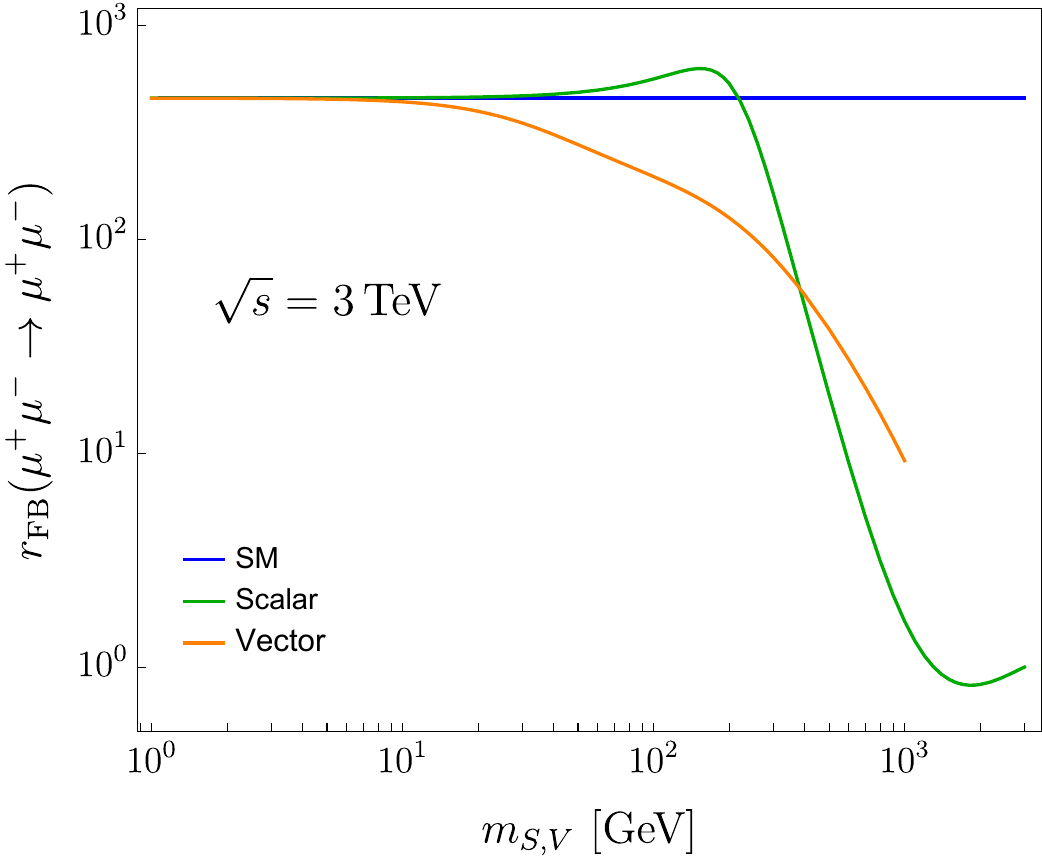}
\end{center}
\vspace{-3ex}
\caption{Prediction for the forward-backward asymmetry variable $r_{\rm FB}$ in Bhabha scattering for Singlet Scenarios at a 215 GeV and 3 TeV MuC. This is independent of $N_{\rm BSM}$.}
\label{r_FB_fig}
\end{figure}

In the Standard Model, Bhabha scattering is mediated by $s$- and $t$-channel exchange of both a photon and a $Z$ boson (Fig. \ref{diag_bhabha}, top). New physics contributions from singlet scalars and vectors have a similar topology (Fig. \ref{diag_bhabha}, bottom) and can produce measurable deviations. 
When the energy of the collisions is close to the mass of the singlets, the distinctive signature of Bhabha scattering is a resonance peak at the mass of the singlet. However, when the energy of the collisions is lower,  one could instead can look for deviations in the total cross section of the process due to contributions from off-shell singlets. 
The potential problem with this approach  is that measurements of total rates for Bhabha scattering are sometimes used to calibrate beams and measure instantaneous luminosity \cite{CarloniCalame:2000pz}. To avoid possible complications in that regard, one can measure deviations in ratio variables similar to a forward-backward asymmetry in parity-violating observables. Ratio variables also have the advantage of mitigating the effect of systematics. 
We therefore define the ratio of the number of forward to backward $\mu^+ \mu^- \to \mu^+ \mu^-$ events:
\begin{equation}
r_{\rm FB}\equiv
\frac{
\displaystyle
\int_{0}^{c_{\theta_0}}\frac{d\sigma}{dc_{\theta}}dc_{\theta}}{
\displaystyle
\int_{-c_{\theta_0}}^{0}\frac{d\sigma}{dc_{\theta}}dc_{\theta}},
\label{r_FB}
\end{equation}
where $c_{\theta}$ is the cosine of the muon scattering angle, $d\sigma/dc_{\theta}$ is the differential cross section of the process $\mu^{-}\mu^{+}\to\mu^{-}\mu^{+}$, and the minimum angle $\theta_0$ is given by the angular acceptance of the MuC detector. 
The  dependence of this variable on singlet mass is illustrated in Fig.~\ref{r_FB_fig} for a 215 GeV (left) and 3 TeV (right) MuC.
For a given mass, the singlet coupling is determined by the value of $\g$. Note that this result again does not depend on $N_{\rm BSM}$ since it depends only on $g_{S,V}^2 N_{\rm BSM}$, which is fixed by $\deltaa = \deltaaexp$.

In Figure~\ref{r_FB_fig}, blue lines represent the SM result. As expected, the number of forward events exceeds that of the backward events by orders of magnitude in the SM. This is typical for Bhabha scattering due to $t$-channel enhancements. 
The contribution of singlets interferes with the SM contribution and reshapes the angular distribution, resulting in deviations from the SM expectation for $r_{\rm FB}$. In particular, near an $s$-channel resonance, $r_{\rm FB} \to 1$, as expected because the singlet-muon coupling is parity-conserving. 
To address the question of how much luminosity is needed to discover deviations from the expected SM behaviour of Bhabha scattering with $5 \sigma$ statistical significance, 
we calculate $r_{\rm FB}$ for the {\it background-only} hypothesis $r_{\rm FB}^{\rm SM}$ and compare it with the {\it background+signal} hypothesis $r_{\rm FB}^{\rm SM+NP}$, obtaining the corresponding $\chi^2$,
\begin{equation}
\chi^{2}=\frac{\left(r_{\rm FB}^{\rm SM+NP}-r_{\rm FB}^{\rm SM}\right)^{2}}{\left(\Delta r_{\rm FB}^{\rm SM+NP}\right)^{2}+\left(\Delta r_{\rm FB}^{\rm SM}\right)^{2}}.
\end{equation}
The uncertainties in the denominator arise from Poisson statistics in the number of forward and backward events expected at each mass and luminosity.

In the right panel of Fig.~\ref{lumi}, solid lines show that a 3 TeV ($1 \ {\rm ab}^{-1}$) MuC will be able to probe singlet masses above 58 GeV for scalars and 14 GeV for vectors through Bhabha scattering. More importantly, a 215 GeV ($0.4 \ {\rm ab}^{-1}$) MuC will probe masses above 17.5 GeV for scalars and 5.5 GeV for vectors. 
The most important role of Bhabha scattering is in enabling a lower-energy 215  GeV muon collider to discover the effects of Singlet Scenarios that solve the $\g$ anomaly over the entire allowed mass range of the singlets (in combination with the inclusive direct search).

\subsection{UV Completion of Scalar Singlet Scenarios}

We close this section by commenting on possible UV completions of Singlet Scenarios.
It is important to keep in mind that the scalar-muon coupling in the singlet scalar model has to be generated by the non-renormalizable operator  $\frac{c_S}{\Lambda} H \mu_L \mu^c S$ after electroweak symmetry breaking. 
There are only a few ways of generating this operator at tree-level using renormalizable interactions.

The simplest  possibility involves the $S H^\dagger H$ operator, which introduces $S$-$H$ mass mixing after electroweak symmetry breaking. Diagonalizing away this mixing induces the $S \mu_L \mu^c$ 
operator which is proportional to both the SM muon Yukawa coupling and $S$-$H$ mixing angle. However, this scenario is experimentally excluded as a candidate explanation
 for $\g$ \cite{Krnjaic:2015mbs} and similar arguments sharply constrain models in which $S$ mixes with the scalar states in a two-Higgs doublet model. 

The singlet-muon Yukawa interaction can also be induced in models where the 
singlet $S$ couples to a vector-like fourth generation of leptons $\psi_i$.
If the $\psi_i$ undergo mass mixing with $L$ and $\mu^c$, the requisite operator 
$S \mu_L \mu^c$ can arise upon diagonalizing the full leptonic
mass matrix after electroweak symmetry breaking.
 In such models, these states inherit the flavour structure of their UV mixing interactions, whose form must be restricted (e.g. by MFV) to ensure that FCNC bounds are not violated. 
 If these additional $\psi_i$ states are sufficiently light ($\lesssim$ few TeV), they may be accessible at future proton and electron colliders, e.g. via 
 established search strategies for heavy new vector-like leptons \cite{Kumar:2015tna}. 
 However, given the multiple dimensionless and dimensionful couplings that these models allow (each with potentially non-trivial flavour structure),  it is also possible for these additional states to be far heavier than the TeV scale,
 and therefore inaccessible at traditional colliders.

 A detailed study of these UV completions is beyond the scope of this paper, but we merely point out that the existence of charged states at or below the TeV scale is not strictly necessary to realize the scalar Singlet Scenario.
 On the other hand, discovering these scalar singlets at a muon collider only relies on the coupling $g_S$ that is determined by solving the $\g$ anomaly.

\section{Analysis of Electroweak Scenarios}
\label{s.ewmodels}

\subsection{SSF and FFS Model Space}
\label{s.EWmodelspace}

In \sref{s.EWmodeldefinition}, we defined the SSF and FFS simplified models, with Lagrangians given in Eqns.~(\ref{e.SSF}) and (\ref{e.FFS}), which we repeat here for convenience 
\begin{eqnarray}
\label{e.SSF2}
\mathcal{L}_\mathrm{SSF}  &\supset& -y_1 F^c L_{(\mu)}  \Phi_A^*  - y_2 F \mu^c  \Phi_B 
- \kappa H \Phi_A^* \Phi_B 
\nonumber \\  &&
- m^2_A |\Phi_A|^2 - m^2_B |\Phi_B|^2 - m_F F F^c
 + h.c. 
 \\ \nonumber \\
\label{e.FFS2}
\mathcal{L}_\mathrm{FFS} &\supset& -y_1 F_A^c L_{(\mu)}   \Phi^*  - y_2 F_B \mu^c  \Phi 
- y_{12} H F_A^c F_B - y_{12}^\prime H^\dagger F_A F_B^c
\nonumber \\ 
&& - m_A F_A F_A^c - m_B F_B F_B^c - m_S^2 |\Phi|^2
 + h.c. \ . \end{eqnarray}
 For $N_{\rm BSM} > 1$, we simply consider multiple degenerate copies of the above field content. 
 In SSF (FFS) models, the fermion $F$ (complex scalar $S$) is in $SU(2)_L$ representation $R$ with hypercharge $Y$, while the two complex scalars $\Phi_{A,B}$ (two fermions $F_{A,B}$) are in representation $R_{A,B}$ with hypercharges $Y_{A,B}$.

 As we discussed in \sref{s.modelindependent}, these two simplified models 
 include the most general form of new one-loop contributions to $\g$, see \fref{f.feynmang-2} (bottom).
In particular, since every particle in the loop is assumed to be a BSM field, the new couplings $y_1, y_2, y_{12}, y_{12}^\prime, \kappa$ are experimentally unconstrained for BSM masses above a TeV or so, and can be chosen to maximize $\deltaa$ subject only to perturbative unitarity (and optionally imposing MFV or naturalness), which in turn allows $\deltaaexp$ to be generated by the heaviest possible BSM states under the assumption of perturbative unitarity and electroweak gauge invariance. 
This allows us to perform the theory space maximization in \eref{e.Mchargedmaxgeneral} by only performing the maximization over the parameter space of all possible SSF and FFS models, as in \eref{e.MchargedmaxSSFFFS}.
The possibilities not covered by these scenarios, like Majorana fermions or real scalars, give smaller $\g$ contributions and hence must feature lighter BSM states than the SSF and FFS scenarios, which does not change the outcome of the theory space maximization. 

Analyzing these two SSF and FFS simplified model classes therefore allows us to find the heaviest possible mass of the lightest new charged state in the theory.
This  dictates the minimum center-of-mass energy a future collider must have to guarantee discovery of new physics by direct Drell-Yan production and visible decay of heavy new states. 
In particular, the discovery of charged states with mass $m \lesssim \sqrt{s}/2$ at lepton colliders is highly robust~\cite{Egana-Ugrinovic:2018roi}, since they have sizeable production rates given by their gauge charge \emph{and} have to lead to visible final states in the detector. This is why our results  allow us to formulate a no-lose theorem for future muon colliders.

\begin{table}
\begin{center}
\footnotesize
\begin{tabular}{|l| |l|l|l| |l|l| |l|l| |l|l| |l|l|}
\hline

& & & &  \multicolumn{8}{c|}{ \bf Highest possible mass (TeV)} \\
& & & &  \multicolumn{8}{c|}{ \bf of lightest charged BSM state} \\
\cline{5-12}
 &&&    
 &\multicolumn{2}{c||}{Unitarity} 
 & \multicolumn{2}{c||}{Unitarity +}  
 & \multicolumn{2}{c|}{Unitarity +} 
  & \multicolumn{2}{c|}{Unitarity +} 
 \\
& & & & 
 \multicolumn{2}{c||}{only} &
  \multicolumn{2}{c||}{MFV} & 
  \multicolumn{2}{c|}{Naturalness}  &
    \multicolumn{2}{c|}{Naturalness + } 
\\
& & & & 
 \multicolumn{2}{c||}{} &
  \multicolumn{2}{c||}{} & 
  \multicolumn{2}{c|}{}  &
    \multicolumn{2}{c|}{MFV} 
\\
& & & & 
 \multicolumn{2}{c||}{$N_{\rm BSM}$:} &
  \multicolumn{2}{c||}{$N_{\rm BSM}$:} & 
  \multicolumn{2}{c|}{$N_{\rm BSM}$:}  &
    \multicolumn{2}{c|}{$N_{\rm BSM}$:} 
\\
Model   & $R$          & $R_A$         & $R_B$    &   
\multicolumn{1}{c}{1}  & \multicolumn{1}{c||}{10} & 
\multicolumn{1}{c}{1}  & \multicolumn{1}{c||}{10} & 
\multicolumn{1}{c}{1}  & \multicolumn{1}{c|}{10}  &
\multicolumn{1}{c}{1}  & \multicolumn{1}{c|}{10}
\\

\hline 
\hline

\multirow{12}{*}{
SSF
}

 & $1 _{-1}$  & $2_{1/2}$  & $1_{0}$    & 65.2 & 241  & 12.9   & 47.1   & 11.5   & 11.5    & 6.54       & 10.1        \\
        & $1 _{-2}$  & $2_{3/2}$  & $1_{1}$    & 85.9 & 321  & 18.1   & 64.8   & 19.2   & 19.2    & 8.41       & 12.3        \\
        & $1 _{0}$   & $2_{-1/2}$ & $1_{-1}$   & 46.2 & 176  & 9.41   & 34.1   & 15.6   & 17.5    & 5.93       & 8.56        \\
        & $1 _{1}$   & $2_{-3/2}$ & $1_{-2}$   & 81.8 & 302  & 17.1   & 63.7   & 19.3   & 19.3    & 8.38       & 12.1        \\

\hhline{|~||===========}

& $2_{-1/2}$ & $3_{0}$    & $2_{-1/2}$ & 21.4 & 107  & 4.2    & 15.5   & 7.47   & 8.99    & 3.23       & 5.0           \\
        & $2_{-3/2}$ & $3_{1}$    & $2_{1/2}$  & 83.7 & 308  & 16.6   & 60.7   & 13.4   & 13.4    & 7.06       & 10.6        \\
        & $2_{1/2}$  & $3_{-1}$   & $2_{-3/2}$ & 95.5 & 356  & 18.3   & 67.8   & 15.6   & 15.6    & 7.75       & 11.3        \\

\hhline{|~||===========}

 & $2_{-1/2}$ & $1_{0}$    & $2_{-1/2}$ & 65.2 & 241  & 12.9   & 47.1   & 11.5   & 11.5    & 6.54       & 10.1        \\
        & $2_{-3/2}$ & $1_{1}$    & $2_{1/2}$  & 85.9 & 321  & 18.1   & 64.8   & 19.2   & 19.2    & 8.41       & 12.3        \\
        & $2_{1/2}$  & $1_{-1}$   & $2_{-3/2}$ & 44.8 & 155  & 8.8    & 32.3   & 10.9   & 10.9    & 5.64       & 8.56        \\

\hhline{|~||===========}

& $3_{-1}$   & $2_{1/2}$  & $3_{0}$    & 95.4 & 359  & 19.4   & 73     & 20.1   & 30      & 7.75       & 11.5        \\
        & $3_{0}$    & $2_{-1/2}$ & $3_{-1}$   & 39.4 & 144  & 7.82   & 28.6   & 10.8   & 15.1    & 4.14       & 6.08        \\

\hline 
\hline

\multirow{12}{*}{
FFS
}

& $1 _{-1}$  & $2_{1/2}$  & $1_{0}$    & 37.3 & 118  & 8.87   & 28     & 12.3   & 18.7    & 4.6        & 7.04        \\
        & $1 _{-2}$  & $2_{3/2}$  & $1_{1}$    & 67.3 & 213  & 15.8   & 50     & 13.5   & 18.8    & 4.86       & 6.93        \\
        & $1 _{0}$   & $2_{-1/2}$ & $1_{-1}$   & 59.1 & 187  & 13.2   & 41.8   & 12.4   & 17.2    & 4.02       & 6.28        \\
        & $1 _{1}$   & $2_{-3/2}$ & $1_{-2}$   & 73.2 & 231  & 17.4   & 55     & 13.9   & 19.7    & 5.04       & 7.25        \\
        
\hhline{|~||===========}

& $2_{-1/2}$ & $3_{0}$    & $2_{-1/2}$ & 40   & 126  & 9.38   & 29.7   & 8.0      & 11.5    & 2.88       & 4.34        \\
        & $2_{-3/2}$ & $3_{1}$    & $2_{1/2}$  & 56.3 & 178  & 13.6   & 42.9   & 11.8   & 16.2    & 4.26       & 6.1         \\
        & $2_{1/2}$  & $3_{-1}$   & $2_{-3/2}$ & 82.3 & 260  & 19.2   & 60.6   & 13.6   & 19      & 4.93       & 7.0           \\
      
\hhline{|~||===========}

& $2_{-1/2}$ & $1_{0}$    & $2_{-1/2}$ & 37.3 & 118  & 8.87   & 28     & 12.3   & 18.7    & 4.6        & 7.04        \\
        & $2_{-3/2}$ & $1_{1}$    & $2_{1/2}$  & 67.3 & 213  & 15.8   & 50     & 13.5   & 18.8    & 4.86       & 6.93        \\
        & $2_{1/2}$  & $1_{-1}$   & $2_{-3/2}$ & 46.2 & 146  & 11.2   & 35.4   & 9.83   & 13.8    & 3.49       & 5.18        \\

\hhline{|~||===========}

& $3_{-1}$   & $2_{1/2}$  & $3_{0}$    & 71   & 225  & 17     & 53.6   & 13.1   & 18.1    & 4.04       & 6.97        \\
        & $3_{0}$    & $2_{-1/2}$ & $3_{-1}$   & 23.4 & 75   & 5.29   & 16.9   & 7.3    & 7.69    & 2.73       & 4.03        \\

\hline 
\hline

\multicolumn{4}{|l||}{
${M^\mathrm{max}_\mathrm{BSM, charged}}$ (max in  each column)
}

  & \bf 95.5 & \bf 359  & \bf 19.4   & \bf 73     & \bf 20.1   & \bf 30      & \bf 8.41       & \bf 12.3       
  \\
\hline
\end{tabular}
\end{center}
\caption{
Summary of all the EW Scenarios we analyze as part of our study. In SSF models, $F \sim R, \Phi_{A,B} \sim R_{A,B}$. In FFS models, $S \sim R, F_{A,B} \sim R_{A,B}$, and the choices of representations are shown in columns 2--4, which covers all unique possibilities satisfying $|Q|\leq 2$ involving $SU(2)_L$ representations up to and including triplets. 
Columns 5--6, 7--8,  9--10 and 11--12 show the highest possible mass in TeV of the lightest BSM state in the spectrum, with the BSM couplings constrained only by unitarity, unitarity + MFV,  unitarity + naturalness and unitarity + naturalness + MFV respectively. For illustration of the $N_{\rm BSM}$ dependence, we show results for a single copy of the BSM states $N_{\rm BSM} = 1$, or for $N_{\rm BSM} = 10$. The highest possible BSM mass scale for unitarity and unitarity + MFV constrained couplings scales as $\sim N_{\rm BSM}^{1/2}$. Adding the naturalness constraint of less than $1\%$ tuning of both the Higgs and muon mass softens this dependence to $\sim N_{\rm BSM}^{1/6}$ (both with and without the MFV constraint). 
%
%
Note that in some scenarios, the lightest charged state does not directly contribute to $\g$, but its existence is nonetheless a requirement of EW gauge invariance. 
The largest possible mass of the lightest new charged state across all the scenarios we examine is shown in the last row, which corresponds to the theory-space maximization in \eref{e.MchargedmaxSSFFFS} and hence \eref{e.Mchargedmaxgeneral}. We do not expect the inclusion of higher $SU(2)_L$ representations to meaningfully increase this upper bound. 
}
\label{t.EWmodels}
\end{table}

Each individual SSF or FFS model is defined by the choice of electroweak representations for the new scalars and fermions. In principle there are infinitely many possibilities that satisfy the requirements in \eref{e.reprequirement}, but theories with very large EW representations lead to issues such as low-energy Landau poles (see \sref{s.landau}) or multiply-charged stable cosmological relics.
We therefore restrict ourselves to models where all new particles have electric charge $|Q| \leq 2$. 
\tref{t.EWmodels} shows a summary of all the EW Scenarios we explicitly analyzed as part of our study, showing the $SU(2)_L \otimes U(1)_Y$ representation of the BSM fields, which are all the unique possibilities with electric charges of 2 or below and representations up to and including triplets of $SU(2)_L$.
This table also lists the highest mass that the lightest charged BSM state in the spectrum can have subject to unitarity, unitarity + MFV, unitarity + naturalness and unitarity + naturalness + MFV constraints. 
For each assumption, the last row contains $M^\mathrm{max}_\mathrm{BSM, charged}$. This constitutes our main result, which we explain in the sections below. Crucially, in some scenarios the lightest charged state does not actually participate in the loop that generates $\deltaaexp$, but its existence is nonetheless required by electroweak gauge invariance.

The requirement of $|Q| \leq 2$ in principle allows for theories featuring  $SU(2)_L$ representations up to and including the $\textbf{5}$. 
 However, we find that the largest possible BSM mass does not appear to increase for higher-rank representations. Therefore, we believe our results for $M^\mathrm{max}_\mathrm{BSM, charged}$ to be robust even though we do not explicitly analyze scenarios involving $\textbf{4}$ and $\textbf{5}$ representations.

\subsection{$\g$ in Electroweak Scenarios}

It is straightforward to compute the general BSM one-loop contribution to $\g$, reproducing results from the literature \cite{Calibbi:2018rzv,Freitas:2014pua}. 
It is convenient to work in the low-energy theory below the scale of electroweak symmetry breaking. 
Consider an effective Lagrangian with a single new Dirac fermion $\Psi_F$ with mass $m_F$ and charge $Q_F$, and a complex scalar $\Phi_S$ with mass $m_S$ and charge $Q_S$ interacting with the muon as follows:
\begin{equation}
\mathcal{L} \supset - \bar \Psi_F (a P_L + b P_R) \mu \Phi_S^*  + h.c. 
\label{e.LIRmaster}
\end{equation}
Note we have temporarily switched to 4-fermion notation for this low-energy calculation; $\mu$ is the muon spinor, and $P_{L,R}$ are the left- and right-chirality projectors.
The contribution of particles $\Psi_F, \Phi_S$ to $\g$ is given by:
\begin{eqnarray}
\nonumber \Delta a_\mu(a,b,m_F,m_S,Q_F,Q_S) = 
- \frac{m_\mu m_F}{8 \pi^2 m_S^2} &&
\left\{
Q_F \left[ \mathrm{Re}(a^* b) I_F(\epsilon, x) + (|a|^2 + |b|^2) \frac{m_\mu}{m_F} \tilde I_F(\epsilon, x) 
\right]
\right.
\\
\nonumber &&
\left.
- Q_S \left[ \mathrm{Re}(a^* b) I_S(\epsilon, x) + (|a|^2 + |b|^2) \frac{m_\mu}{m_F} \tilde I_S(\epsilon, x) 
\right]
\right\}
\\
\label{e.deltaamumaster}
\end{eqnarray}
where $\epsilon = m_\mu/m_S$, $x = m_F^2/m_S^2$ and the loop integrals are:
\begin{eqnarray}
I_F(\epsilon, x) &=& \int_0^1 dz \frac{(1-z)^2}{(1-z)(x - z \epsilon^2) + z}
\\ \nonumber \\
\tilde I_F(\epsilon, x) &=& \frac{1}{2} \int_0^1 dz \frac{z (1-z)^2}{(1-z)(x - z \epsilon^2) + z}
\\ \nonumber \\
I_S(\epsilon, x) &=& \int_0^1 dz \frac{z (1-z)}{(1-z)(1 - z \epsilon^2) + z x}
\\ \nonumber \\
\tilde I_S(\epsilon, x) &=& \frac{1}{2} \int_0^1 dz \frac{z (1-z)^2}{(1-z)(1 - z \epsilon^2) + z x}
\end{eqnarray}

\eref{e.deltaamumaster} makes it straightforward to calculate $\g$ for all the EW Scenarios in \tref{t.EWmodels} (which may involve several scalar-fermion combinations coupling to the muon and contributing to $\deltaa$), after solving for the BSM spectrum after EWSB. 
In FFS models, 
\begin{equation}
\label{e.deltaaFFS}
\deltaa \  \sim \ N_{\rm BSM} \ y_{1, 2}^2   \frac{y_{12}^{(\prime)} v \ m_\mu}{m_{\rm BSM}^2} \ ,
\end{equation}
where $m_{\rm BSM}$ is some combination of the BSM particle masses, while for FFS models,
\begin{equation}
\label{e.deltaaSSF}
\deltaa \  \sim \  N_{\rm BSM} \ y_{1, 2}^2  \frac{\kappa v  \ m_\mu}{m_{\rm BSM}^3}.
\end{equation}
Once upper bounds on the BSM couplings from unitarity or other considerations are determined, we can therefore find upper bounds on the  BSM mass scale under the assumption that $\deltaa = \deltaaexp$.

\subsection{Constraining the BSM Mass Scale with Perturbative Unitarity}
\label{s.ew-unitarity}

As discussed in \sref{s.unitarity}, the BSM couplings in SSF and FFS theories have to satisfy perturbative unitarity. Deriving the upper bounds for the new Yukawa couplings is straightforward. We constrain the Yukawa couplings $y_1$ and $y_2$ in the SSF models from the process $\mu^-(\lambda_\pm) F(\lambda_\mp) \to \mu^-(\lambda_\pm) F(\lambda_\mp)$. The same Yukawas in the FFS models were constrained from processes $\mu^-(\lambda_\pm) S \to \mu^-(\lambda_\pm) S$, whereas for the extra Yukawas $y_{12}$ and $y_{12}'$ we used the processes $f_i(\lambda_\pm) f_j(\lambda_\pm) \to f_k(\lambda_\pm) f_l(\lambda_\pm)$, where $f_i$ are the mass eigenstates of the two fermions in the model after mixing.
For scalar-fermion scattering, the intermediate fermion propagator scales at large $s$ as $1/\sqrt{s}$ for the $+ \to +$ helicity-preserving amplitude, and $M/s$ for the helicity-violating $+ \to -$ amplitude, where $M$ is the mass of the intermediate fermion. After taking into account the normalization of the initial-and final-state spinors, we find that the $+ \to +$ amplitudes are independent of energy (and give constraints $y \simeq \mathcal{O}(1) \times \sqrt{4\pi}$ where $y$ is a Yukawa coupling), while the $+ \to -$ amplitudes are largest at small $s$. 
For the SSF and FFS model respectively, the constraints are:
\begin{eqnarray}
\label{e.SSFunitarityyukawa}
|y_1|, |y_2| &\leq& \sqrt{16 \pi}  \approx 7.09 \ \ \ \ \ \ \mbox{(SSF unitarity bound)}
\\ 
\nonumber \\
\label{e.FFSunitarityyukawa}
|y_1|, |y_2| &\leq& \sqrt{8 \pi} \approx 5.01 \ \ \ \ \ \ \mbox{(FFS unitarity bound)} 
\\
|y_{12}|, |y_{12}^\prime| &\leq& \sqrt{4 \pi} \approx 3.55
\nonumber 
\end{eqnarray}
independent of $N_{\rm BSM} $.

Obtaining a unitarity bound for the dimensionful coupling $\kappa$ in SSF models is slightly more involved. It has to satisfy $|\kappa| < \kappa_{\rm max}$, where parametrically,
\begin{equation}
\label{e.kappamax}
\kappa_{\rm max} = d(m_A, m_B, m_F) \frac{m_A m_B}{v}.
\end{equation}
The dimensionless factor $d$ is a function of BSM mass parameters with size $d \sim \mathcal{O}(0.1 - 1)$ if there is large hierarchy between $m_A$ and $m_B$, asymptoting to $d \ll 1$ as $m_A \to m_B$.
This upper bound on the size of $\kappa$ is far more restrictive than the requirement that none of the new scalars acquire VEVs. 
The derivation is as follows. Scalar-scalar amplitudes are a sum of 3-point and 4-point diagrams; the latter are independent of energy, but the former scale as $\kappa^2/s$. Thus the amplitude will be largest, and hence the strongest constraints on $\kappa$ will generally be obtained, at the smallest $s$ which is kinematically accessible, which in principle motivates focusing on the scattering channels with the smallest initial- and final-state masses, namely $h S_i \to h S_j$. However, these processes include cases where $s$-, $t$-, and $u$-channel singularities appear. The $s$-channel poles appear due to the exchange of a scalar $S_k$ whose mass is above the threshold $s=(m_h+m_{S_i})^2$. We can avoid dealing with such poles by considering the scattering of the lightest scalars $S_i$ through $s$- and $t$-channel exchange of a Higgs boson. This way, neither of the $s,t,u$ channel singularities appear when calculating the constraints given by Eqn. (\ref{UniCond}). In this sense, our constraints are conservative, but they avoid defining arbitrary ways to deal with singularities (a fully correct treatment would be model-dependent), and is sufficient to find a conservative but useful estimate of $M^\mathrm{max}_\mathrm{BSM, charged}$.

The scattering amplitude for the process $S_i S_i \to S_i S_i$ is given by 
\begin{equation}
\mathcal{M}=-4\lambda_{{\rm eff}}-\kappa_{{\rm eff}}^{2}\left(\frac{1}{s-m_{h}^{2}}+\frac{1}{t-m_{h}^{2}}\right),
\end{equation}
where the coefficients $\lambda_{{\rm eff}}$ and $\kappa_{{\rm eff}}$ are functions of mixing angles, self-quartics for the scalars $S_A$, $S_B$, quartics between different scalars and/or the Higgs (indicated by subscripts):
\begin{equation}
\lambda_{{\rm eff}}=\cos\theta^4\lambda_A+\cos\theta^2\sin\theta^2\lambda_{AB}+\sin\theta^4\lambda_B,
\end{equation}
\begin{equation}
\kappa_{{\rm eff}}=-\sqrt{2}\cos\theta\sin\theta\kappa+\cos\theta^2v\lambda_{AH}+\sin\theta^2v\lambda_{BH}.
\end{equation}
From this process, the lowest-order partial wave is given by
\begin{multline}
a_{0}=-\frac{1}{32\pi}\left\{ \sqrt{1-\frac{4m_{S_{i}}^{2}}{s}}\left(8\lambda_{{\rm eff}}+\frac{2\kappa_{{\rm eff}}^{2}}{s-m_{h}^{2}}\right) + \frac{2\kappa_{{\rm eff}}^{2}}{\sqrt{s(s-4m_{S_{i}}^{2})}}\log\left[\frac{m_{h}^{2}}{m_{h}^{2}+(s-4m_{S_{i}}^{2})}\right]\right\}.
\end{multline}
The unitarity bound on $\kappa < \kappa_{max}$ corresponds to the maximum value that for a given set of parameters (couplings, masses, etc.), satisfies the condition
\begin{equation}
\left|{\rm Re}(a_{0})\right|\leq\frac{1}{2},
\end{equation}
for large $s$, but since the constraint asymptotes rapidly above threshold, this corresponds to requiring consistency of the theory close to (a factor of a few above) threshold $s \gtrsim 4m_{S_i}^2$. To marginalize over the dependence of scalar quartic couplings, we maximized $\kappa_{max}$ with respect to the unknown quartics, subject to these quartics themselves obeying perturbative unitarity.

\begin{figure}
\begin{center}
\begin{tabular}{ccc}
& 
\footnotesize \phantom{aaaa} SSF, all BSM fields charged
&
\footnotesize \phantom{aaaa}  SSF, charged and neutral fields
\\
\begin{turn}{90}\footnotesize \phantom{aaaaaaaaa}Unitarity only\end{turn}
& \includegraphics[width=0.3\textwidth]{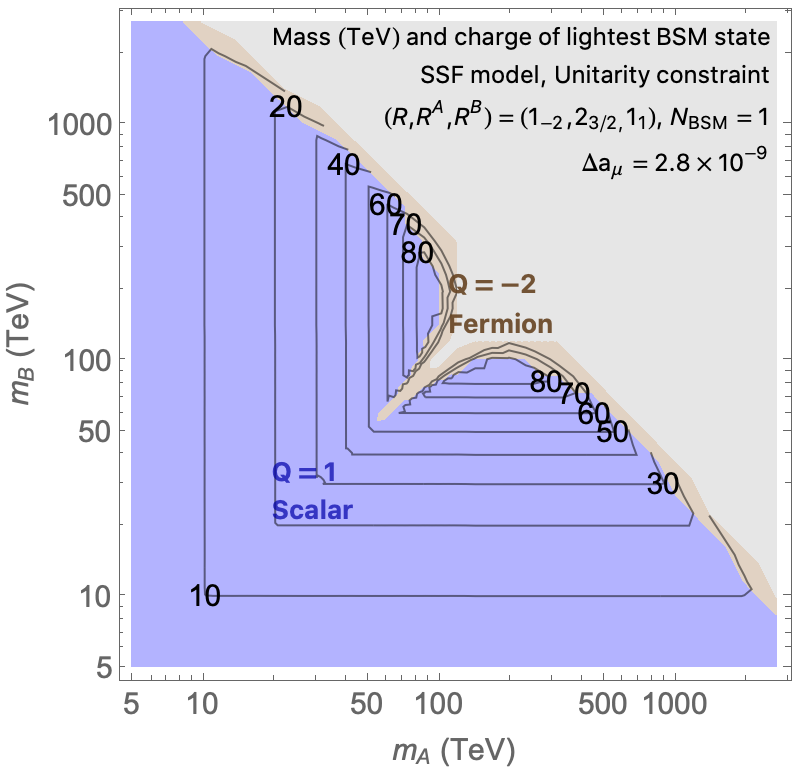}
& 
\includegraphics[width=0.3\textwidth]{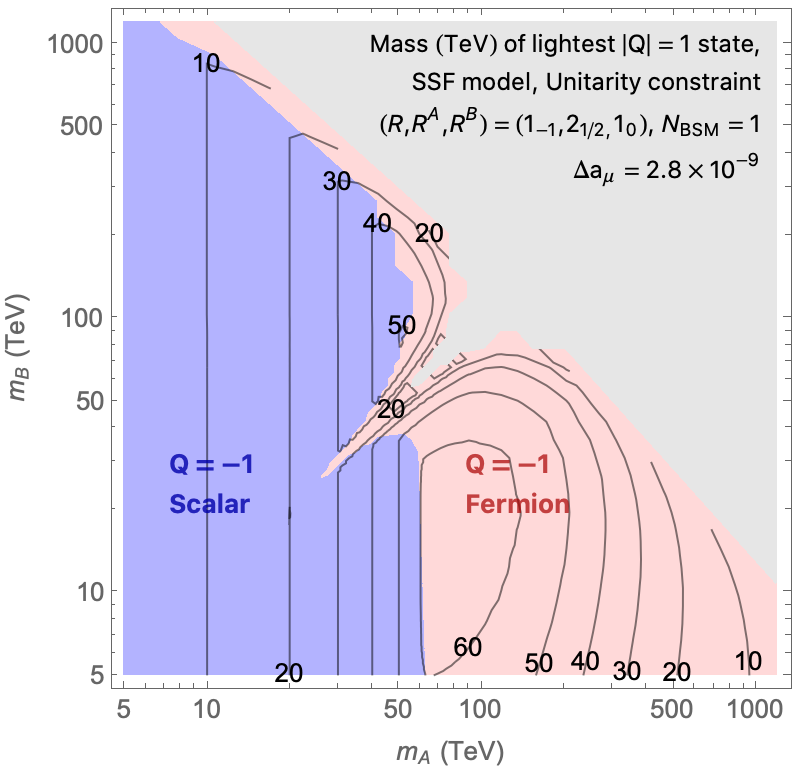}
\\
\begin{turn}{90}\footnotesize \phantom{aaaaaaaa}Unitarity + MFV\end{turn}
& \includegraphics[width=0.3\textwidth]{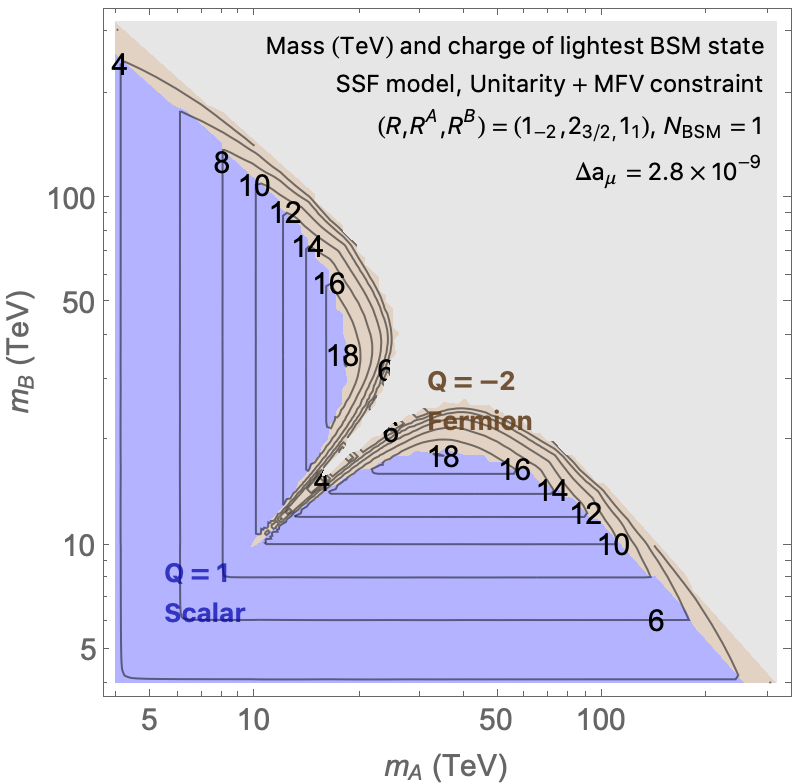}
& 
\includegraphics[width=0.3\textwidth]{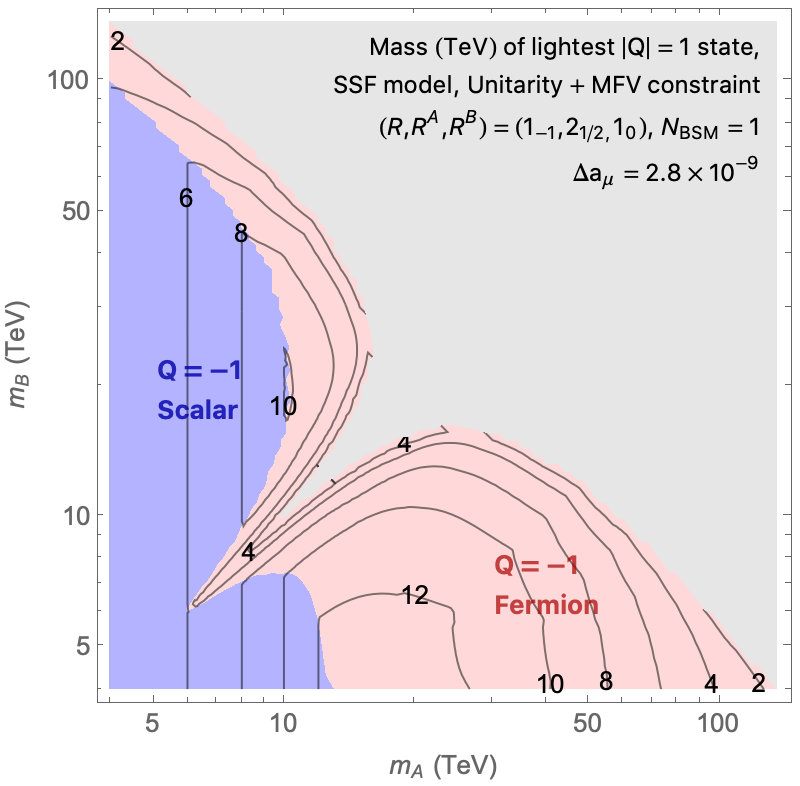}
\\
\begin{turn}{90}\footnotesize \phantom{aalaa}Unitarity + Naturalness\end{turn}
& \includegraphics[width=0.3\textwidth]{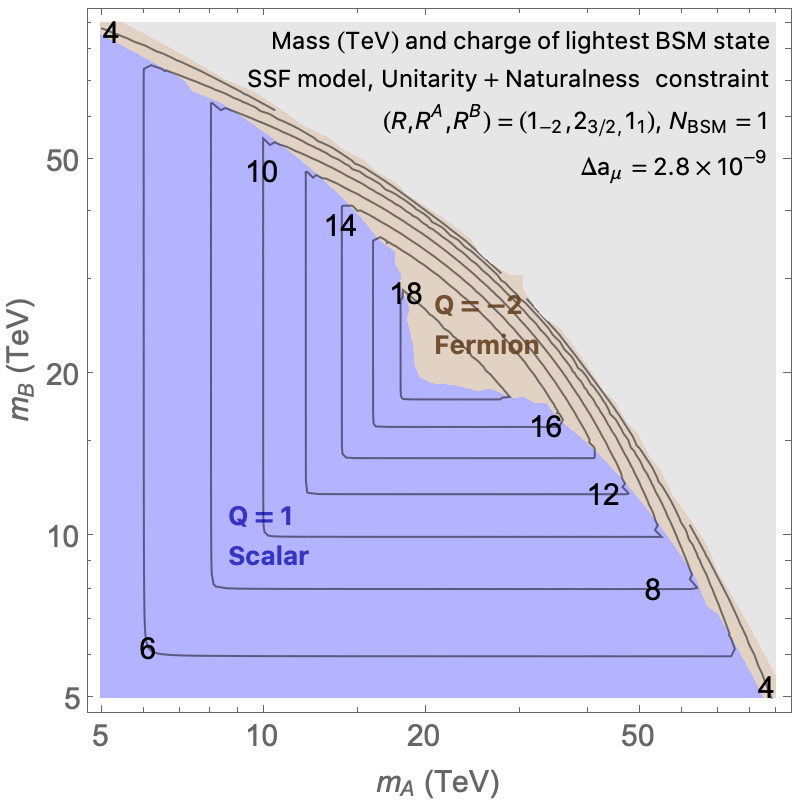}
& 
\includegraphics[width=0.3\textwidth]{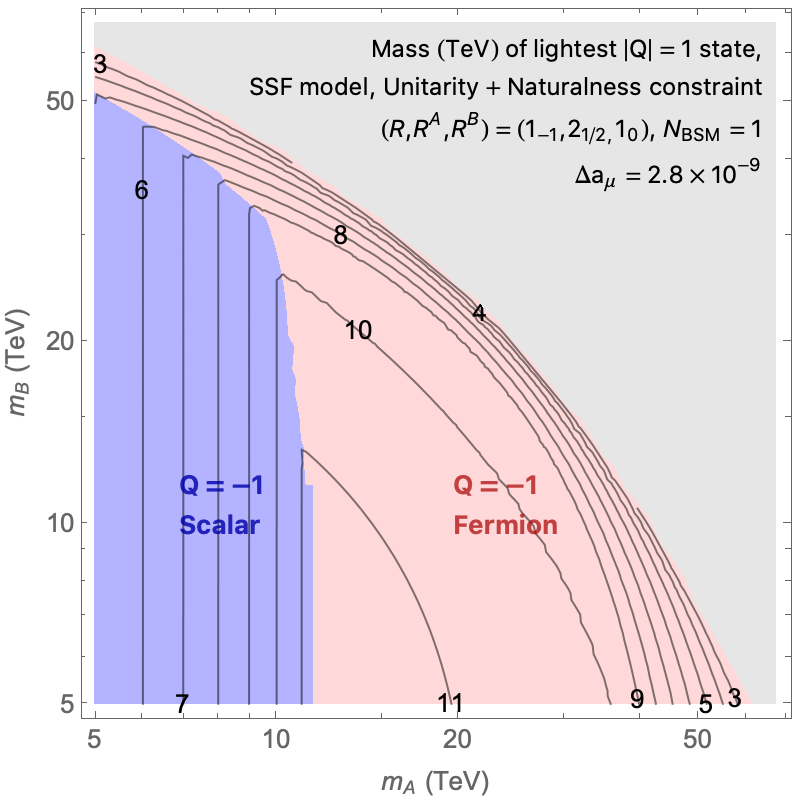}
\\
\begin{turn}{90}\footnotesize \phantom{.}Unitarity + Naturalness + MFV\end{turn}
& \includegraphics[width=0.3\textwidth]{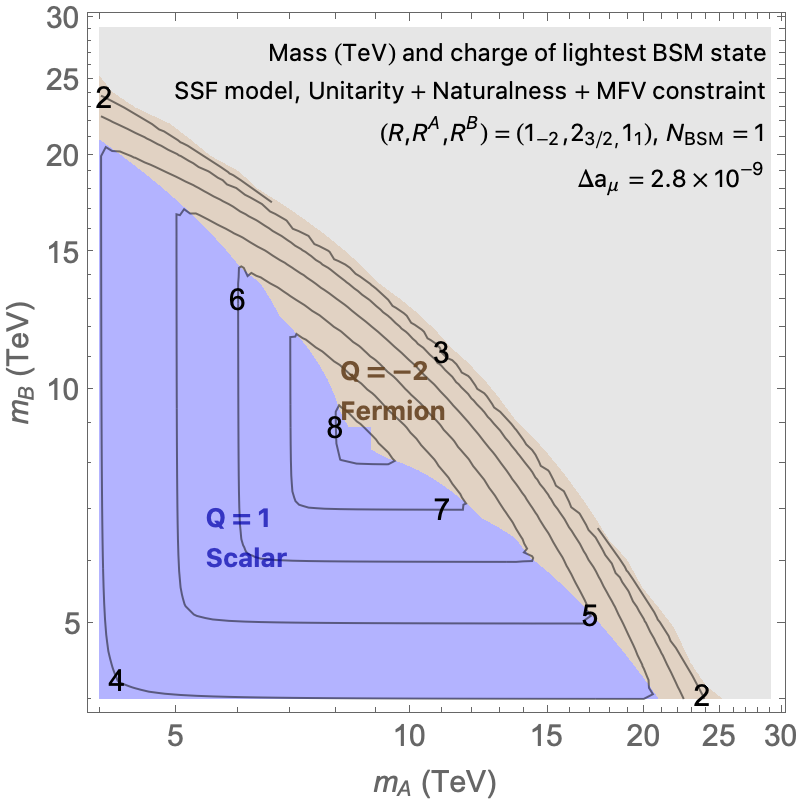}
& 
\includegraphics[width=0.3\textwidth]{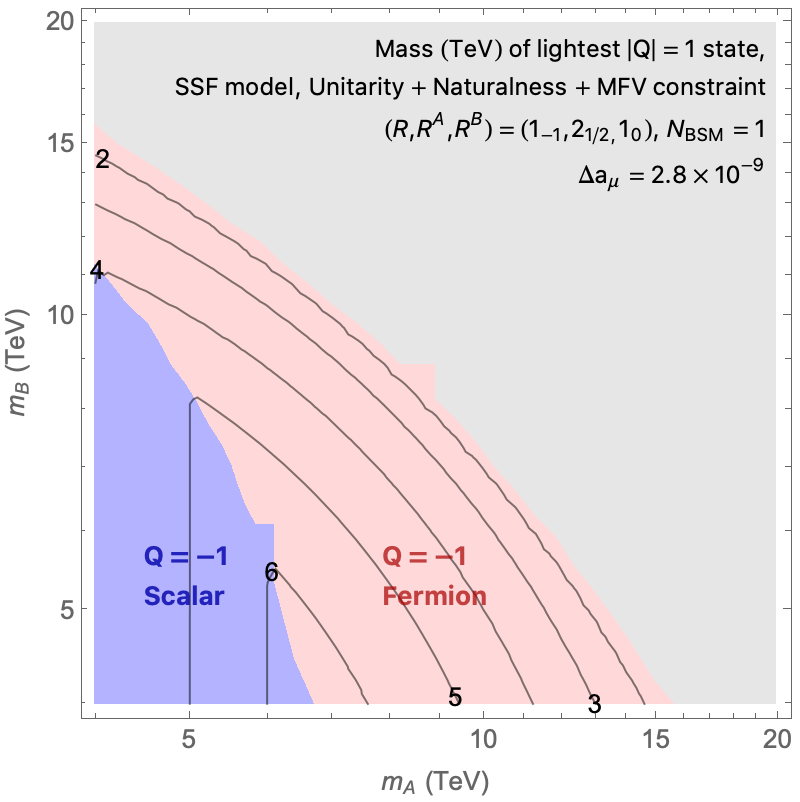}
\end{tabular}
\end{center}
\vspace*{-5mm}
\caption{
Contours show mass in TeV of lightest charged state in two representative SSF models with $N_{\rm BSM} = 1$ as a function of scalar masses $m_A, m_B$. 
The largest possible fermion mass $m_F$ was determined by $\Delta a^\mathrm{BSM} = \deltaaexp$, with the couplings $y_1, y_2, \kappa$ chosen to maximize $\g$ while obeying the constraint from perturbative unitarity (1st row), unitarity + MFV (2nd row), unitarity + naturalness (3rd row) or unitarity + naturalness + MFV (4th row)
On the left, $(R, R^A, R^B) = (1_{-2}, 2_{3/2}, 1_1)$, and all fields contributing to $\g$ are charged. On the right, $(R, R^A, R^B) = (1_{-1}, 2_{1/2}, 1_0)$, and the scalars in the $\g$ loop are neutral but since $\Phi_A$ is an EW doublet, there is a charged scalar with mass $m_A$. 
}
\label{f.EWresults}
\end{figure}

We can now find the upper bound on the  BSM particle masses in each model, under the assumption that $\deltaa = \deltaaexp$. 
For each SSF (FFS) model in \tref{t.EWmodels} the explicit steps in the calculation are the following:
\begin{enumerate}
\item For a given choice of scalar (fermion) mass parameters $m_A, m_B$ and coupling $\kappa$ ($y_{12}, y_{12}^\prime$), find the masses and effective muon couplings of all the mass eigenstates. The $\deltaa$ contribution can then be found using \eref{e.deltaamumaster}. 
\item Find largest fermion mass $m_F$ (scalar mass $m_S$) that can still  generate $\deltaaexp$, under the assumption that the BSM couplings $y_1, y_2, \kappa $ ($y_1, y_2, y_{12}, y_{12}^\prime$) are chosen to maximize $\deltaa$ subject only to the above unitarity bounds. 
\item With the fermion (scalar) mass fixed to this maximum value and the couplings chosen to maximize $\deltaa$, the entire  BSM spectrum of the theory is fully determined as a function of just the two  scalar (fermion) masses $m_A, m_B$. As expected, we find that $\deltaaexp$  can be generated only in a compact region of the $(m_A, m_B)$-plane. 
\item We can then ask at each point in this plane what the mass of the lightest charged BSM state is. This is shown in \fref{f.EWresults} (1st row) for two representative SSF models. Importantly, in some theories, the lightest charged state does not contribute to $\g$, but its existence and mass is determined by  gauge invariance in the given SSF or FFS model.
\item Since the region of parameter space that can account for $\deltaaexp$ is compact, we can determine the \emph{highest possible mass} of the \emph{lightest charged BSM state} that is consistent with this particular EW Scenario accounting for the $\g$ anomaly.
\end{enumerate}
In effect, this procedure allows us to explore the ``maximum-BSM-mass boundary'' of each EW Scenario's parameter space, subject to the requirement that $\deltaa = \deltaaexp$ and the BSM couplings obey perturbative unitarity. 
The resulting highest possible mass of the lightest BSM state in the spectrum for each EW Scenario we examine is listed in columns 5 and 6 of \tref{t.EWmodels} for $N_{\rm BSM}  = 1$ and 10 respectively.

Obviously, the result for a given model in \tref{t.EWmodels} is not particularly illuminating, since it is by definition model-dependent. 
However, obtaining this maximum allowed mass of the lightest new charged state for different possible choices of EW gauge representations in both SSF and FFS models allows us to perform the theory space maximization in \eref{e.MchargedmaxSSFFFS}, and hence obtain $M^\mathrm{max}_\mathrm{BSM, charged}$ for all possible perturbative solutions of the $\g$ anomaly:
\begin{equation}
\label{e.Mchargedmaxunitarity}
M^\mathrm{max, unitarity}_\mathrm{BSM, charged}  \ \ \ 
\equiv \ \ \ 
\max_{\tiny \deltaa = \deltaaexp \ , \  \mathrm{perturbative\ unitarity}} \ \  \left\{ \ \  \min_{\tiny i  \ \in \  \mathrm{BSM\ spectrum}} \left( m_{\rm charged}^{(i)}   \right)  \ \ \right\}
\end{equation}
where we have added the `unitarity' superscript to distinguish this bound from subsequent results with additional assumptions.
We can perform this maximization by taking the largest values from columns 5 and 6 in \tref{t.EWmodels}, which are shown in the last row. We therefore present the final result of our perturbative unitarity analysis of EW Scenarios:
\begin{equation}
\label{e.Mchargedmaxunitarity2}
M^\mathrm{max, unitarity}_\mathrm{BSM, charged} 
 \ \approx \ 
  \left\{
\begin{array}{ll}
100 \tev & \mbox{for $N_{\rm BSM} = 1$}
\\
360 \tev & \mbox {for $N_{\rm BSM} = 10$}
\end{array}
\right\}
 \ \approx \ 
 (100 \tev)  \ \cdot \ N_{\rm BSM}^{1/2}\ .
\end{equation}
The $N_{\rm BSM}$ scaling arises due to the linear dependence of $\deltaa$ on $N_{\rm BSM}$. For FFS models, this is clearly seen from \eref{e.deltaaFFS}, while for SSF models this relationship is obscured by the detailed form of the unitarity bound on $\kappa$, but we verified the approximate $\sqrt{N_{\rm BSM}}$ scaling empirically. 
New charged states therefore have to appear at or below the 100 TeV scale unless $N_{\rm BSM}$ is truly enormous, a scenario which is disfavoured not just by theoretical parsimony but also by avoiding Landau poles close to the BSM mass scale, see \sref{s.landau}. 

It is important to keep in mind that realizing this upper bound from unitarity would also require extreme alignment of the non-muonic BSM couplings to avoid CLFV decay bounds, see \sref{s.unitarityMFV}. This can be regarded as a severe form of tuning of the BSM lepton couplings before mass diagonalization, which disfavours the unitarity-only assumption.

\subsection{Constraining the BSM Mass Scale with Unitarity + MFV}
\label{EWflavor}

As discussed in \sref{s.unitarityMFV}, the MFV assumption is motivated for EW Scenarios by severe experimental bounds on CLFV decays. 
Adopting this ``Unitarity + MFV'' assumption significantly reduces the maximum allowed BSM mass scale. 
We repeat verbatim the unitarity-only analysis from \sref{s.ew-unitarity}, with the additional step of lowering the perturbativity bound on either $y_1$ or $y_2$ by $m_\mu/m_\tau$, whichever gives higher BSM masses at that point in parameter space. (In practice there is almost no difference between these two possibilities since $\deltaa \propto y_1 y_2$ up to tiny corrections.)
The resulting largest possible mass of the lightest BSM charged state for two representative SSF models is shown in \fref{f.EWresults} (2nd row), with the results for all EW Scenarios we examine summarized in 
columns 7 and 8 of \tref{t.EWmodels} for $N_{\rm BSM} = 1$ and 10 respectively.
We can therefore define, for all possible perturbative solutions of the $\g$ anomaly that obey MFV: 
\begin{equation}
\label{e.MchargedMFV}
M^\mathrm{max, MFV}_\mathrm{BSM, charged}  \ \ \ 
\equiv \ \ \ 
\max_{\tiny \deltaa = \deltaaexp \ , \  \mathrm{unitarity} \ , \ \mathrm{MFV}} \ \  \left\{ \ \  \min_{\tiny i  \ \in \  \mathrm{BSM\ spectrum}} \left( m_{\rm charged}^{(i)}   \right)  \ \ \right\}
\end{equation}
where the outer theory-space maximization is now constrained by unitarity as well as MFV, and can again be performed by taking the largest values from columns 7 and 8 in \tref{t.EWmodels}, which are shown in the last row. This gives:
\begin{equation}
\label{e.MchargedMFV2}
M^\mathrm{max, MFV}_\mathrm{BSM, charged} 
 \ \approx \ 
  \left\{
\begin{array}{ll}
20 \tev & \mbox{for $N_{\rm BSM} = 1$}
\\
73 \tev & \mbox {for $N_{\rm BSM} = 10$}
\end{array}
\right\}
 \ \approx \ 
 (20 \tev)  \ \cdot \ N_{\rm BSM}^{1/2} \ .
\end{equation}
The reduction in BSM mass scale compared to the unitarity-only assumption is very significant, and could be within reach of future muon collider proposals.

\subsection{Constraining the BSM Mass Scale with Unitarity + Naturalness}

The physical concreteness of the Higgs and muon mass corrections in EW Scenarios, see Eqns.~(\ref{e.deltamh}) - (\ref{e.deltaymu}), means that confirmation of the $\g$ anomaly \emph{and} confirmed non-existence of the required new charged states up to some scale $M_{exp}$ means that these states \emph{must} exist at some scale $M_{\rm BSM} > M_{exp}$, which implies a certain amount of tuning in the Lagrangian. 
Such an empirical confirmation of fine-tuning would have profound consequence for our thinking about the hierarchy problem or cosmological vacuum selection. 
It is therefore worth quantifying how heavy the new charged states could be \emph{without} inducing such physical fine-tuning. 

We therefore define a very conservative ``naturalness'' criterion by requiring the tuning in both the Higgs mass and the muon Yukawa coupling to not exceed $1\%$, which amounts to imposing
\beq
\label{e.naturalness}
\Delta \equiv \mathrm{max}\left(\frac{\Delta m_H^2}{m_H^2}, \frac{\Delta y_\mu}{y_\mu}\right) < 100 \ .
\eeq
We repeat verbatim the unitarity-only analysis from \sref{s.ew-unitarity}, with the above naturalness bound applied in addition to the unitarity bound. In practice, this means that both the Higgs and muon masses are tuned at the $1\%$ level for the largest BSM masses we find, since maximizing all couplings relevant for $\Delta  a_\mu$ saturates both tuning bounds.

The largest possible mass of the lightest BSM charged state for two representative SSF models under this ``unitarity + naturalness'' assumption is shown in \fref{f.EWresults} (3rd row), with the results for all EW Scenarios we examine summarized in 
columns 9 and 10 of \tref{t.EWmodels} for $N_{\rm BSM} = 1$ and 10 respectively.
We can therefore define, for all possible perturbative solutions of the $\g$ anomaly that obey our conservative naturalness requirement \eref{e.naturalness}, the  largest possible mass of the lightest BSM states:
\begin{equation}
\label{e.Mchargedmaxnaturalness}
M^\mathrm{max, naturalness}_\mathrm{BSM, charged}  \ \ \ 
\equiv \ \ \ 
\max_{\tiny \deltaa = \deltaaexp \ , \  \mathrm{unitarity} \ , \ \mathrm{\Delta < 100}} \ \  \left\{ \ \  \min_{\tiny i  \ \in \  \mathrm{BSM\ spectrum}} \left( m_{\rm charged}^{(i)}   \right)  \ \ \right\}
\end{equation}
where again the superscript indicates the additional naturalness constraint on the theory space maximization, and we can perform this maximization by taking the largest values from columns 9 and 10 in \tref{t.EWmodels}, which are shown in the last row. This gives:
\begin{equation}
\label{e.Mchargedmaxnaturalness2}
M^\mathrm{max, naturalness}_\mathrm{BSM, charged} 
 \ \approx \ 
  \left\{
\begin{array}{ll}
20 \tev & \mbox{for $N_{\rm BSM} = 1$}
\\
30 \tev & \mbox {for $N_{\rm BSM} = 10$}
\end{array}
\right\}
 \ \approx \ 
 (20 \tev)  \ \cdot \ N_{\rm BSM}^{1/6}\ .
\end{equation}
The reduction in BSM mass scale compared to the unitarity-only analysis is even more dramatic than for the MFV assumption.
The unusual $N_{\rm BSM}$ scaling was empirically determined, but arises because unlike the unitarity constraint, the tuning constraint on the couplings becomes more severe with increasing BSM multiplicity, which mostly cancels the increased contribution to $\deltaa$.\footnote{In fact, for many SSF models the maximum BSM mass is realized in regions of parameter space where the maximum allowed value for \emph{all} BSM couplings is set by the naturalness constraint. In that case the $N_\mathrm{BSM}$ dependence cancels exactly, but this does not affect the model-exhaustive upper bound, since it is not the case for all SSF models, and is never  the case for FFS models (which have an additional BSM coupling, meaning that there is always a coupling combination that can saturate unitarity).}

\subsection{Constraining the BSM Mass Scale with Unitarity + Naturalness + MFV}

Given how strongly CLFV decay bounds motivate the MFV ansatz, it is reasonable to ask how high the BSM mass scale could be if solutions to the $\g$ anomaly have to respect both naturalness and MFV. 
We investigate this by imposing both constraints simultaneously in our analysis.\footnote{Note that under the MFV assumption, there may be additional states generating contributions to the Higgs mass or the other lepton Yukawas. Since these depend on the representations chosen under the flavour group we do not include them in our tuning measure, making our analysis conservative.}
 The largest possible mass under this combined assumption for two representative SSF models is shown in  \fref{f.EWresults} (4th row), with the results for all EW Scenarios we examine summarized in 
columns 11 and 12 of \tref{t.EWmodels} for $N_{\rm BSM} = 1$ and 10 respectively.

This allows us to define, for all possible perturbative, natural and MFV-respecting solutions of the $\g$ anomaly, the  largest possible mass of the lightest BSM states:
\begin{equation}
\label{e.MchargedmaxnaturalnessMFV}
M^\mathrm{max, naturalness, MFV}_\mathrm{BSM, charged}  \ \ \ 
\equiv \ \ \ 
\max_{\tiny \deltaa = \deltaaexp \ , \  \mathrm{unitarity} \ , \ \mathrm{\Delta < 100,\ \mathrm{MFV}}} \ \  \left\{ \ \  \min_{\tiny i  \ \in \  \mathrm{BSM\ spectrum}} \left( m_{\rm charged}^{(i)}   \right)  \ \ \right\}
\end{equation}
We can perform this  maximization by taking the largest values from columns 11 and 12 in \tref{t.EWmodels}, which are shown in the last row. This gives our strongest constraint:
\begin{equation}
\label{e.MchargedmaxnaturalnessMFV2}
M^\mathrm{max, naturalness, MFV}_\mathrm{BSM, charged} 
 \ \approx \ 
  \left\{
\begin{array}{ll}
9 \tev & \mbox{for $N_{\rm BSM} = 1$}
\\
12 \tev & \mbox {for $N_{\rm BSM} = 10$}
\end{array}
\right\}
 \ \approx \ 
 (9 \tev)  \ \cdot \ N_{\rm BSM}^{1/6}\ .
\end{equation}
The $N_\mathrm{BSM}$ scaling, similar to the Naturalness-only constraint, was empirically determined and is obeyed to  very good precision for $N_\mathrm{BSM} \lesssim 100$. 
This result strongly reinforces the notion that any ``theoretically reasonable'' BSM solution to the $\g$ anomaly must give rise to charged states at or below the 10 TeV scale. 

\subsection{Electroweak Landau Poles}
\label{s.landau}

Apart from flavour and naturalness considerations, the parameter space for Electroweak Scenarios may be restricted by imposing the requirement that the SU(2)$_L$ and U(1)$_Y$ gauge couplings do not hit low-lying Landau poles. In this section, we demonstrate parametrically that such considerations disfavour truly enormous values of the BSM multiplicity $\NBSM$, which is relevant since our upper bounds on the BSM scale increase with $\NBSM$.

Since new matter of mass $\MBSM$ with electroweak charges only contributes to the running of gauge couplings at scales $\mu > \MBSM$, a muon collider which is only barely able to produce new states on-shell cannot easily probe the threshold corrections to the gauge coupling.
However, in the spirit of our flavour and naturalness discussions to find the most ``reasonably theoretically motivated'' parts of parameter space, we will impose the modest requirement that both of the electroweak gauge couplings remain finite up to a scale $\Lambda = 10 \MBSM$, where $\MBSM$ here represents the largest mass of all the new states. For this simple estimate, we set $\MBSM = 100 \tev$, inspired by the upper bounds from unitarity. We also consider the effect of avoiding Landau poles all the way up to the Planck scale. 
This allows us to obtain approximate bounds on  $\NBSM$ which depend on the electroweak representations of the new states in SSF and FFS models.

\begin{table}
\begin{center}
\hspace*{-13mm}
\footnotesize
\begin{tabular}{|l| |l|l|l| |l|l| |l|l| |l}
\hline

Model   & $R$          & $R_A$         & $R_B$      & $\NBSM$ (U(1)$_Y$) & $\NBSM$ (SU(2)$_L$) & $\bm{N}_{\rm \textbf{BSM}}$ \textbf{(min)} \\

\hline
\hline

\multirow{12}{*}{
SSF
}

   & $1 _{-1}$  & $2_{1/2}$  & $1_{0}$    & 170 (8) & 571 (60) & \textbf{170 (8)} \\
        & $1 _{-2}$  & $2_{3/2}$  & $1_{1}$     & 37 (1) & 571 (60) & \textbf{37 (1)} \\
        & $1 _{0}$   & $2_{-1/2}$ & $1_{-1}$    & 580 (28) & 571 (60) & \textbf{571 (28)} \\
        & $1 _{1}$   & $2_{-3/2}$ & $1_{-2}$    & 70 (3) & 571 (60) & \textbf{70 (3)} \\

\hhline{|~||======}

    & $2_{-1/2}$ & $1_{0}$    & $2_{-1/2}$   & 580 (28) & 114 (12) & \textbf{114 (12)} \\
        & $2_{-3/2}$ & $1_{1}$    & $2_{1/2}$   & 70 (3) & 114 (12) & \textbf{70 (3)} \\
        & $2_{1/2}$  & $1_{-1}$   & $2_{-3/2}$  & 170 (8) & 114 (12) & \textbf{114 (12)} \\

\hhline{|~||======}

    & $2_{-1/2}$ & $3_{0}$    & $2_{-1/2}$   & 580 (28) & 63 (6) & \textbf{63 (6)} \\
        & $2_{-3/2}$ & $3_{1}$    & $2_{1/2}$  & 70 (3) & 63 (6) & \textbf{63 (3)} \\
        & $2_{1/2}$  & $3_{-1}$   & $2_{-3/2}$  & 170 (8) & 63 (6) &  \textbf{63 (6)} \\

\hhline{|~||======}

    & $3_{-1}$   & $2_{1/2}$  & $3_{0}$    & 170 (8) & 27 (2) & \textbf{27 (2)} \\
        & $3_{0}$    & $2_{-1/2}$ & $3_{-1}$   & 580 (28) & 27 (2) & \textbf{27 (2)} \\

\hline 
\hline

\multirow{12}{*}{
FFS
}

    & $1 _{-1}$  & $2_{1/2}$  & $1_{0}$    & 362 (17) & 142 (15) & \textbf{142 (15)} \\
        & $1 _{-2}$  & $2_{3/2}$  & $1_{1}$    & 42 (2) & 142 (15) & \textbf{42 (2)} \\
        & $1 _{0}$   & $2_{-1/2}$ & $1_{-1}$    & 145 (7) & 142 (15) & \textbf{142 (7)} \\
        & $1 _{1}$   & $2_{-3/2}$ & $1_{-2}$   & 27 (1) & 142 (15) & \textbf{27 (1) } \\
        
\hhline{|~||======}

    & $2_{-1/2}$ & $1_{0}$    & $2_{-1/2}$  & 580 (28) & 114 (12) & \textbf{114 (12)} \\
        & $2_{-3/2}$ & $1_{1}$    & $2_{1/2}$   & 100 (4) & 114 (12) & \textbf{100 (4)} \\
        & $2_{1/2}$  & $1_{-1}$   & $2_{-3/2}$  & 54 (2) & 114 (12) & \textbf{54 (2)} \\

\hhline{|~||======}

    & $2_{-1/2}$ & $3_{0}$    & $2_{-1/2}$  & 580 (28) & 27 (2) & \textbf{27 (2)} \\
        & $2_{-3/2}$ & $3_{1}$    & $2_{1/2}$   & 100 (4) & 27 (2) & \textbf{27 (2)} \\
                & $2_{1/2}$  & $3_{-1}$   & $2_{-3/2}$  & 54 (2) & 27 (2) &  \textbf{27 (2)} \\

\hhline{|~||======}

    & $3_{-1}$   & $2_{1/2}$  & $3_{0}$   & 362 (17) & 23 (2) & \textbf{23 (2)} \\
        & $3_{0}$    & $2_{-1/2}$ & $3_{-1}$   & 145 (7) & 23 (2) & \textbf{23 (2)} \\

\hline 
\end{tabular}
\end{center}
\caption{
Approximate maximum values of $\NBSM$ for each of the models in Tab.~\ref{t.EWmodels}, obtained using Eqn.~(\ref{e.LandauB}) by requiring that each model avoid a Landau pole below 1 PeV (or the Planck scale, $M_{\rm Pl} = 2.4 \times 10^{18} \ {\rm GeV}$ in parentheses) in the hypercharge (4th column) and SU(2)$_L$ (5th column) gauge coupling. The last column is the minimum of the two $\NBSM$ values for the two EW gauge groups.
}
\label{t.LandauPoles}
\end{table}

The 1-loop SU(2)$_L$ and U(1)$_Y$ beta functions are $\beta_{Y,L} = \frac{1}{16\pi^2} b_{Y,L} \, g_{Y,L}^3$, where
\begin{align}
b_Y & = \frac{41}{6} + \frac{1}{3} \sum_S Y_S^2 + \frac{2}{3} \sum_F Y_F^2, \\
b_L & = -\frac{19}{6} + \frac{1}{3} \sum_S T(R_S) + \frac{2}{3} \sum_F T(R_F).
\end{align}
The first term in $b_Y$ and $b_L$ represents the SM contribution, and the remaining terms give the contributions from complex scalars  $S$ and 2-component fermions $F$, respectively. In $b_L$, $T(R)$ is the index of the representation, equal to $(d+1)(d)(d-1)/12$ for the $d$-dimensional representation of SU(2). A positive $b_L$ or $b_Y$ indicates a coupling which grows with increasing energy, hitting a Landau pole at the scale $\Lambda$ when
\begin{equation}
\ln \left(\frac{\Lambda_{Y,L}}{\mu}\right) = \frac{2\pi}{\alpha_{Y,L}(\mu) b_{Y,L}}.
\end{equation}
Using the measured values of the couplings at $\mu = m_Z$, evolving them with the SM beta functions up to $\mu = 100$~TeV, and imposing the absence of a Landau pole at 1 PeV (Planck scale $M_{\rm Pl} = 2.4 \times 10^{18} \ {\rm GeV}$) 
requires
\begin{equation}
b_Y < 249 \  (18.6), \qquad b_L  < 92 \ (6.9).
\label{e.LandauB}
\end{equation}
Since the BSM states do not all have the same mass, these bounds are approximate but sufficient for a useful estimate. 
Applying these constraints to the 24 models in Table~\ref{t.EWmodels}, we find the maximum values of $\NBSM$ shown in Table~\ref{t.LandauPoles}. 
The maximum allowed BSM multiplicity decreases for larger electroweak representations, with the strongest constraint being  $\NBSM \leq 27 \ (23)$ for the highest-representation SSF (FFS) models to avoid PeV-scale Landau poles. Avoiding Planck-scale Landau poles requires $\NBSM \leq 28 \ (15)$ for \emph{all} models, with the strongest constraint requiring some SSF or FFS models to have $\NBSM = 1$.

Given the very modest scaling of our mass bounds with $\NBSM$, and the severity of the Planck-scale constraints, this suggests that
\begin{equation}
N_{\rm BSM} \lesssim \mathcal{O}(10)
\end{equation}
represents the most reasonably motivated BSM parameter space.
It also justifies our choice to restrict our numerical model-exhaustive analysis of SSF/FFS models to representations up to and including triplets. Models with larger representations hit Landau poles for much lower BSM multiplicities, lowering the maximum possible BSM mass compared to models that account for the $\g$ anomaly with smaller EW representations.

\subsection{EW Scenarios with fewer than 3 new BSM states}
\label{s.edgecases}

The SSF and FFS scenarios we study are engineered to generate $\deltaaexp$ with the maximum possible masses for all the BSM particles in the $\g$ loop, which  justifies concentrating on these simplified models to determine the largest possible BSM mass scale. 
However, for the purposes of finding $M^\mathrm{max}_\mathrm{BSM, charged}$, one could imagine the following loophole to our argument: imagine replacing one of the charged BSM states by a SM particle, specifically the Higgs or the muon. 
In that case, $\deltaa$ would be generated by new diagrams involving one or more SM particles and two or fewer BSM particles in the loop. Since the charged SM particle does not count as a new discoverable charged state despite its low mass, it might be possible for the BSM charged states to  be much heavier than our  $M^\mathrm{max}_\mathrm{BSM, charged}$ upper bound. 
In this section, we show that this is not the case.

Our exhaustive analysis of SSF and FFS scenarios covers all possible EW representations that could generate new $\deltaa$ contributions (up to and including triplets). We can re-use this classification and identify scenarios where  some of the BSM scalars/fermions can be replaced by the Higgs/muon, subject to our assumption that no new significant sources of electroweak symmetry breaking are introduced (which would give rise to other experimental signatures). We categorize them as follows:
\begin{itemize}
\item FFH models, which are FFS models where $S \sim 2_{\pm {1}/{2}}$ is replaced by $H$ or $\tilde H$, where $\tilde H_i = \epsilon_{ij} H^{*j}$. 
\item $\mu$FS models, which are FFS models where $F_A^c$ or $F_B \sim 2_{-{1}/{2}}$ or $1_1$ and is therefore replaced by $L_{(\mu)}$ or $\mu^c$. In that case, no new  $F_A$ (or $F_B^c$) field is added, and there is no $y_{12}'$-type interaction. 
\item HSF models, which are SSF models where $S_A$ or $S_B \sim 2_{\pm {1}/{2}}$ is replaced by $H/\tilde H$.
\end{itemize}
Replacing the $F$ in SSF models by a muon field would require introducing a vector partner for the muon, which would introduce a new charged state at much lower masses than our $M^\mathrm{max}_\mathrm{BSM, charged}$ upper bound.  The above are therefore all the relevant modifications of the SSF/FFS models where one BSM particle is replaced by a Higgs or a muon. 
Replacing two BSM particles by SM fields is not relevant to our discussion, since 
there are no SSF (FFS) scenarios where both scalars (fermions) have the correct EW representation to be replaced by the Higgs doublet (muon spinors). One could consider replacing one BSM fermion and one scalar by the Higgs and muon respectively in the FFS scenario, but this would identify one of the $y_{1,2}$ couplings with the small muon Yukawa, suppressing $\deltaa$ and guaranteeing a small BSM mass scale. To ensure that our derivation of $M^\mathrm{max}_\mathrm{BSM, charged}$ is correct, we therefore only have to consider the FFH, $\mu$FS and HSF cases.

We systematically explored the entire allowed parameter space of all 3 possible FFH scenarios, 5 $\mu$FS scenarios and  5 HSF scenarios, for $N_{BSM} \geq 1$. None of them give rise to larger charged particle masses than the full SSF/FFS scenarios, meaning they have no bearing on the $M^\mathrm{max}_\mathrm{BSM, charged}$ upper bounds derived above. 
In most cases, it is easy to understand why this is the case. 

In $\mu$FS models, there are two new $\deltaa$ contributions: one with the muon-dominated mass eigenstate in the loop, and one with the new heavy fermion in the loop. Only the latter is chirally enhanced by the large BSM fermion mass $m_F$, but since there is no $y_{12}'$ coupling, it is suppressed by a very small mixing $\sim v y_{12} m_\mu/m_F^2$, effectively reintroducing the same parametric suppression by $m_\mu$ as in Singlet Scenarios. Therefore, $\mu$FS models that account for the $\g$ anomaly and respect perturbative unitarity always require new charged states below a few TeV. 

HSF models are most easily analyzed in Feynman-t'Hooft gauge, where the charged and neutral Higgs goldstone modes are kept in the spectrum with masses $m_W, m_Z$ respectively. This allows our calculations of $\g$ and radiative corrections to be applied almost verbatim. 
The unitarity limit on the $\kappa$-type coupling, see Eqn.~\ref{e.kappamax}, now becomes $\kappa_{max} \sim 0.3 m_S$, where $m_S$ is the mass of the BSM scalar.\footnote{The $\kappa H H S$-type coupling also leads to a tiny vev for the BSM scalar, but since its EWSB contribution is aligned with the Higgs in cases where $S$ carries EW charge, it does not meaningfully affect our discussion.}
In HSF scenarios where the scalar is a SM singlet, this ensures that the charged fermion cannot be made so heavy as to violate our $M^\mathrm{max}_\mathrm{BSM, charged}$ upper bound: generating $\deltaaexp$ with a very heavy charged fermion mass would require a relatively light SM singlet scalar, but such low values of $m_S$ forbid the $\kappa$ couplings required to generate $\deltaaexp$. 
We therefore find that all HSF models require new charged states below 25 TeV from perturbative unitarity alone, and much lower masses once MFV assumptions are included. 
HSF models also contain additional large radiative corrections to the Higgs mass $\sim y_1^2 m_F^2/8 \pi^2$. This makes naturalness constraints even more severe than in regular SSF models, requiring new charged states far below our calculated upper bound for all assumptions.

Finally, we discuss the FFH models, which introduce no parametric suppressions for $\deltaa$. There are two cases where both vector-like BSM fermions carry EW charge: $(F_A, F_B) \sim (3_0, 2_{-1/2})$ or $(1_{-1}, 2_{-3/2})$. For unitarity-only or unitarity + MFV assumptions, we find that the upper bound on the lighter charged particle mass is almost the same as for the corresponding FFS model. On the other hand, any naturalness constraint leads to much lower allowed charged masses, since like HSF models, FFH models include additional large finite Higgs mass contributions $(y_1^2 m_A^2 + y_2^2 m_B^2)/8 \pi^2$. Any such models obeying our naturalness criterion  hence require new charged states below a few TeV. Therefore, these FFH scenarios can never violate our $M^\mathrm{max}_\mathrm{BSM, charged}$  upper bounds. 

The one case that requires further discussion is the FFH model with $(F_A, F_B)  \sim (1_0, 2_{-1/2})$. Because $F_A$ is a SM singlet, the only  BSM charged state is $F_B$. 
Imposing the naturalness criterion for even just the Higgs mass still guarantees $m_B \lesssim \mathcal{O}(\tev)$,  well within our upper bound. However, if we do not impose the naturalness constraint, it is naively possible to generate $\deltaaexp$ for a relatively light singlet $F_A$, large $y_1, y_1, y_{12}$ couplings ($y_{12}'$ does not contribute in this limit), and very heavy charged $F_B$, driving up the maximum allowed charged mass to $\mathcal{O}(1000) \tev$ and $\mathcal{O}(100) \tev$ for the unitarity and unitarity  + MFV assumptions respectively. 
Fortunately, such an extreme scenario violates electroweak precision constraints. 
This FFH model contains the coupling
\begin{equation}
-y_1 F_A^c \nu_\mu H^0
\end{equation}
which does not contribute to $\g$ directly but is a requirement of $SU(2)_L \times U(1)_Y$ gauge invariance. This gives rise to a mixing between the active muon neutrino and the heavy sterile $F_A$ fermion $\theta_{\nu_\mu A} \sim y_1 v /m_A$, generating deviations from SM predictions for the $Z\nu\nu$ coupling, see e.g.~\cite{Bolton:2019pcu}. Imposing the constraints $|V_{\mu N}|^2 \lesssim 10^{-3}$ for the unitarity assumption and $|V_{\tau N}|^2 \lesssim 10^{-2}$ for the unitarity + MFV assumption forbids the extreme case of very light $F_A$ and very large $y_1$ which would permit a very heavy charged $F_B$ mass. Including electroweak precision constraints, the heaviest possible charged mass for the unitarity (unitarity + MFV) assumption in this FFH model is smaller than 65 TeV (10 TeV) for $\NBSM = 1$. Larger values of $\NBSM$ also do not violate our upper bound. 

In summary, the $M^\mathrm{max}_\mathrm{BSM, charged}$ upper bounds we calculate using the FFS and SSF simplified models also apply to scenarios with fewer BSM fields in non-trivial EW representations, and hence to all possible EW scenarios in general.

\subsection{Muon Collider Signatures}
\label{s.ewMuCsignals}

We focus on the simplest and most robust signature of EW Scenarios at muon colliders: direct production of new heavy charged states. 
Such a state $X$ would be pair-produced in Drell-Yan processes independent of its direct couplings to muons, with a pair production cross section similar to SM EW $2\to 2$ processes above threshold, $\sigma_{XX} \sim \mathrm{fb} \, (10 \tev/\sqrt{s})^2$~\cite{Delahaye:2019omf}, as long as $\sqrt{s} > 2 m_X$.
At high energies far above a TeV, the same is true of electrically neutral states carrying weak quantum numbers, which are also present in EW Scenarios.
 However, charged states must either decay to visible SM final states, or are themselves visible if they are detector-stable. As a result, conclusive discovery of such heavy states should be possible in the clean environment and known center-of-mass-frame of a lepton collider regardless of their detailed phenomenology.\footnote{Note that the large Drell-Yan cross sections imply that a discovery is possible even at a considerably lower luminosity than ab$^{-1}$, which may provide some practical advantages.}

In the discussions of the next section, we can therefore simply assume a muon collider will be able to discover any heavy BSM charged state with $m_X \lesssim \frac{1}{2} \sqrt{s}$. As we have seen, for reasonable BSM solutions  to the $\g$ anomaly, this will call for an $\mathcal{O}(10\tev)$ muon collider (or an electron collider, if it could be built at such high energies).

The complications particular to a muon collider, like the shielding cone necessary to reduce beam-induced background, do not affect this argument for heavy charged states.  
Of course, it is always possible to imagine very unusual scenarios where details of the model conspire to make discovery much harder than generically expected. However, such edge cases do not invalidate a no-lose theorem. 
For example, while models that could hide the Higgs boson at the LHC were certainly considered prior to its discovery (see e.g.~\cite{Falkowski:2010cm}), this did not invalidate the fact that the combination of EWSB and basic unitarity requires the production of new states at the LHC. Indeed, if such a scenario had come to pass, the no-lose theorem for the Higgs would have motivated herculean analysis efforts to tease the hidden signals out of the data. 
(Furthermore, production and observation of new charged states via gauge couplings is much more robust than production of neutral scalars.)
Our no-lose theorem serves a similar function: it motivates the construction of colliders that can produce the predicted new charged states, and in case those states are not found right away, it will hopefully provide similar emotional fortification for future experimentalists looking to uncover the new physics behind the by then well-established $\g$ anomaly.

In some EW Scenarios there is an electrically neutral or even complete SM singlet state that is lighter than the lightest charged state. As we discussed in our first study~\cite{Capdevilla:2020qel}, this can also be discovered in a mono-photon search if the new state escapes as missing energy, where VBF-enhanced SM backgrounds can be effectively vetoed with a high-momentum-cut on the recoiling photon. However, while this signature is interesting in its own right, it is not our focus in this study. Across the whole space of possible EW Scenarios and hence all theories that solve the $\g$ anomaly, assuming that all kinematically accessible BSM states can be discovered versus only assuming that charged states can be discovered does not actually lower the resulting minimum required energy of the muon collider necessary to guarantee discovery of new physics. We can therefore focus on charged BSM states without being unduly conservative.

\section{No-Lose Theorem for the Muon Collider Program}
\label{s.implications}

We now synthesize the results of our model-exhaustive analysis to understand the concrete implications for a future muon collider program, and use them to derive our no-lose theorem for the discovery of new physics. 

One-loop perturbative solutions to the $\g$ anomaly can be classified as either Singlet Scenarios or EW Scenarios, based simply on whether the new physics contributions in the loop are only SM singlets or if there are any particles with SM gauge quantum numbers.
Direct discovery of Singlet Scenarios requires observation of the SM singlet, while EW Scenarios can be discovered by producing the lightest new charged state at lepton colliders.

BSM theories that only generate $\g$ at higher-loop order necessarily feature lower
mass scales relative to those found in one-loop models and are thus easier to discover.
Furthermore,
 strongly coupled BSM scenarios involving composite new states in the $\g$ loop are parametrically covered by our analysis, since we consider BSM multiplicity of states $N_{\rm BSM} > 1$ and large couplings at the unitarity limit.\footnote{While we considered large BSM couplings that are borderline non-perturbative to derive upper bounds on new particle masses,  the existence and production of the new EW states at colliders is a consequence of gauge invariance and only involves perturbative couplings, making our signal predictions robust.}

 If  Singlet Scenarios explain the $\g$ anomaly, the maximum possible mass of BSM states based on perturbative unitarity only is 3 TeV, and only 200 GeV  if we impose MFV, as  motivated by CLFV decay bounds.
We performed a careful analysis of direct singlet production at muon colliders via the same coupling that generates $\deltaa$, which is completely inclusive with respect to the singlet stability or decay mode. 
We find that a 3 TeV muon collider with $1 \ \mathrm{ab}^{-1}$ integrated luminosity would be able to discover all Singlet Scenarios that solve the $\g$ anomaly, provided the mass of the singlet is larger than $\sim 10 \gev$. 
A 215 GeV muon collider with $0.4 \ \mathrm{ab}^{-1}$ would not be able to probe the highest possible singlet masses, but could discover singlets heavier than 2 GeV. However, such a lower-energy muon collider would also be able to observe deviations in Bhabha scattering $\mu^+ \mu^- \to \mu^+ \mu^-$ at the 5$\sigma$ level to indirectly discover the effects of these singlets with masses as high as the unitarity limit. 
These results are independent of $N_{\rm BSM}$ because all observables scale with  $g_{\rm BSM} N_{\rm BSM}$, the same combination of parameters that determines $\deltaa$.

On the other hand, EW Scenarios are the most general way to solve the $\g$ anomaly at one-loop, hence resulting in much higher possible BSM mass scales. %
We defined the following highest possible mass for the lightest BSM charged state in the spectrum:
\begin{equation}
\label{e.MchargedmaxX}
M^\mathrm{max, X}_\mathrm{BSM, charged}  \ \ \ 
\equiv \ \ \ 
\max_{\tiny \deltaa = \deltaaexp \ , \  X} \ \  \left\{ \ \  \min_{\tiny i  \ \in \  \mathrm{BSM\ spectrum}} \left( m_{\rm charged}^{(i)}   \right)  \ \ \right\} \ .
\end{equation}
The outer $\max$ represents a maximization over theory space subject to assumptions $X$, where we examined four possibilities:
\begin{equation}
X  \ \ = \ \  \left\{
\begin{array}{l}
\mbox{perturbative unitarity*} 
\\
\mbox{unitarity + MFV} 
\\
 \mbox{unitarity + naturalness*} 
 \\
 \mbox{unitarity + naturalness + MFV}
 \end{array}
 \right. \ .
\end{equation}
The last three assumptions include perturbative unitarity but are more restrictive.
 MFV avoids CLFV decay bounds
 and assumes that the SM Yukawas are the only source of flavour violation in whatever new physics solves the flavour puzzle, which lowers the unitarity bound on some of the BSM muon couplings, since the corresponding BSM tau coupling must obey perturbative unitarity.
  Naturalness is defined to require that both the muon and Higgs mass, which both become technically unnatural in EW Scenarios due to calculable new loop corrections, are tuned to no more than 1\%. 
  The star (*) indicates that assumptions without MFV implicitly rely on some coincidence or unknown mechanism to suppress CLFVs while allowing the muonic BSM couplings to be pushed up to the unitarity (or naturalness) limit.

We can perform this theory space maximization using our SSF and FFS simplified models to obtain the highest possible mass of the lightest new charged state as a consequence of resolving the $\g$ anomaly:
\begin{equation}
\label{e.MchargedmaxXresult}
M^\mathrm{max, X}_\mathrm{BSM, charged} 
\approx
\left( \frac{2.8 \times 10^{-9}}{\deltaaexp} \right)^{\frac{1}{2}} \times
\left\{
\begin{array}{rcl}
(100 \tev)  \ N_{\rm BSM}^{1/2} &   \mathrm{for}  & X = \mbox{(unitarity*)}
\\ \\ 
(20 \tev)  \ N_{\rm BSM}^{1/2} &   \mathrm{for} & X = \mbox{(unitarity+MFV)} 
\\ \\ 
(20 \tev)  \ N_{\rm BSM}^{1/6} & \mathrm{for}    & X =  \mbox{(unitarity+naturalness*)}
\\ \\ 
(9 \tev)  \ N_{\rm BSM}^{1/6} &  \mathrm{for} & X =  \mbox{(unitarity+naturalness+MFV)}
\end{array}
\right.
\end{equation}
We include the scaling of these mass bounds with $\deltaaexp$ so they can be easily adapted to updated measurements of $\g$.\footnote{The dependence of the naturalness bounds on SM masses could make this scaling less than completely trivial, but we have verified that it holds within a factor of a few of the BNL measurement \eref{delta-amu}.}
The presence of required CLFV suppression is  again indicated with a star. In light of CLFV decay bounds, the two MFV results are the most theoretically and experimentally motivated.
Furthermore, avoiding relatively low-lying Landau poles motivates $N_{\rm BSM} \lesssim \mathcal{O}(10)$.

Since charged states of mass $m$ are efficiently produced by a lepton collider with $\sqrt{s} \gtrsim 2 m$ and have to leave visible signals in the detector, we assume that any such BSM state would be discovered at a sufficiently energetic muon collider. 
Specifically, a $\sqrt{s} \sim 30 \tev$ muon collider would be able to discover any high-scale, MFV-respecting solution to the $\g$ anomaly that avoids introducing two new hierarchy problems and has BSM multiplicity up to  $N_{\rm BSM} \lesssim 10$.
Such a collider would also be able to indirectly confirm the existence of the effective BSM operator responsible for generating $\deltaa$ via $h \gamma$ measurements~\cite{Buttazzo:2020eyl, Yin:2020afe}. 
This makes a 30 TeV muon collider a highly ambitious but highly motivated benchmark goal for the discovery of new physics.

High-scale solutions to the $\g$ anomaly which evade discovery at a 30 TeV machine are extremely strange: they would have to have a high BSM multiplicity, resulting in possible Landau poles below the Planck or even the PeV scale; or violate the assumptions of MFV while avoiding CLFV decay bounds; or be highly tuned in an explicitly calculable way.
Therefore, non-observation of new states at a 30 TeV muon collider (alongside confirmation of the new BSM operator via $h \gamma$ measurement) would force the $\g$ solution into theoretically extreme territory, which still has to satisfy the bounds of unitarity with charged states below a few hundred TeV. 
Such a scenario would constitute empirical proof that nature is fine-tuned, and/or refute the MFV ansatz for the solution of the flavour puzzle, which would now be much more severe since unknown mechanisms have to suppress naively large CLFV contributions. 
This in itself would be highly meaningful and new information about the fundamental nature of our universe, the selection of its vacuum, and the origin of flavour.

These results allow us to formulate the no-lose theorem for future muon colliders, which we already stated in \sref{s.intro}, but we repeat the chronological progression here for completeness:
\begin{enumerate}

\item \textbf{Present day confirmation:} 

Assume the $\g$ anomaly is real.

\item \textbf{Discover or falsify low-scale Singlet Scenarios $\mathbf{\lesssim}$ GeV:}

 If Singlet Scenarios with BSM masses below $\sim \gev$ generate the required $\deltaaexp$ contribution~\cite{Pospelov_2009}, multiple fixed-target and $B$-factory experiments are projected to discover new physics in the coming decade \cite{Gninenko_2015,Chen_2017,Kahn_2018,kesson2018light,Berlin_2019,Tsai:2019mtm,Ballett_2019,Mohlabeng_2019,Krnjaic_2020}. 

\item \textbf{Discover or falsify all Singlet Scenarios $\mathbf{\lesssim}$ TeV:} 

If fixed-target experiments do not discover new BSM singlets that account for $\deltaaexp$, a 3 TeV muon collider with $1~\mathrm{ab}^{-1}$ would be guaranteed to directly discover these singlets if they are heavier than $\sim 10 \gev$.

Even a lower-energy machine can be useful: a  215 GeV  muon collider with $0.4~\mathrm{ab}^{-1}$ could directly observe singlets as light as 2 GeV under the conservative assumptions of our inclusive analysis, while indirectly observing the effects of the singlets for all allowed masses via Bhabha scattering. 

Importantly, for singlet solutions to the $\g$ anomaly, only the muon collider is guaranteed to discover these signals since the only required new coupling is to the muon.

\item   \textbf{Discover non-pathological Electroweak Scenarios ($\mathbf\lesssim$  10 TeV):}

If TeV-scale muon colliders do not discover new physics, the $\g$ anomaly \emph{must} be generated by EW Scenarios. 
In that case, all of our results indicate that in most reasonably motivated scenarios, the mass of new charged states cannot be higher than few $\times$ 10 TeV. However, such high masses are only realized by the most extreme boundary cases we consider. 
Therefore, a muon collider with $\sqrt{s} \sim 10 \tev$ is incredibly motivated, since it will have excellent coverage for EW Scenarios in most of their reasonable parameter space.

A very strong statement can be made for future muon colliders with $\sqrt{s} \sim 30 \tev$: such a machine can discover via pair production of heavy new charged states \emph{all} EW Scenarios that avoid CLFV bounds by satisfying MFV and avoid generating two new hierarchy problems, with $N_{\rm BSM}  \lesssim 10$.

\item \textbf{Unitarity Ceiling ($\mathbf\lesssim$  100 TeV):}

Even if such a high energy muon collider does not produce new BSM states directly, the recent investigations by~\cite{Buttazzo:2020eyl, Yin:2020afe} show that a 30 TeV machine would detect deviations in $\mu^+ \mu^- \to h \gamma$, which probes the same effective operator generating $\g$ at lower energies. This would provide high-energy confirmation of the presence of new physics.

In that case,  our results guarantee the presence of new states below $\sim 100 \tev$ by perturbative unitarity, and the lack of direct BSM particle production at $\sqrt{s} \sim 30 \tev$ will prove that the universe violates MFV and/or is highly fine-tuned to stabilize the Higgs mass and muon mass, all while suppressing CLFV processes. 
\end{enumerate}
As we already argued in \sref{s.intro}, if the $\g$ anomaly is confirmed, this should serve as supremely powerful motivation for an ambitious muon collider program, from the test-bed or Higgs-factory scale of $\mathcal{O}(100 \gev)$ to energies in excess of 10 TeV.
It would of course also be interesting to understand if and how proposed future hadron or electron colliders could explore the same physics.


\vspace{5mm}
\textbf{Acknowledgements:} 
We thank Pouya Asadi, Jared Barron, Brian  Batell, Nikita Blinov, 
Ayres Freitas, Chris Tully,
Aida El-Khadra,
Tao Han,
Shirley Li, Patrick Meade, Federico Meloni, 
Jessie Shelton, Raman Sundrum, and Jos\'{e} Zurita
for helpful conversations.
We are also grateful to Radovan Dermisek for discussions that prompted us to carefully examine EW Scenarios with just two new BSM particles. 
The research of RC and DC was supported in part by a Discovery Grant
from the Natural Sciences and Engineering Research Council of Canada, and by the Canada
Research Chair program.
The work of RC was supported in part by the Perimeter Institute for Theoretical Physics (PI). Research at PI is supported in part by
the Government of Canada through the Department of
Innovation, Science and Economic Development Canada
and by the Province of Ontario through the Ministry of
Colleges and Universities.
The work of YK was supported in part by US Department of Energy grant DE-SC0015655. 
 This manuscript has been authored by Fermi Research Alliance, LLC under Contract No. DE-AC02-07CH11359
with the U.S. Department of Energy, Office of High Energy Physics.

\bibliographystyle{JHEP}
\bibliography{References}

\end{document}